\RequirePackage{ifpdf}
\documentclass[12pt,letterpaper]{JHEP3}
\usepackage{amssymb,amsfonts}
\usepackage{amsmath}

\usepackage{epsfig}

\newcommand{\nn}{\nonumber}
\newcommand{\bec}{\begin{center}}
\newcommand{\eec}{\end{center}}
\newcommand{\beq}{\begin{equation}}
\newcommand{\eeq}{\end{equation}}
\newcommand{\bea}{\begin{eqnarray}}
\newcommand{\eea}{\end{eqnarray}}

\newcommand{\hf}{\frac{1}{2}}
\newcommand{\qtr}{\frac{1}{4}}

\newcommand{\bx}{{\bf x}}
\newcommand{\by}{{\bf y}}
\newcommand{\bp}{{\bf p}}
\newcommand{\bq}{{\bf q}}

\newcommand{\psib}{\overline{\psi}}
\newcommand{\Qb}{\overline{Q}}
\usepackage{listings}
\usepackage{slashed}

\title{Markov Chain Monte Carlo Methods in Quantum Field Theories: A Modern Primer}

\author{Anosh Joseph \\ 
Department of Physical Sciences \\ 
Indian Institute of Science Education and Research (IISER) Mohali \\ 
Knowledge City, Sector 81 \\ 
SAS Nagar, Punjab 140306, India \\

Email: \email{anoshjoseph@iisermohali.ac.in}

\vspace{7cm}

These lecture notes are based on the three lectures given at the {\bf 2019 Joburg School in Theoretical Physics: Aspects of Machine Learning}, Mandelstam Institute for Theoretical Physics, The University of the Witwatersrand, Johannesburg, South Africa (November 11 - 15, 2019). These lecture notes have been published as part of the SpringerBriefs in Physics book series by Springer Nature. DOI: https://doi.org/10.1007/978-3-030-46044-0.}
\preprint{December 2019}

\abstract{

\vspace{0.5cm}

We introduce and discuss Monte Carlo methods in quantum field theories. Methods of independent Monte Carlo, such as random sampling and importance sampling, and methods of dependent Monte Carlo, such as Metropolis sampling and Hamiltonian Monte Carlo, are introduced. We review the underlying theoretical foundations of Markov chain Monte Carlo. We provide several examples of Monte Carlo simulations, including one-dimensional simple harmonic oscillator, unitary matrix model exhibiting Gross-Witten-Wadia transition and a supersymmetric model exhibiting dynamical supersymmetry breaking.}

\begin{document}

\newpage

\subsection*{\center{{\bf Acknowledgements}}}

{\ }\\
I would like to thank the organizers, the lecturers and the students of the {\it 2019 Joburg School in Theoretical Physics: Aspects of Machine Learning}, at the Mandelstam Institute for Theoretical Physics, The University of the Witwatersrand, Johannesburg, South Africa (November 11 - 15, 2019), for an inspiring atmosphere at the School. I would also like to thank Pallab Basu, Dimitrios Giataganas, Vishnu Jejjala, Robert de Mello Koch, Joao Rodrigues, Jonathan Shock, Giacomo Torlai and Konstantinos Zoubos for many useful discussions. 

I am indebted to my students 
Navdeep Singh Dhindsa, 
Roshan Kaundinya,
Arpith Kumar, and 
Vamika Longia 
for providing valuable suggestions on improving the manuscript.

The work on these lecture notes was supported in part by the Start-up Research Grant (No. SRG/2019/002035) from the Science and Engineering Research Board (SERB), Government of India, and in part by a Seed Grant from the Indian Institute of Science Education and Research (IISER) Mohali.    

\newpage

\section{Introduction}

Quantum field theory is a tool to understand a vast array of perturbative and non-perturbative phenomena found in physical systems. Some of the most interesting features of quantum field theories, such as spontaneous symmetry breaking, phase transitions, and bound states of particles, demand computational tools beyond the machinery of ordinary perturbation theory. Monte Carlo methods using Markov chain based sampling algorithms provide powerful tools for carrying out such explorations. 

We can use lattice regularized quantum field theories and simulation algorithms based on Monte Carlo methods to reveal the non-perturbative structure of many interesting quantum field theories, including Quantum Chromodynamics (QCD), the theory of strong interactions. The rapidly developing field of Machine Learning could provide novel tools to find phase structures and order parameters of systems where they are hard to identify. 

These lecture notes contain ten sections. In Section \ref{sec:Monte-Carlo-method-for-integration} we discuss various simple methods of numerical integration, including the rectangle rule, midpoint rule, trapezoidal rule and Simpson's rule. Random numbers are introduced next. We discuss pseudo-random numbers and how they can be generated on a computer using a seed number. After that we move on to discuss Monte Carlo method for numerical integration. We also discuss how to compute the error in Monte Carlo integration, the questions on when Monte Carlo is useful for integration and when it can fail. In Section \ref{sec:Monte-Carlo-with-importance-sampling} we discuss Monte Carlo with importance sampling and how it reduces the variance of the Monte Carlo estimate of the given integral. In Section \ref{sec:Markov-chains} we introduce Markov chains and discuss their properties and convergence to the unique equilibrium distribution when the chain is irreducible and aperiodic. In Section \ref{sec:Markov-chain-Monte-Carlo} we introduce Markov chain Monte Carlo. Concepts such as Metropolis algorithm and thermalization of Markov chains are introduced. In Section \ref{sec:Monte-Carlo-and-Feynman-path-integrals} we discuss the connection between Markov chain Monte Carlo and Feynman path integrals of Euclidean quantum field theories. We also numerically study a zero-dimensional quantum field theory that undergoes dynamical supersymmetry breaking, one-dimensional simple harmonic oscillator, and a unitary matrix model that undergoes Gross-Witten-Wadia phase transition. In Section \ref{sec:Reliability-of-simulations} we discuss the reliability of Monte Carlo simulations and introduce the idea of auto-correlation time in the observables. The method of Hybrid (Hamiltonian) Monte Carlo is discussed next in Section \ref{sec:Hybrid-(Hamiltonian)-Monte-Carlo}. There, we look at the properties of Hamiltonian dynamics and how Leapfrog integration method can be used to evolve the system in simulation time. We then apply Hamiltonian Monte Carlo to a Gaussian model and a zero-dimensional supersymmetric model. In Section \ref{sec:MCMC-and-quantum-field-theories-on-a-lattice} we briefly discuss how Markov chain Monte Carlo can be used to extract physics from quantum field theories formulated on a spacetime lattice. In Section \ref{sec:Machine-learning-and-QFT} we discuss how Machine Learning and quantum field theory can work together to further our understanding of the nature of the physical systems we are interested in. These lecture notes end with several appendices containing various C++ programs that were used to generate data and numerical results provided in various sections. 

\section{Monte Carlo Method for Integration}
\label{sec:Monte-Carlo-method-for-integration}

This Section is a brief introduction to numerical method for integration. We start with the traditional deterministic rules for numerical integration, based on Newton-Cotes quadrature formulas. After that random numbers are introduced. We then discuss briefly the Monte Carlo method of integration, based on independent sampling. At the end of the Section we address two questions on when Monte Carlo method is a good method for integration, and the situations in which this method can fail. 

\subsection{Numerical Integration}
\label{sec:Numerical-integration}

In many places, we encounter situations where analytical methods fail to compute the values of integrals. The method of numerical integration may be used in such situations to compute the integrals reliably. The term numerical integration consists of a broad family of algorithms for computing numerical values of definite integrals. If we consider a smooth function $f(x)$, of one variable $x$, in the interval $[x_i, x_f]$, then the goal of numerical integration is to approximate the solution to the definite integral
\beq
I = \int_{x_i}^{x_f} f(x) \;dx,
\eeq
to a given degree of accuracy. We can also generalize the concept of numerical integration to several variables.

There exist several methods for approximating the integral in question to a desired precision. They belong to a class of formulas known as {\it Newton-Cotes quadrature formulas}\footnote{They are named after Sir Isaac Newton (1642 - 1727) and Roger Cotes (1682 - 1716). Numerical integration in one dimension is referred to as quadrature.}.

The simplest of Newton-Cotes quadrature formulas is obtained by considering the function $f(x)$ as a constant within the given interval $[x_i, x_f]$. This results in the numerical integration formula
\beq
\label{eq:rectangle-rule}
I = \int_{x_i}^{x_f} f(x) \;dx \simeq \left( x_f - x_i \right) f(x_i) + \hf f'(\eta) \left( x_f - x_i \right)^2,
\eeq
where the last term denotes the truncation error, and $\eta$ is a real number, $x_i < \eta < x_f$.

This approximation is referred to as the {\it rectangle rule}.

If we choose $x_m = (x_i + x_f)/2$, which is the midpoint of the interval $[x_i, x_f]$, we get
\beq
\label{eq:mid-point-rule}
I = \int_{x_i}^{x_f} f(x) \;dx \simeq \left( x_f - x_i \right) ~ f(x_m) + \frac{1}{24} f''(\eta) \left( x_f - x_i \right)^3.
\eeq 

This gives the {\it midpoint rule}.

\begin{figure}[t]
\centering
\includegraphics[width=9cm]{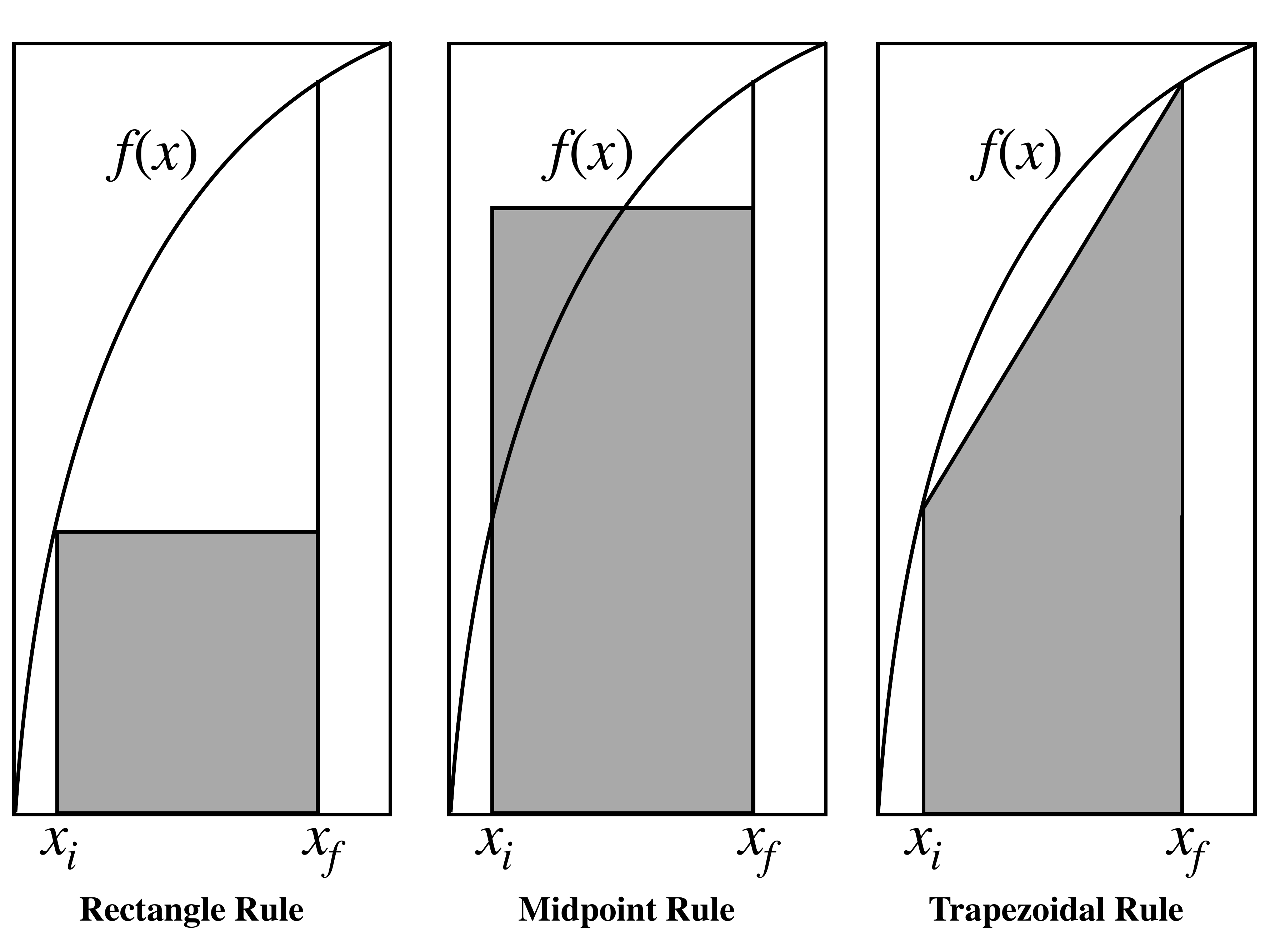}
\caption{The three basic rules for numerical integration - midpoint rule, rectangle rule and trapezoidal rule.}
\label{fig:r_m_t_rules}
\end{figure}

From Fig. \ref{fig:r_m_t_rules}, we see that the midpoint rule is more accurate than the rectangle rule. The extra area included in the rectangle compensates for the area not included, to some extent, in the midpoint rule.

Let us approximate the function $f(x)$ by a straight line passing through the two end points $x_i$ and $x_f$. We get the following approximation to the integral
\bea
\label{eq:trapezoidal-rule}
I &=& \int_{x_i}^{x_f} f(x) \;dx \nn \\
&\simeq& \hf (x_f - x_i) \left[ f(x_i) + f(x_f) \right] - \frac{1}{12} f''(\eta) \left( x_f - x_i \right)^3.
\eea

This formula is referred to as the {\it trapezoidal rule}, since the integral is approximated by a trapezium. 

In Fig. \ref{fig:r_m_t_rules} we show the rectangle rule, mid-point rule and trapezoidal rule applied for a function say, $f(x)$.

The integration formula with one more level of sophistication, and thus leading to an approximation to the integral with more accuracy, is the three-point Newton-Cotes quadrature rule or the {\it Simpson's rule}\footnote{This rule is named after the mathematician Thomas Simpson (1710-1761). Johannes Kepler (1571 - 1630) used similar formulas over 100 years prior, and for this reason, this rule is also referred to as Kepler's rule.}.

In Simpson's rule we approximate function $f(x)$ in the interval $[x_i, x_f]$ using three equidistant interpolation points $\left( x_i, x_m, x_f \right)$, where $x_m = (x_i + x_f)/2$ is the midpoint. This leads to the following approximation to the integral
\bea
\label{eq:Simpsons}
I &=& \int_{x_i}^{x_f} f(x) \;dx \nn \\
&\simeq& \frac{1}{6} (x_f - x_i) \left[ f(x_i) + 4 f(x_m) + f(x_f) \right] - \frac{1}{2880} ~f^{(4)}(\eta)~(x_f - x_i)^5.
\eea

\begin{figure}[t]
\centering
  \includegraphics[width=9cm]{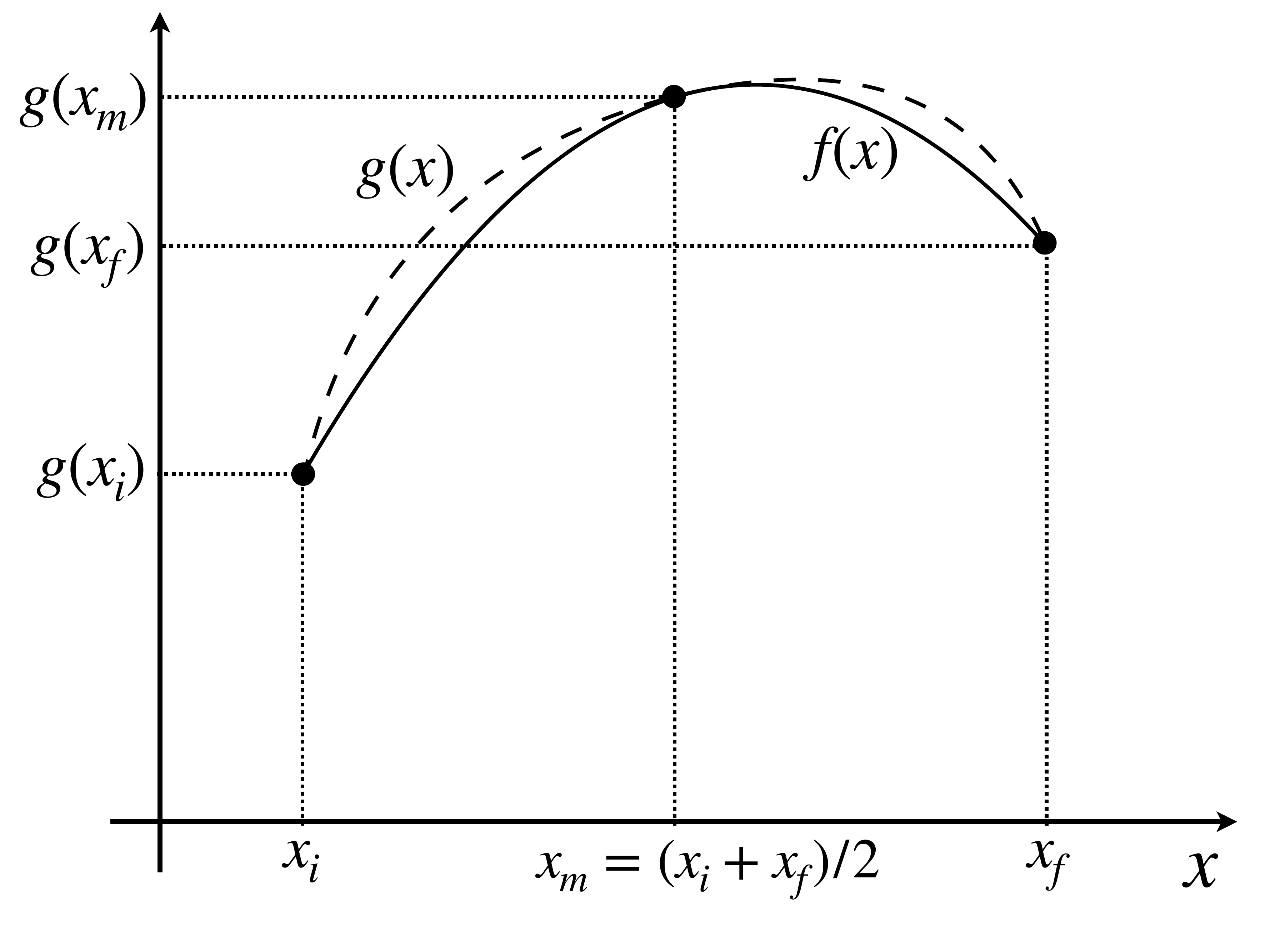}
  \caption{\label{fig:simpsons-one-third}Simpson's rule for numerical integration. In this rule a quadratic interpolant $g(x)$ can be used to approximate the function $f(x)$.}
\end{figure}

This rule is also sometimes referred to as {\it Simpson's one-third rule}\footnote{The factor one-third appears in the first term if we introduce a step size $h = (x_f - x_i)/2$.}. In Fig. \ref{fig:simpsons-one-third} we show an illustration of this rule.

\subsection{Composite Formulas for Numerical Integration}
\label{sec:Composite-formulas-for-numerical-integration}

In order to achieve better accuracy in numerical integration we almost always consider breaking up the integral into several parts. That is,
\beq
\int_{x_i}^{x_f} f(x) \; dx = \sum_{r = 1}^m \int_{x_{r - 1}}^{x_r} f(x) \; dx,
\eeq
with $x_r$ say, equally spaced, $x_r = x_i + r h$, $h = (x_f - x_i)/m$, $x_0 = x_i$ and $x_m = x_f$. After this, we can apply the quadrature formulas on each of these $m$ sub-intervals. The resulting formulas are known as {\it composite formulas} or {\it compound rules}.

\subsubsection{Composite Rectangle Rule}
\label{sec:Composite-rectangle-rule}

From the rectangle rule given in Eq. \eqref{eq:rectangle-rule} we can construct a composite formula. Let us approximate $f(x)$ by a piecewise constant step function, with a jump at each point $x_r = x_i + r h$, $r = 1, 2, \cdots, m$.

This leads to the formula for composite rectangle rule 
\beq
\int_{x_i}^{x_f} f(x) \; dx \simeq h \sum_{r = 1}^m ~ f\left( x_r \right) + \hf h f'(\eta) (x_f - x_i).
\eeq

Since the error term is proportional to $h$, the composite rectangle rule is first order accurate.

\subsubsection{Composite Midpoint Rule}
\label{sec:Composite-midpoint-rule}

From the midpoint rule given in Eq. \eqref{eq:mid-point-rule} we can construct a composite formula. We approximate $f(x)$ by a piecewise constant step function, with a jump at each point $x_i + r h$, $r = 1, 2, \cdots, m$.

Defining $x_r \equiv x_i + (r - \hf) h$, we have
\beq
\int_{x_i}^{x_f} f(x) \; dx \simeq h \sum_{r = 1}^m ~ f\left( x_r \right) + \frac{1}{24} h^2 f''(\eta) (x_f - x_i).
\eeq

Thus composite midpoint rule is second order accurate.

\subsubsection{Composite Trapezoidal Rule}
\label{sec:Composite-trapezoidal-rule}

We can apply the trapezoidal rule Eq. \eqref{eq:trapezoidal-rule} on each $m$ sub-interval, to get the composite trapezoidal rule
\beq
\int_{x_i}^{x_f} f(x) \; dx = \hf h \left( f(x_i) + 2 \sum_{r = 1}^{m - 1} f(x_r) + f(x_f) \right) - \frac{1}{12} h^2 f''(\eta) (x_f - x_i),
\eeq
where $h = (x_f - x_i)/m$. Thus composite trapezoidal rule is also second order accurate.

\begin{figure}[t]
\centering
  \includegraphics[width=9cm]{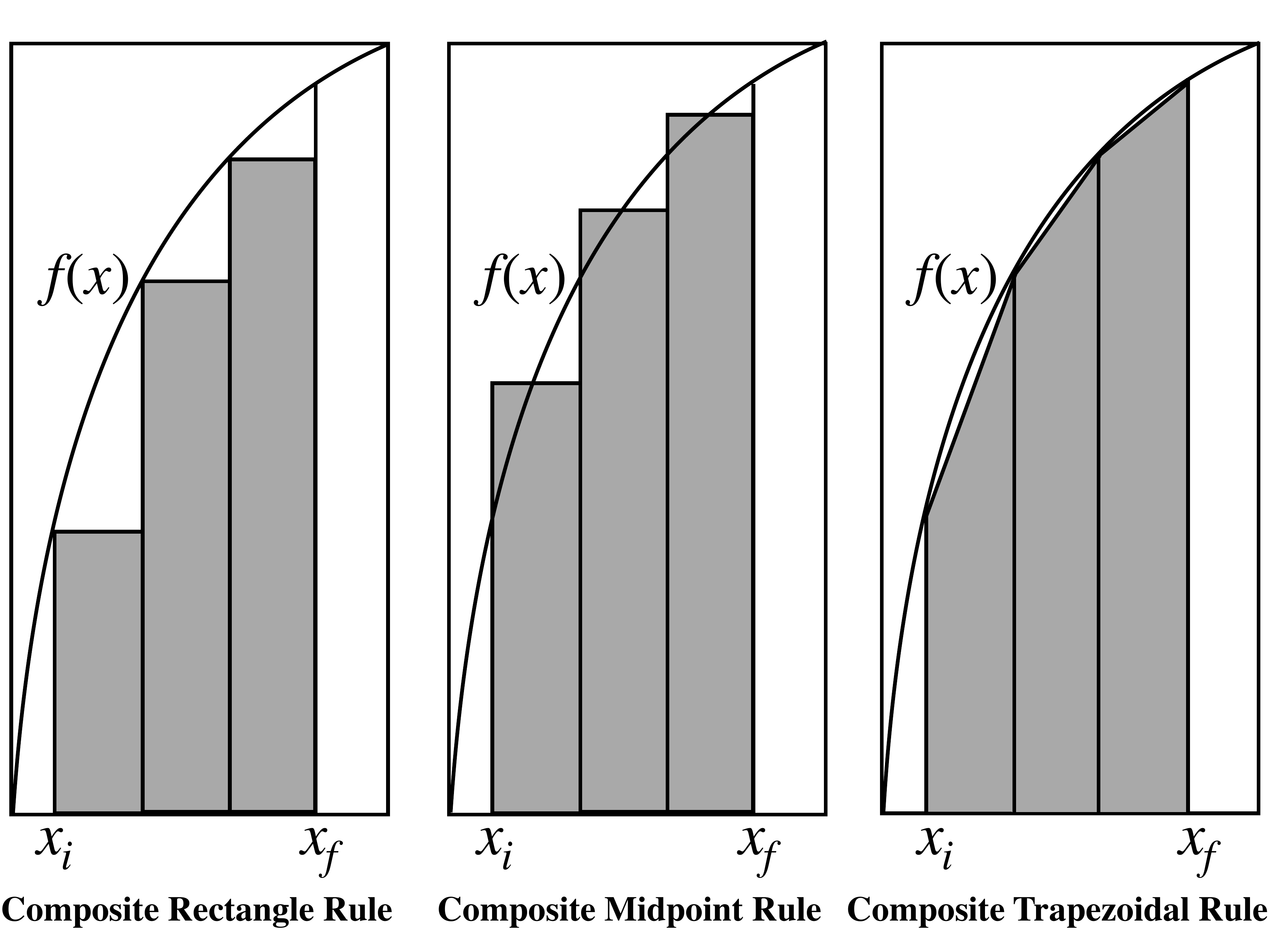}
  \caption{\label{fig:composite-rule}Composite (or compound) rules for numerical integration.}
\end{figure}

In Fig. \ref{fig:composite-rule} we show the composite rules - rectangle, midpoint and trapezoidal - for numerical integration.

\subsubsection{Composite Simpson's Rule}
\label{sec:Composite-Simpsons-one-third-rule}

Let us look at the composite version of Simpson's rule given in Eq. \eqref{eq:Simpsons}. Take $m/2$ as the number of sub-intervals on which Simpson's rule is applied. Taking $h = (x_f - x_i)/m$ and $x_r = x_i + r h$, we have the composite Simpson's rule\footnote{This is also known as composite Simpson's one-third rule.}
\bea
\int_{x_i}^{x_f} f(x) \; dx &\simeq& \frac{h}{3} \left( f(x_i) + f(x_f) + 4 \sum_{r = 1}^{\frac{m}{2}} f(x_{2r - 1}) + 2 \sum_{r = 1}^{\frac{m}{2} -1} f(x_{2r}) \right) \nn \\
&& - \frac{1}{180} h^4 f^{(4)}(\eta) (x_f - x_i).
\eea

\begin{figure}[t]
\centering
  \includegraphics[width=9cm]{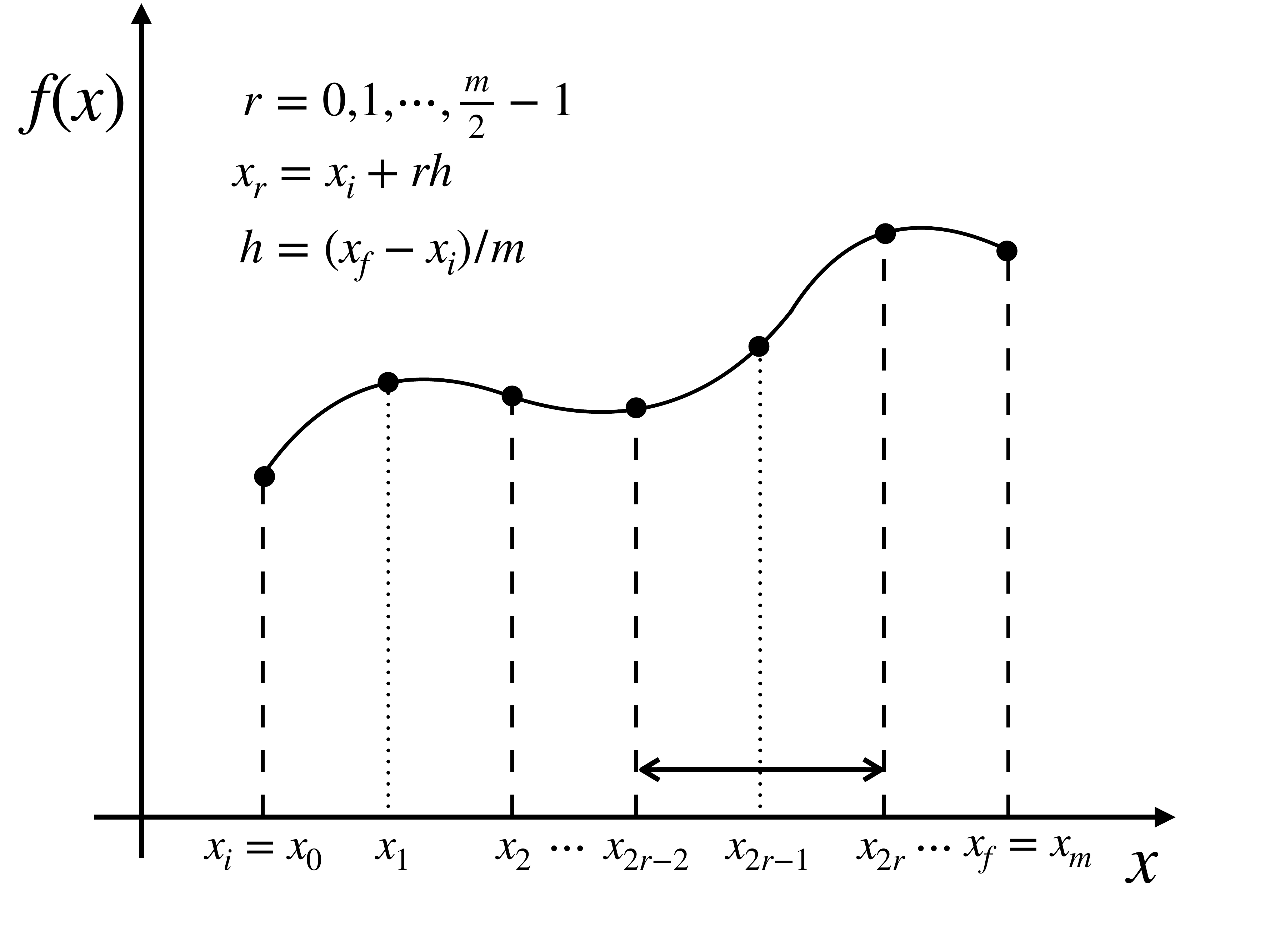}
  \caption{\label{fig:simpsons-compound}Composite version of Simpson's rule.}
\end{figure}

We see that composite Simpson rule is fourth-order accurate. In Fig. \ref{fig:simpsons-compound} we show composite Simpson's rule for numerical integration.

\subsection{Random Numbers}
\label{sec:Random-numbers}

Let us now proceed to understand another method of numerical integration, where random numbers play a crucial role in computing the integral. This method, known as Monte Carlo method, makes use of random numbers to scan the interval $[x_i, x_f]$ to get values of $x_{MC}$ and the corresponding values of $f(x_{MC})$ to numerically approximate the integral. 

The random numbers used in Monte Carlo simulations are usually generated using a deterministic algorithm. They exhibit sufficiently random-like properties. The numbers produced this way are known as {\it pseudo-random numbers}\footnote{Whenever we use a pseudo-random number generator, let us keep in mind the dictum by John von Neumann (1903 - 1957) ``Anyone who considers arithmetical methods of producing random digits is, of course, in a state of sin."}. 

There exist several methods to obtain pseudo-random numbers in the context of Monte Carlo simulations. We should keep in mind that pseudo-random numbers come from a deterministic process and thus they may not be foolproof. A random number generator that is considered good today may turn out to be bad tomorrow\footnote{For instance, RANDU, a random number generator designed by IBM, was in use since the 1960s and it turned out to be incorrect. As a result of the wide use of RANDU in the early 1970s, many simulation results from that time are seen as suspicious.}.

\subsubsection{Physical Random Numbers}
\label{sec:Physical-random-numbers}

The other choice is generating {\it physical random numbers} or {\it true random numbers}. They are generated from some truly random physical process such as radioactive decay, atmospheric noise, thermal noice, roulette wheel etc.  A physical random number generator is based on an essentially random atomic or subatomic physical phenomenon. Thus, we can trace the unpredictability of the sequence of numbers to the laws of quantum mechanics. 

True random numbers are not very useful for Monte Carlo simulations due to the following reasons: (i.) the sequence is not repeatable, (ii.) the random number generators are often slow, and (iii.) the distribution of random numbers may be biased. Unfortunately, the physical phenomena and tools used to measure them may not be free from asymmetries and systematic biases, and that make the numbers generated not uniformly random.

\subsubsection{Pseudo-random Numbers}
\label{sec:Pseudo-random-numbers}

Almost all of the Monte Carlo calculations make use of pseudo-random numbers. Typically the pseudo-random number generators (PRNGs) produce a random integer (with a definite number of {\it bits}), which is then converted to a {\it floating point number} $X \in [0, 1]$ by multiplying with a suitable constant. Before using the sequence, usually a {\it seed} number sets the initial state of the generator. The seed is typically an integer value or values.

We can consider a random number generator in hand as a good choice if it meets the following essential criteria.
\begin{itemize}
\item[1.] {\it Randomness} - the random numbers should be drawn from a uniform distribution and they must be {\it uncorrelated}. The uniform distribution could be, for example, within an interval $[0, 1)$. Note that generating uncorrelated random numbers is a very difficult task. No pseudo-random sequence is truly independent.
\item[2.] {\it Long period} - the sequence of pseudo-random numbers must repeat itself after a finite period since the generators have a finite amount of internal state information. This finite period is called {\it full cycle}\footnote{In a PRNG, a full cycle or full period is the behavior of a PRNG over its set of valid states. In particular, a PRNG is said to have a full cycle if, for any valid seed state, the PRNG traverses every valid state before returning to the seed state.}. Preferably, the full cycle should be much longer than the amount of numbers needed for the Monte Carlo calculation.
\item[3.] {\it Repeatability} - the generator is initialized with a {\it seed} number before generating random numbers. The same seed produces the same sequence of random numbers. This can help us with debugging the simulation program.
\item[4.] {\it Insensitive to seeds} - the period and randomness properties should not depend on the seed.
\item[5.] {\it Fast} - the algorithm must generate random numbers fast enough.
\item[6.] {\it Portability} - the algorithm must produce the same results on different computers.
\end{itemize}

There are several top quality PRNGs available on the market. Let us look at some of the interesting and important ones.

\begin{itemize}
\item[1.] {\bf Middle-square method:} This was proposed by Jon von Neumann in 1946. It is of poor quality and we should look at it only from the perceptive of historical interest. 

\item[2.] {\bf Linear congruential generator (LCG):} This was proposed in 1949 and this is historically the most influential and studied generator\footnote{LCG was proposed by a mathematician, Derrick Lehmer (1905-1991), while he was using the ENIAC computers for number theory.}. For example, the {\bf rand()} function, which is the ``standard" random number routine in ANSI C, is based on this algorithm. This generator is not good for serious Monte Carlo computations since it has a short full cycle. The full cycle is approximately $10^9$ for {\bf rand()}. The UNIX function {\bf drand48()} generates uniformly distributed PRNGs using an LCG algorithm and 48-bit integer arithmetic. Its full cycle is approximately $10^{14}$ and it is good enough for most of the tasks. 

\item[4.] {\bf Lagged Fibonacci generator (LFG):} This class of random number generators was devised by Mitchell and Moore in 1958. It is based on an improvement on the LCG and a generalization of the Fibonacci sequence. 

\item[3.] {\bf Rule 30:} It was introduced in 1983 by Stephen Wolfram. This PRNG is based on {\it cellular automata}. This rule was used in the Mathematica\texttrademark $~$software package for creating random integers.

\item[5.] {\bf Subtract-with-borrow (SWB):} It was introduced in 1991. It is a modification of {\it Lagged-Fibonacci generators}. An SWB generator is the basis for the {\bf RANLUX} generator \cite{Luscher:1993dy}. It has a full cycle of about $10^{171}$ and higher (depending on the ``luxury level" needed). It is widely used in elementary particle physics simulations, especially in lattice QCD simulations. 

\item[6.] {\bf Mersenne Twister (MT):} It was introduced in 1998. Probably it is the most commonly used modern PRNG. It is the default generator in the Python language (starting from version 2.3). This PRNG has a huge full cycle, about $10^{6000}$. We can definitely consider this as a good generator for Monte Carlo simulations.

\end{itemize}

\subsubsection{Random Numbers Using UNIX Function drand48()}
\label{sec:Random-numbers-using-UNIX-drand48()}

\begin{figure}[t]
\centering
  \includegraphics[width=9cm]{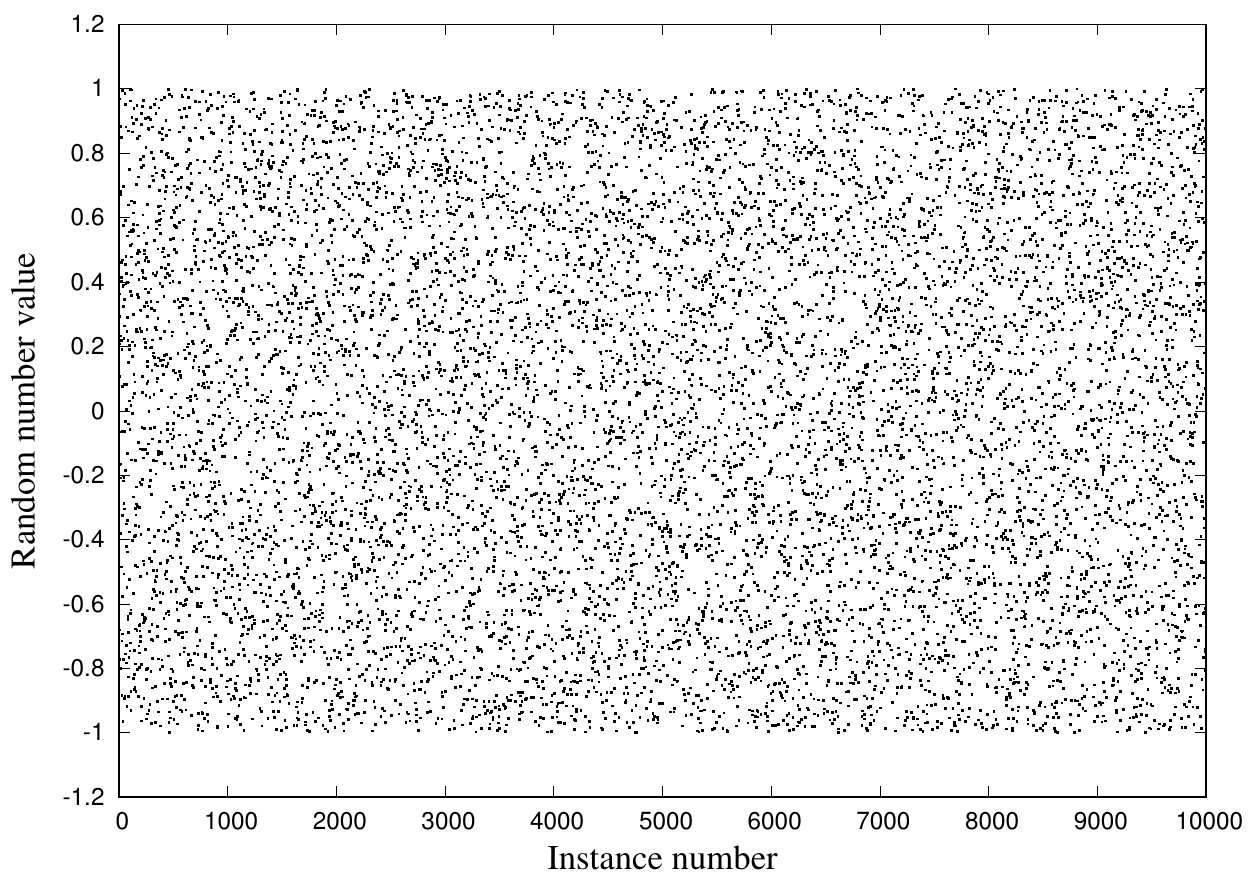} \\
  \vspace{0.5cm}
  \includegraphics[width=9cm]{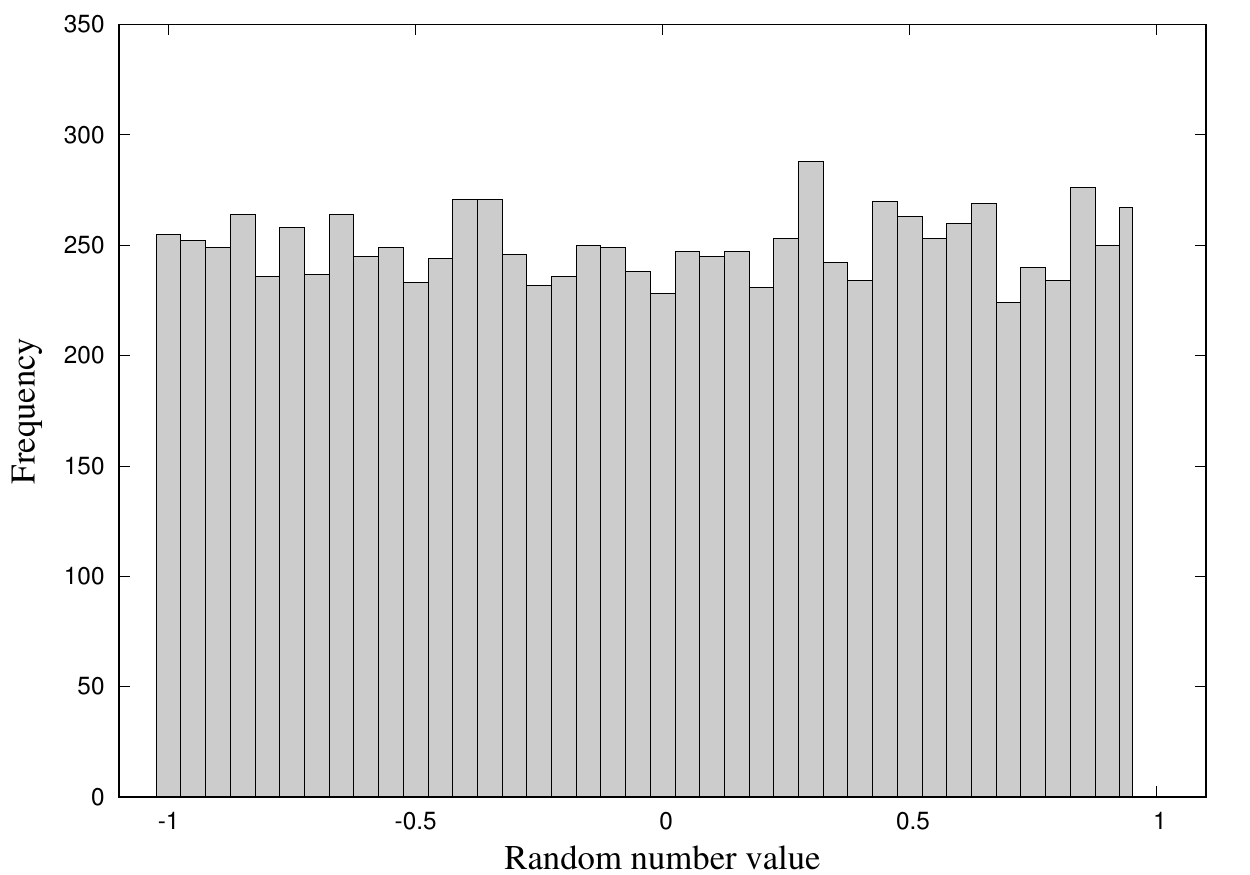}
  \caption{\label{fig:data-random}A sequence of random numbers uniformly distributed in the interval $[-1, +1)$. It is produced using the function drand48(). A total of 10,000 instances have been generated. The mean value is $-0.00015$ with a standard error of $0.01160$. A C++ program to generate this sequence is provided in Appendix \ref{sec:uniform-m-one-p-one}. (Top) Instance number against the value of the random number. (Bottom) The histogram of the random numbers generated shows that they are uniformly distributed within the given interval. }
\end{figure}

Simulation results we discuss here are mostly produced based on the random numbers generated using the UNIX function drand48(). This function generates uniformly distributed PRNGs using an LCG algorithm and 48-bit integer arithmetic. The numbers generated are non-negative, double-precision, floating-point values, and uniformly distributed over the interval $[0,1)$. In Fig. \ref{fig:data-random} we show a uniform distribution of random numbers in the interval $[-1, +1)$. It is produced using drand48() with its default seed, which is 1. A total of 10,000 instances have been generated. The mean value is $-0.00015$ with a standard error of $0.01160$. A C++ program to generate these random numbers is provided in Appendix \ref{sec:uniform-m-one-p-one}. 

\subsubsection{Random Numbers Using a Seed}
\label{sec:Random-numbers-using-a-seed}

We can use srand48() function to set a starting point for producing a series of PRNGs. If srand48() is not called, then the drand48() seed is set as if srand48() was called at program start with seed 1. Any other value for the seed sets the generator to a different starting point.

\begin{figure}[h]
\centering
  \includegraphics[width=9cm]{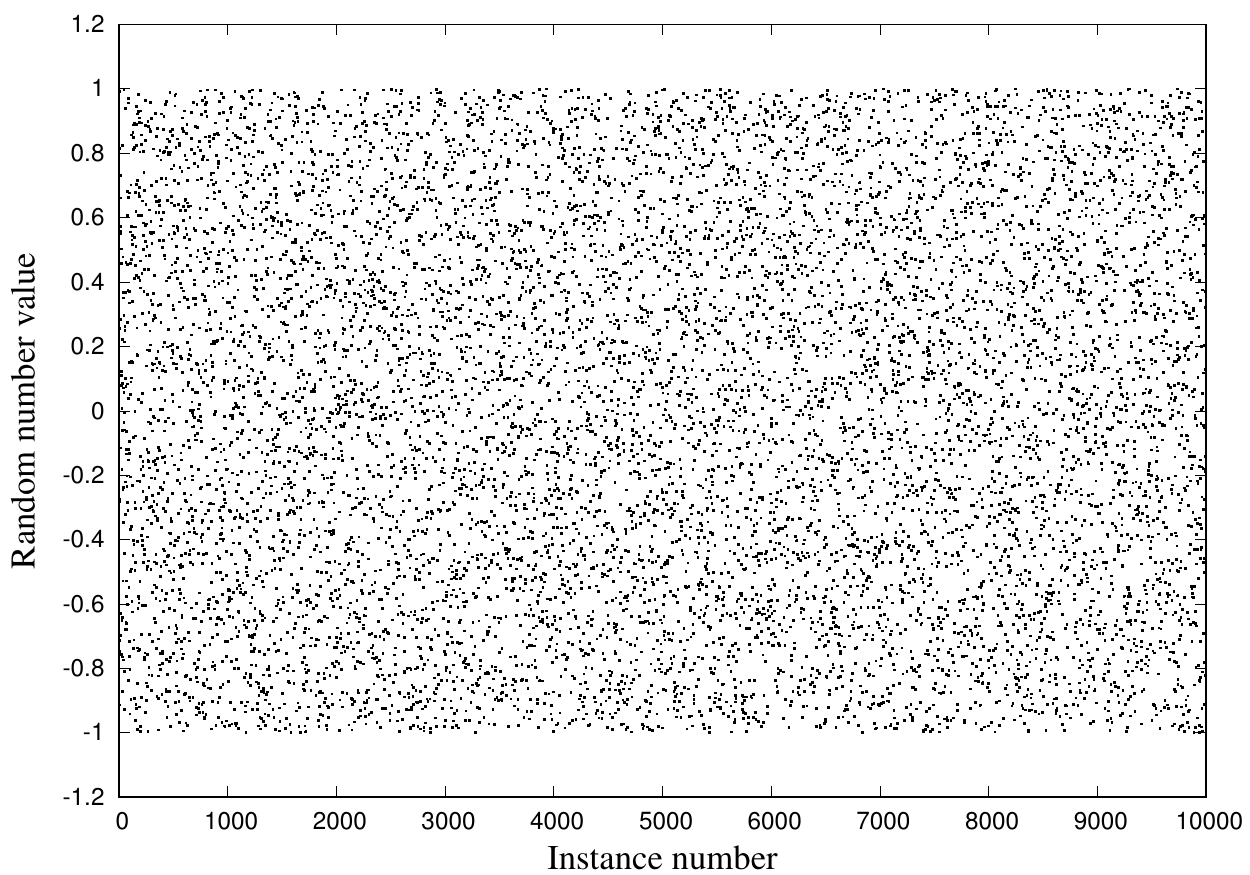} \\
   \vspace{0.5cm}
   \includegraphics[width=9cm]{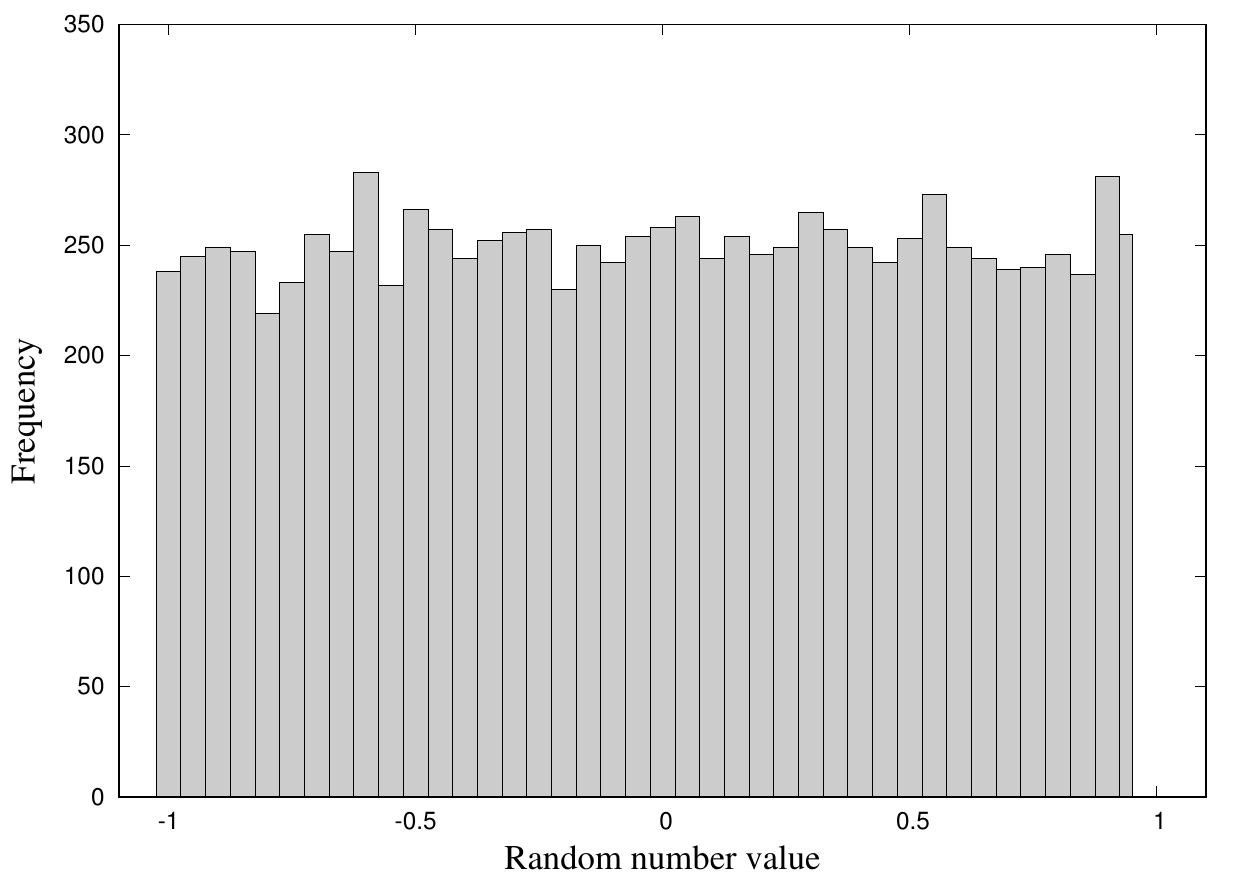}
  \caption{\label{fig:data-random-w-seed}A uniform distribution of random numbers in the interval $[-1, +1)$. It is produced using drand48() with the seed function srand48(), and the seed set to 41. A total of 10,000 instances have been generated. The mean value is $0.00008$ with a standard error of $0.01160$.}
\end{figure}

In Fig. \ref{fig:data-random-w-seed} we show a uniform distribution of random numbers in the interval $[-1, +1)$. It is produced using drand48() with the seed function srand48() and the seed set to 41. A total of 10,000 instances have been generated. The mean value is $0.00008$ with a standard error of $0.01160$. A C++ program to generate these random numbers is provided in Appendix \ref{sec:uniform-m-one-p-one-seed}. 

\subsubsection{Random Numbers from Non-uniform Distributions}
\label{sec:Random-numbers-from-non-uniform-distributions}

In most of the cases we need random numbers from a non-uniform distribution. In such cases the raw material from the PRNGs should be transformed into the desired distributions.

Suppose the distribution we are interested in is the Gaussian distribution
\beq
p(x) = \frac{1}{\sqrt{2 \pi}} e^{- \hf x^2}.
\eeq 

We can use the {\it Box-Muller method} to generate Gaussian random numbers using two-dimensional Gaussian distributions
\beq
p(x, y) = \frac{1}{2 \pi} e^{-(x^2 + y^2)}.
\eeq 

The obvious way to handle this product of 2 one-dimensional distributions is by changing the variables to polar coordinates, $(x, y) \to (r, \theta)$, and proceeding from there. However, in this case, instead of generating polar coordinates we can directly generate Cartesian coordinates from a uniform distribution inside a circle using the {\it rejection method}\footnote{The {\it rejection method} or {\it rejection sampling} is a technique used to generate observations from a given distribution. This method is also known as the {\it acceptance-rejection method} or {\it accept-reject algorithm}. We can apply this method successfully to any distribution in ${\mathbf R}^n$ with a density. The idea of rejection method is based on the observation that to sample a random variable in one dimension, we can perform a uniform random sampling of the two-dimensional Cartesian graph, and keep the samples in the region under the graph of its density function. We can apply the same idea to $n$-dimensional functions.}. Here, the idea is to generate two random numbers, $v_i \in (-1, +1)$ and then accept if $R^2 = v_1^2 + v_2^2 < 1$ and, otherwise, we go back to the previous step. 

\begin{figure}[h]
\centering
  \includegraphics[width=9cm]{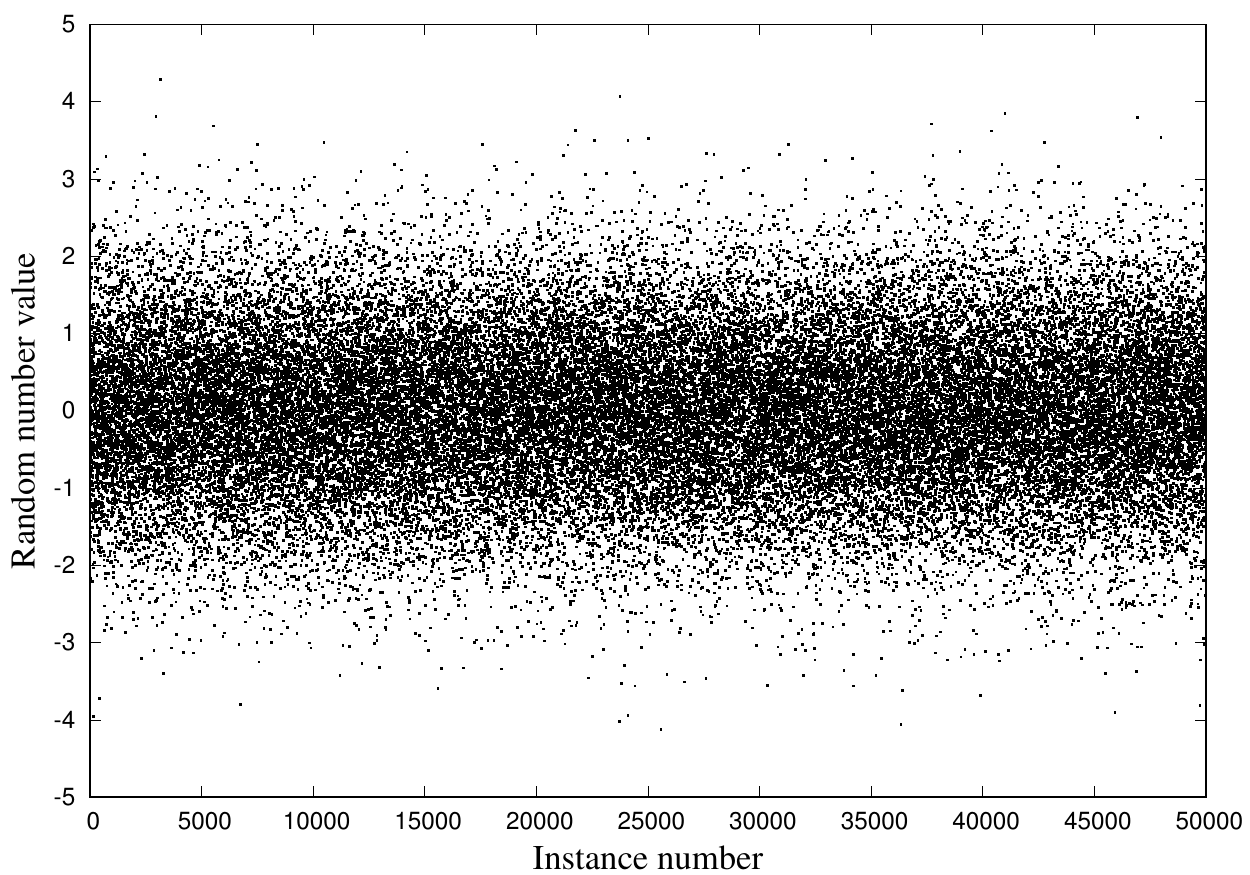} \\
   \vspace{1cm}
   \includegraphics[width=9cm]{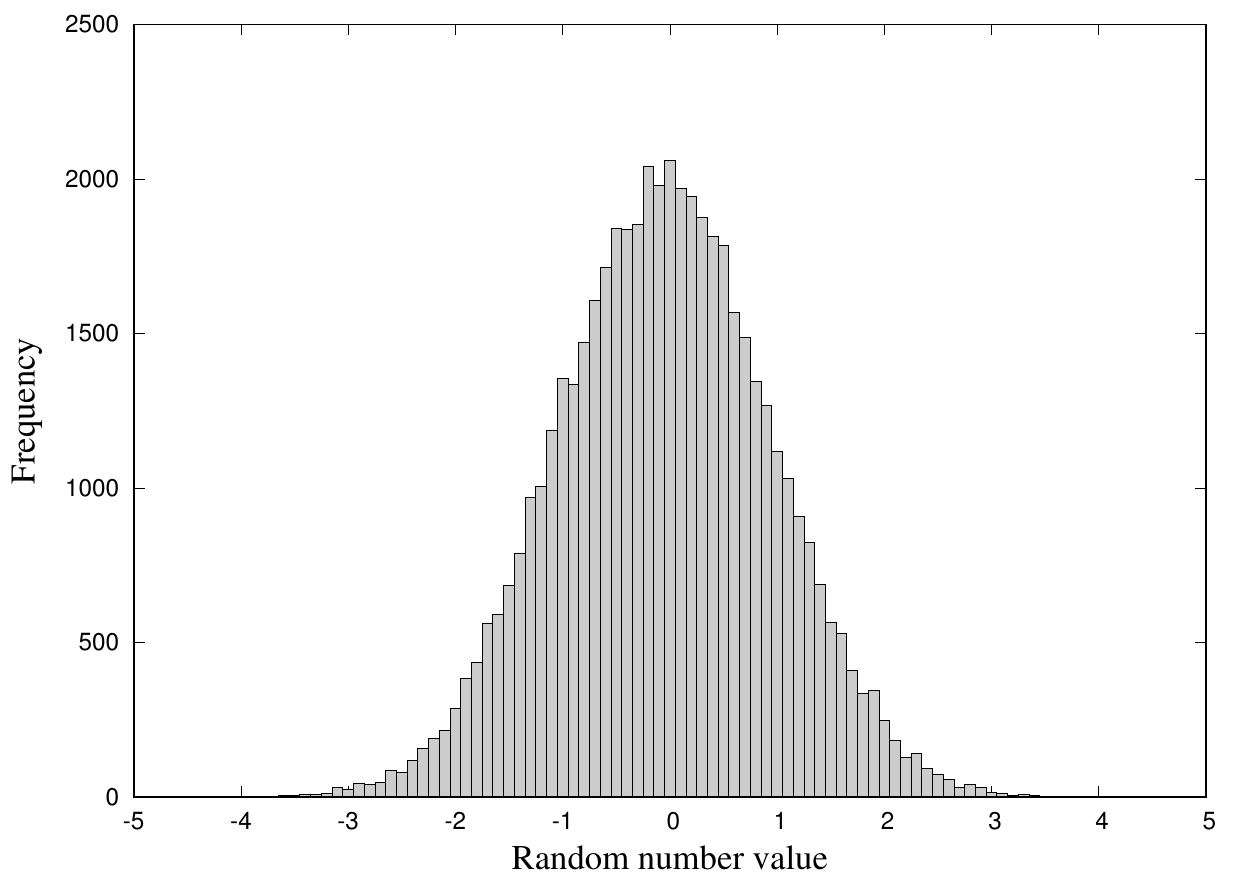}
  \caption{\label{fig:data-gauss-random-w-seed}Random numbers drawn from a Gaussian distribution with mean 0 and width 1. It is produced using the rand() function and then applying a Box-Muller transformation to produce the Gaussian distribution. The seed function srand() is used to set the seed to the value 41. A total of 50,000 instances have been generated. The mean value is $-0.000025$ with a standard error of $0.008930$. (Top) Number of instances against the value of the Gaussian random number at that instance. (Bottom) Histogram of the random numbers generated.}
\end{figure}

In Fig. \ref{fig:data-gauss-random-w-seed} we show the random numbers generated from a Gaussian distribution with mean 0 and width 1. The numbers are produced using Box-Muller transformation of uniform random numbers produced using rand() function\footnote{The rand() function returns a pseudo-random number in the range between 0 and RAND\_MAX.} and seed set to 41. A total of 50,000 instances have been generated. The mean value is $-0.000025$ with a standard error of $0.008930$. A C++ program to generate these random numbers is provided in Appendix \ref{sec:gauss-random}. 

In general it can be very difficult to devise the distributions from which we want to sample the random numbers. It can become prohibitively difficult as the number of dimensions of numerical integration increases. We need to resort to Markov chain Monte Carlo (MCMC) method as an alternative strategy.  

\subsection{Monte Carlo Method}
\label{sec:Monte-Carlo-method}

Monte Carlo method was invented by Nicholas Metropolis \cite{Metropolis:1953am} and popularized by the pioneers in the field; Nicholas Metropolis, Stanislaw Ulam, Enrico Fermi and John von Neumann in the 1940s and 1950s. The term Monte Carlo refers to the inherent randomness present in this method of numerical integration.

Let us again consider a well-behaved function $f(x)$ of single variable $x$. The definite integral of the function, with the lower and upper limits $x = x_i$ and $x = x_f$ respectively, is
\beq
I = \int_{x_i}^{x_f} f(x) \; dx.
\eeq

The mean value of $f(x)$ over interval is $M = (x_f - x_i)^{-1} I$. If $x_1, x_2, \cdots, x_n$ are $n$ points in the interval, then the average value of $f$ over this sample is
\beq
\overline{f}_n = \frac{1}{n} \sum_{r=1}^n f(x_r).
\eeq 

If points are distributed uniformly over the interval, we expect that
\beq
\overline{f}_n \simeq \frac{I}{(x_f - x_i)} = M
\eeq 
and
\beq
I = \int_{x_i}^{x_f} f(x) \; dx \simeq (x_f - x_i) \overline{f}_n = (x_f - x_i) ~\frac{1}{n} \sum_{r=1}^n f(x_r).
\eeq

Note that this expression is similar to the quadrature formulas we encountered earlier. If random values are used for $x_r$, then the resulting method is called the Monte Carlo (MC) method.

\subsubsection{Worked Example - Composite Midpoint Rule}
\label{sec:Worked-example-Composite-midpoint-rule}

Let us look at an example where the composite midpoint rule is in action. The composite midpoint rule to compute the integral of a function has the form
\beq
I = \int_{x_i}^{x_f} f(x) \; dx \simeq h \sum_{r=1}^m f \left( x_i + (r - \hf) h \right) + {\cal O}(h^2).
\eeq

\begin{figure}[t]
\bec
  \includegraphics[width=9cm]{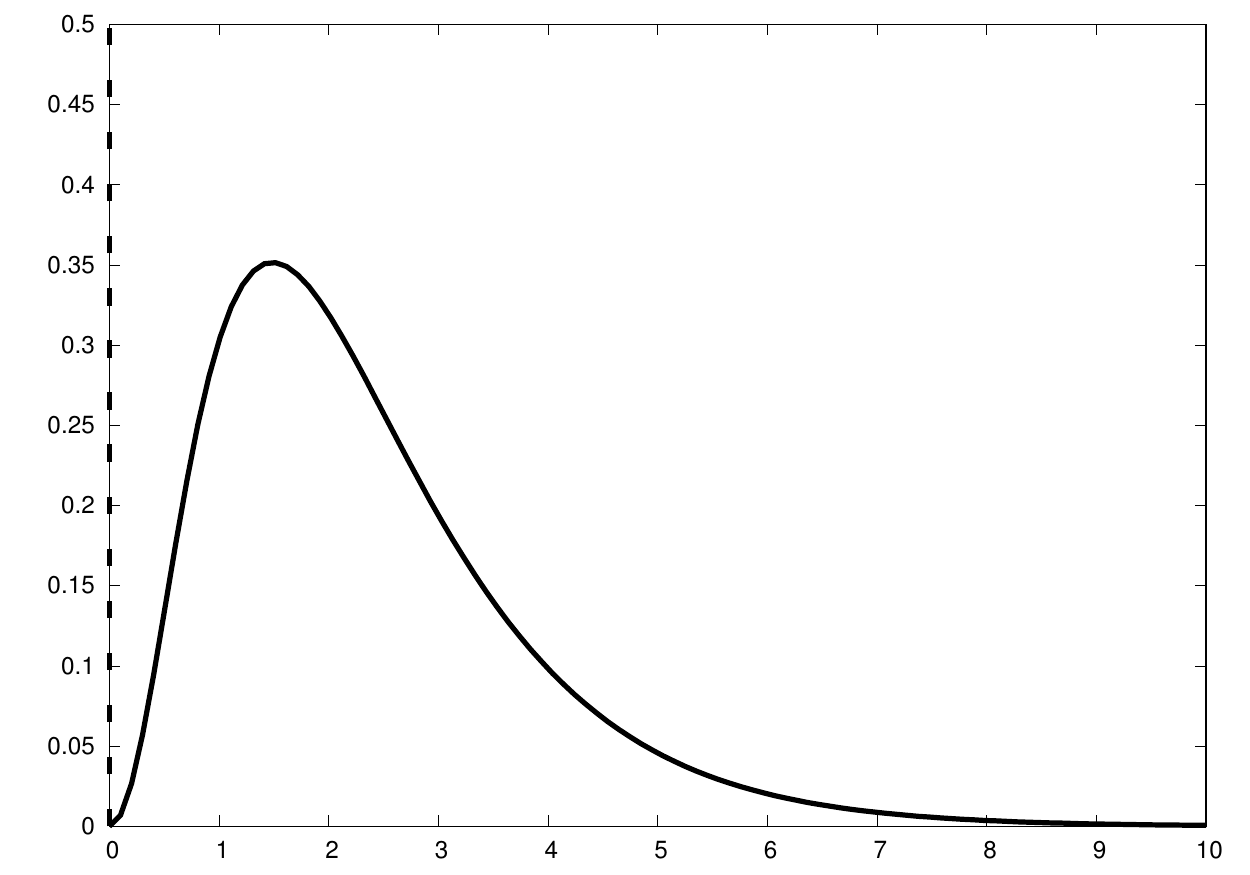}
  \caption{The function $f(x) = \frac{27}{2 \pi^2} \left( \frac{1 - e^{-x}}{1 + e^{-3x}} \right) x e^{-x}$ against $x$. The value of the integral is $I = \int_0^{10} f(x) ~dx \simeq 1$.}
  \label{fig:plot-function-1}
  \eec
\end{figure}

Let us use this rule to evaluate the following integral 
\beq
\label{eq:integral}
I = \int_0^{10} f(x) \; dx,
\eeq
where the function $f(x)$ has the form (see Fig. \ref{fig:plot-function-1})
\beq
\label{eq:function}
f(x) = \frac{27}{2 \pi^2} ~\left( \frac{1 - e^{-x}}{1 + e^{-3x}} \right) ~x ~e^{-x}.
\eeq

From the table of integrals, we see that the exact value of the integral is 
\beq
\label{eq:exact_val_int}
I = \int_0^\infty f(x) \;dx = 1.
\eeq

In Appendix \ref{sec:num-int-compo-midpoint} we provide a C++ program that computes the integral given in Eq. \eqref{eq:integral}, using composite midpoint rule. Table \ref{table:composite-mid-point-rule} shows the values computed for $I$ for various $h$ values. We get a number close to the value given in Eq. \eqref{eq:exact_val_int} as $h$ is decreased.

\begin{table}[h]
\bec
\begin{tabular}{ c  c  c }
\hline
  $h$ & $m = 0.5 + (x_f - x_i)/h$ & $I(h)$ \\ \hline \hline
  10  & 1.5 & 0.4577  \\ 
  5    & 2.5 & 1.3159 \\ 
  2    & 5.5 & 1.1062  \\
  1    & 10.5 & 1.0035 \\
  0.5 & 20.5 & 0.9993  \\
  0.1 & 100.5 & 0.9993  \\ \hline
\end{tabular}
\caption[Table]{Computing the integral given in Eq. \eqref{eq:integral} using composite midpoint rule. As $h$ is decreased the estimated value approaches the exact value, which is approximately 1.}
\label{table:composite-mid-point-rule}
\eec
\end{table}

\subsubsection{Worked Example - Composite Simpson's Rule}
\label{sec:Worked example-Composite-Simpsons-one-third-rule}

Let us evaluate the same integral, Eq. \eqref{eq:integral}, using composite Simpson's (one-third) rule. The formula for composite Simpson's rule is
\bea
\int_{x_i}^{x_f} f(x) dx &\simeq& \frac{h}{3} \left( f(x_i) + f(x_f) + 4 \sum_{r = 1}^{\frac{m}{2}} f(x_{2r - 1}) + 2 \sum_{r = 1}^{\frac{m}{2} -1} f(x_{2r}) \right),~~
\eea
where $h = (x_f - x_i)/m$.

A C++ program that computes the integral given in Eq. \eqref{eq:integral}, using composite Simpson's rule, is provided in Appendix \ref{sec:num-int-compo-simpsons}. Table \ref{table:compo-simpsons} shows the results from numerical integration. We get a number close to the value given in Eq. \eqref{eq:exact_val_int} as $m$, the number of sub-intervals, is increased.

\begin{table}[h]
\bec
\begin{tabular}{ c  c  c }
\hline
  $m$ & $h$ & $I(m)$ \\ \hline \hline
  10  & 1 & 0.7567  \\
  25  & 0.4 & 0.9715  \\
  50  & 0.2 & 0.9959  \\
  100 & 0.1 & 0.9989  \\
  1000 & 0.01 & 0.9993  \\ \hline
\end{tabular}
\caption[Table]{Computing the integral given in Eq. \eqref{eq:integral} using composite Simpson's $1/3$ rule. As $m$ is increased the estimated value approaches the exact value, which is approximately 1.}
\label{table:compo-simpsons}
\eec
\end{table}

\subsubsection{Worked Example - Monte Carlo Integration}
\label{sec:Worked-example-Monte-Carlo-integration}

\begin{figure}[t]
\bec
  \includegraphics[width=9cm]{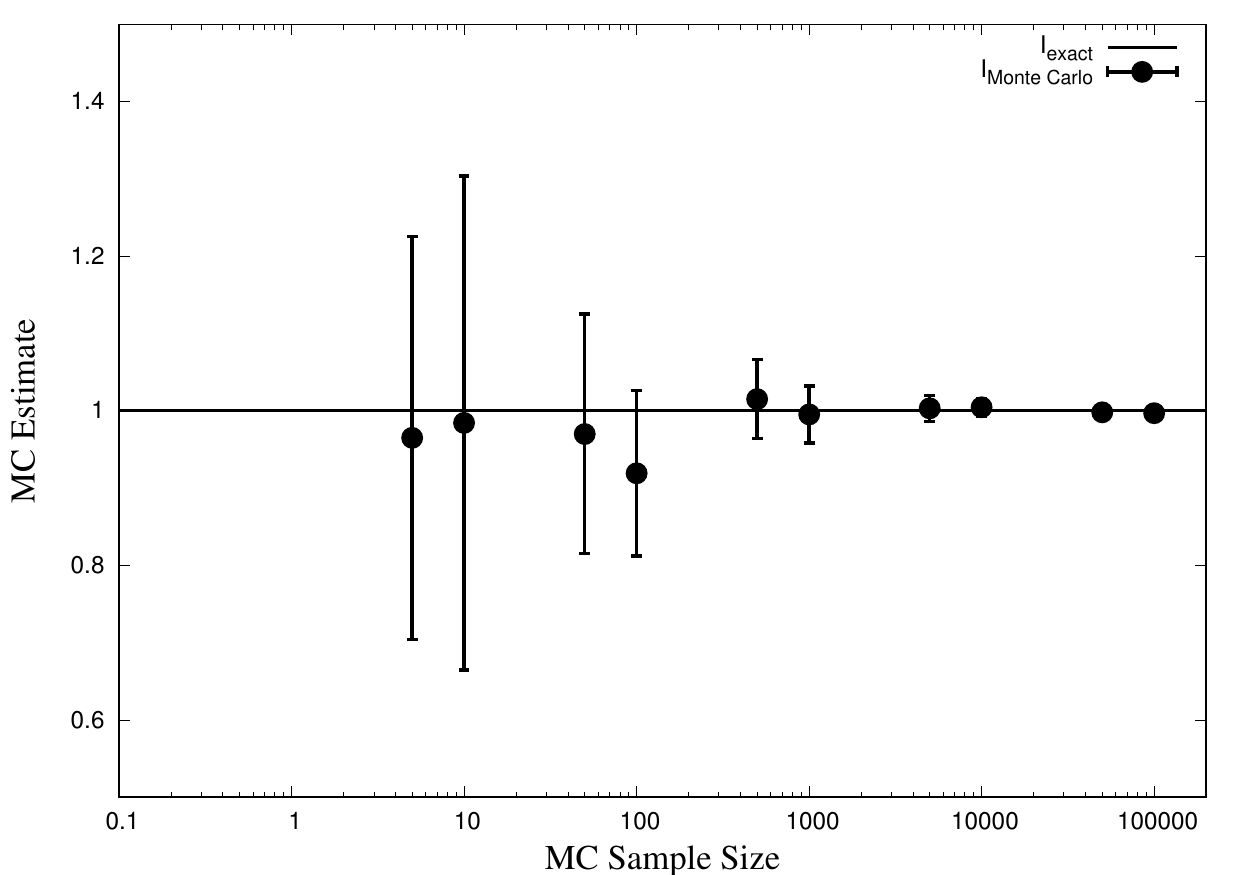}
  \caption{Computing the integral of the function given in Eq. \eqref{eq:function} using Monte Carlo method. The integral converges to the analytical value for large sample sizes.}
  \label{fig:plot-mc-int}
  \eec
\end{figure}

We can evaluate the integral given in Eq. \eqref{eq:integral} using Monte Carlo method. The method we are using here is a naive (or independent) Monte Carlo sampling method since we are not focusing on how efficiently the integral is being computed. In Appendix \ref{sec:num-int-monte-carlo} we provide a C++ program that computes the integral given in Eq. \eqref{eq:integral}, using Monte Carlo method. Table \ref{table:mc-method-integral} shows the data obtained using Monte Carlo calculation (with the corresponding 1-$\sigma$ error). In Fig. \ref{fig:plot-mc-int} we show that the Monte Carlo estimate converges to the analytical value of the integral for large sample sizes.

\begin{table}[h]
\bec
\begin{tabular}{ c c }
\hline
  $n$ & $I(n) \pm \delta I(n)$ \\ \hline \hline
  5 & $0.9651 \pm 0.2608$ \\
  10  & $0.9843 \pm 0.3197$  \\
   50 & $0.9701 \pm 0.1550$ \\
  100 & $0.9193 \pm 0.1071$  \\
   500 & $1.0152 \pm 0.0511$ \\
  1000 & $0.9953 \pm 0.0368$  \\
   5000 & $1.0032 \pm 0.0167$ \\
  $10^4$ & $1.0045 \pm 0.0118$  \\
   $5 \times 10^4$ & $0.9982 \pm 0.0052$ \\
  $10^5$ &  $0.9970 \pm 0.0037$ \\ \hline
\end{tabular}
\caption[Table]{Computing the integral given in Eq. \eqref{eq:integral} using Monte Carlo method. As the sample size $n$ is increased the Monte Carlo estimate of the integral converges to the exact value.}
\label{table:mc-method-integral}
\eec
\end{table}

\subsection{Error in Monte Carlo Integration}
\label{sec:Error-in-Monte-Carlo-integration}

Integrating the function $f(\vec{x})$ in a volume $V$
\beq
I = \int_V f(\vec{x}) \; d^dx,
\eeq
using Monte Carlo method, with $n$ samples, $\vec{x}_1$, $\vec{x}_2$, $\cdots$, $\vec{x}_N$, chosen independently and randomly, throughout the $d$-dimensional volume $V$, leads to the following error estimate
\beq
\label{eq:one-sig-err}
\sigma_N = V \sqrt{\frac{\langle f^2 \rangle - \langle f \rangle^2}{N}}.
\eeq

In general, we do not know the expectation values $\langle f \rangle $ and $\langle f^2 \rangle $ beforehand. Thus, we use the corresponding Monte Carlo estimates
\beq
\langle f \rangle \simeq \frac{1}{N} \sum_r f(\vec{x}_r),~~ \langle f^2 \rangle \simeq \frac{1}{N} \sum_r f^2(\vec{x}_r).
\eeq

In order not to underestimate the error we must also divide by $\sqrt{N-1}$ instead of $\sqrt{N}$ in Eq. \eqref{eq:one-sig-err}. The error given in Eq. \eqref{eq:one-sig-err} is called the 1-$\sigma$ error\footnote{1-$\sigma$ error is the most common error value quoted in the literature.}. That is, the error is the width of the Gaussian distribution of
\beq
f_N = \frac{1}{N} \sum_r f(\vec{x}_r),
\eeq
which is when the true value (or exact value) is within $V \langle f \rangle \pm \sigma_N$ with $68\%$ probability. 

Thus we have an expression for the Monte Carlo estimate of the integral
\beq
\int_V f(\vec{x}) \; d^dx = V f_N \pm \sigma_N.
\eeq

\subsection{When is Monte Carlo Good for Integration?}
\label{sec:When-is-Monte-Carlo-good-for-integration?}

When can we say that we are certainly gaining something by using Monte Carlo integration method compared to the traditional numerical integration methods? In order to answer this question let us see how errors behave in traditional deterministic integration methods such as the trapezoidal and Simpson's rules. 

Let us consider a $d$-dimensional system, ${\mathbf R}^d$, and divide each axis in $n$ evenly spaced intervals. Thus we have a total number of points $N = n^d$.

Then the error is
\beq
\propto \frac{1}{n^2}~~({\rm for~Trapezoidal~rule}) ~~ {\rm and} ~~\propto \frac{1}{n^4}~~({\rm for~Simpson's~rule}). \nn
\eeq

When $d$ is small, Monte Carlo integration has much larger errors compared to the deterministic methods. Let us check when Monte Carlo method is as good as Simpson's rule. Noting that the error in Monte Carlo method goes down at a rate ${\cal O}(1/\sqrt{N})$, independent of the dimensionality of the system, we have
\beq
\frac{1}{n^4} = \frac{1}{N^\frac{4}{d}} = \frac{1}{\sqrt{N}} ~~\implies~~d = 8.
\eeq

This tells us that when we are dealing with $d < 8$ Simpson's rule is much better than Monte Carlo. But for $d > 8$ Simpson's rule is much worse than Monte Carlo.

As an example let us consider a system with 10 points (or nodes) per axis of ${\mathbf R}^d$. Application of Simpson's rule would need $N = 10^d$ integration points. Unfortunately, this becomes an unachievably huge number on computers when $d \gtrsim 10$.

\subsection{When Does Monte Carlo Fail?}
\label{sec:When-does-Monte-Carlo-fail?}

Monte Carlo  integration method is not always foolproof: it can go wrong in several ways.

One such case is when the mean of the function we are trying to integrate does not exist. The standard Cauchy distribution (Lorentz distribution or Breit-Wigner distribution) 
\beq
f(x) = \frac{1}{\pi(1 + x^2)}
\eeq
looks similar to a normal distribution but it has much heavier tails. The mean and standard deviation of the Cauchy distribution are undefined. What this means is that accumulating 10 data points and $10^6$ data points would give similar accuracy when estimating the mean and standard deviation. 

Another case is when the mean of the function is finite but its variance is infinite. In this case, the integral converges to the right answer but not at the ${\cal O}(1/\sqrt{N})$ rate. 

Let us consider the function
\beq
f(x) = \frac{1}{\sqrt{x}}.
\eeq

This gives the integral
\beq
I = \int_0^1 f(x) \; dx = 2.
\eeq

The variance is 
\beq
\langle f^2 \rangle = \int_0^1 ~x^{-1} \; dx = \infty.
\eeq

It is still possible to perform Monte Carlo but the error, $\sigma_N$, is formally infinite. We can compute the error, but it suffers from a lot of fluctuations.

Another example is integrals of the following type
\beq
\int_{-1}^1 \frac{1}{x} \; dx = 0,
\eeq
which is an ill-defined expression. This integral can be defined in the Cauchy principal value sense
\beq
\lim_{\epsilon \to 0+} \left( \int_{-1}^\epsilon \frac{1}{x} \; dx + \int_\epsilon^1 \frac{1}{x} \; dx \right) = 0.
\eeq

However, naive Monte Carlo methods cannot handle the above integral.

Let us note that Monte Carlo method works most efficiently when the functions are flat, and becomes most problematic when the integrand oscillates rapidly or is sharply peaked.

\section{Monte Carlo with Importance Sampling}
\label{sec:Monte-Carlo-with-importance-sampling}

In this Section we discuss a method that can increase the efficiency of Monte Carlo integration. This technique is called importance sampling. It is one of the several available {\it variance reduction} techniques, in the context of Monte Carlo integration. 

\subsection{Naive Sampling and Importance Sampling}
\label{sec:Naive-sampling-and-importance-sampling}

One question we can ask now is how the efficiency of Monte Carlo integration can be improved. It would be more useful if somehow we can make the function under consideration more flat. The method of importance sampling tries to increase the efficiency of the Monte Carlo method by choosing a function that is more flat. This is also a simple generalization of the Monte Carlo method using a weight function. 

The random numbers, $x_r$, are selected according to a probability density (or weight) function, $w(x)$. The weight function is normalized to unity
\beq
\int_{x_i}^{x_f} w(x) \; dx= 1.
\eeq

The integral is then computed as
\beq
I = \int_{x_i}^{x_f} w(x) f(x) \; dx \simeq \frac{1}{N} \sum_{r =1}^N f(x_r).
\eeq

It is possible to extend this method easily to multiple integrals. 

A drawback of Monte Carlo method with importance sampling is that it is not possible to give any theoretical bound on the truncation error. What we can say is that the calculated average value of $f(x)$ is between the smallest and the largest value of the function in the given integration region. If a large number of sample points are used, then we can talk about a {\it probable error} rather than the error bound.

We can obtain the probable error using the {\it Central Limit Theorem} (CLT) of statistics. For the integral given above, the variance $\sigma$ is defined as
\bea
\sigma^2 &=& \int_{x_i}^{x_f} \left( f(x) - I \right)^2 w(x) \; dx = \int_{x_i}^{x_f} \left[ f^2(x) w(x) - I^2 \right] \; dx.
\eea

We have assumed that the function is square integrable over the required integration region. In this case, the CLT tells us that
\beq
{\rm prob} \left( \left| \frac{1}{N} \sum_{r = 1}^N f(x_r) - I \right| \leq \frac{\lambda \sigma}{\sqrt{N}} \right) = \alpha( \lambda ) + {\cal O} \left( \frac{1}{\sqrt{N}} \right),
\eeq
where $\alpha( \lambda )$ is the probability integral (Gauss' error function)
\beq
\alpha( \lambda ) \equiv {\rm erf} \left( \frac{\lambda}{\sqrt{2}} \right) = \frac{1}{\sqrt{2 \pi}} \int_{-\lambda}^\lambda dx~ e^{-\hf x^2}.
\eeq

Thus we see that for a fixed $\lambda$ ({\it level of confidence}) the error estimate $\lambda \sigma/ \sqrt{N}$ varies directly as $\sigma$ and inversely as $\sqrt{N}$. This is the typical rate of convergence for Monte Carlo method. This appears to be slow, but pleasingly it is independent of the dimension or the smoothness of the integrand. Let us also note that Monte Carlo method is quite effective for integration over irregular regions or when the number of dimensions is quite large. 

For all practical applications of the Monte Carlo method, we can write the variance $\sigma$ as
\beq
\sigma^2 \simeq \left[ \frac{1}{N} \sum_{r = 1}^N f^2(x_r) - \left( \frac{1}{N} \sum_{r = 1}^N f(x_r) \right)^2 \right].
\eeq

Apart from this error estimate, we also need an algorithm to select the $x_r$ values. Normally we use a sequence produced by a PRNG. For numerical integration in one dimension, the only requirement is that the numbers should be uniformly distributed, that is, $w(x) = 1$. 

Coming back to the idea of importance sampling, let us note that in order to reduce the Monte Carlo error we can either increase the sample size $N$ or decrease the variance $\sigma$. In the first case, we require a very large number of sample points $N$ to decrease the error substantially. In the second case we can reduce the variance using some variance reduction techniques. Importance sampling is one of the several variance reduction methods available on the market\footnote{Other methods for variance reduction include antithetic variables, control variates and stratified sampling. For a description of variance-reduction techniques see Ref. \cite{Hammersley:1964}.}.

Let us consider the integral of a well-behaved function $f(x)$
\beq
I = \int_0^1 f(x) \; dx,
\eeq
and rewrite it in the following way, after ``a multiplication by 1"
\beq
I = \int_0^1 dx~ \left[ \frac{f(x)}{p(x)}\right]~ p(x).
\eeq
Here $p(x) > 0$ and 
\beq
\int_0^1 dx~ p(x) = 1.
\eeq

We can treat $p(x)$ as the weight function, and use random numbers with a probability density distribution $p(x)$ on $0 \leq x \leq 1$. In this case, we can approximate the integral by
\beq
I \simeq \frac{1}{N} \sum_{r = 1}^N \frac{f(x_r)}{p(x_r)}.
\eeq

The variance is
\beq
\label{eq:imp-samp-choose-p}
\sigma^2 = \int_0^1 \frac{f^2(x)}{p^2(x)} ~p(x)~dx - \left( \int_0^1 \frac{f(x)}{p(x)} ~p(x)~dx \right)^2.
\eeq

Let us assume that $f(x) > 0$, (if not we can always add a constant) and choose $p(x)$ as the following
\beq
p(x) = \frac{1}{Z} f(x), ~~{\rm with}~~ Z = \int_0^1 f(x) \; dx.
\eeq

This choice leads to $\sigma^2 = 0$ in Eq. \eqref{eq:imp-samp-choose-p}. This would be an ideal choice, that is, 
\beq
p(x) \propto \left| f(x) \right|,
\eeq
but unfortunately it requires a prior knowledge of the integral we are trying to evaluate. Instead, as a promising strategy, we could also select $p(x)$ as some approximation to the above function. Then also we will end up with a small variance.

The main problem with importance sampling method is the difficulty of generating random numbers from a probability density function $p(x)$, particularly in several dimensions. If we use a $p(x)$ whose integral is known and whose behavior approximates that of $f(x)$, then we can expect to reduce the variance. Thus, importance sampling is choosing a {\it good distribution} from which to simulate our random variables. It involves multiplying the integrand by ``1" to give an expectation of a quantity that varies less than the original integrand, over the region of integration.

A good importance sampling function $p(x)$ should have the following properties:
\begin{enumerate}
\item $p(x) > 0$ whenever $f(x) \neq 0$
\item $p(x)$ should be close to being proportional to $\left | f(x) \right |$
\item it should be easy to draw values from $p(x)$
\item it should be easy to compute the density $p(x)$ for any value $x$ that we might realize.
\end{enumerate}

As the dimensionality of the integral increases, where Monte Carlo techniques are most useful, fulfilling the above set of criteria can turn out to be quite a non-trivial endeavor.

\subsection{Worked Example - Importance Sampling}
\label{sec:Worked-example-Importance-sampling}

As an example of importance sampling in action let us consider the following function
\beq
f(x) = \exp \left( - \hf x^2 + \qtr x - \frac{1}{32} \right),
\eeq
and the integral
\beq
\label{eq:fun-imp-samp}
I = \int_{-10}^{10} dx ~f(x).
\eeq
The true value of the integral is about $\sqrt{2 \pi} \simeq 2.5066$.

We can use Monte Carlo with naive sampling, by using random numbers generated from a uniform distribution $u(x)$ in the interval $[-10, 10]$, and look at the sample
mean of $20 f(x_i)$. Notice that this is equivalent to importance sampling with the importance function $p(x) = u(x)$.

The function $f(x)$ given in Eq. \eqref{eq:fun-imp-samp} is peaked around 0.25 and decays quickly elsewhere. Thus, under the uniform distribution, many of the points are contributing very little to this expectation. Something more like a Gaussian function with peak at 0 and small variance say, 1 would provide a greater precision.

Let us rewrite the integral, after ``multiplying by 1" as
\beq
I = \int_{-10}^{10} dx \left[ \frac{f(x)}{p(x)} \right] p(x),
\eeq
where 
\beq
p(x) = \frac{1}{\sqrt{2 \pi}} \exp\left( - \hf x^2 \right).
\eeq

The {\it importance function} we are using is
\beq
\sqrt{2 \pi} \exp \left( \hf x^2 \right).
\eeq

\begin{figure}[h]
\bec
  \includegraphics[width=9cm]{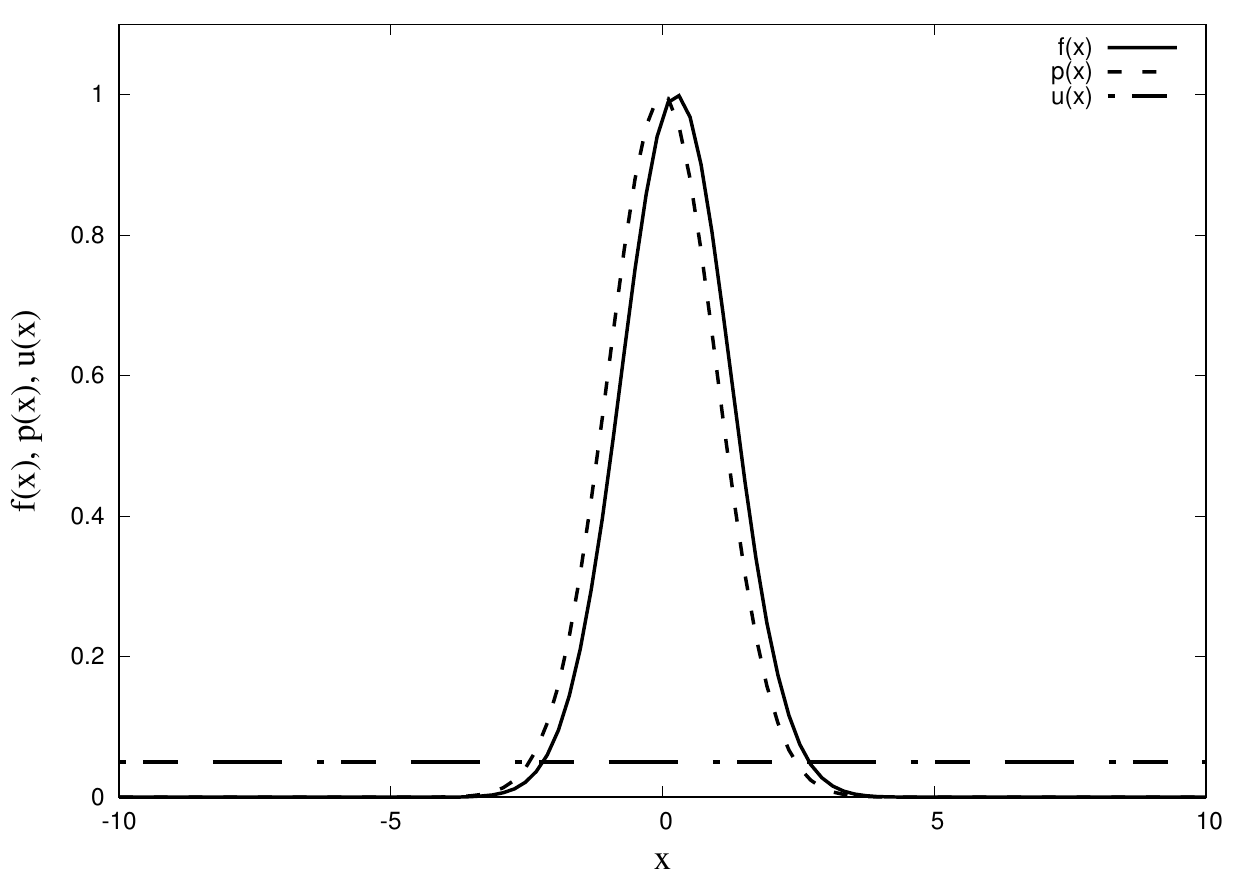}
  \caption{The integrand $f(x) = \exp \left( - \hf x^2 + \qtr x - \frac{1}{32} \right)$, the weight function used for importance sampling $p(x) = \exp \left( - \hf x^2 \right)$, and the weight function used for naive sampling $u(x) = {\rm Uniform} \; (-10, 10)$.}
  \label{fig:imp-sample-fun}
  \eec
\end{figure}

\begin{figure}[h]
\bec
  \includegraphics[width=9cm]{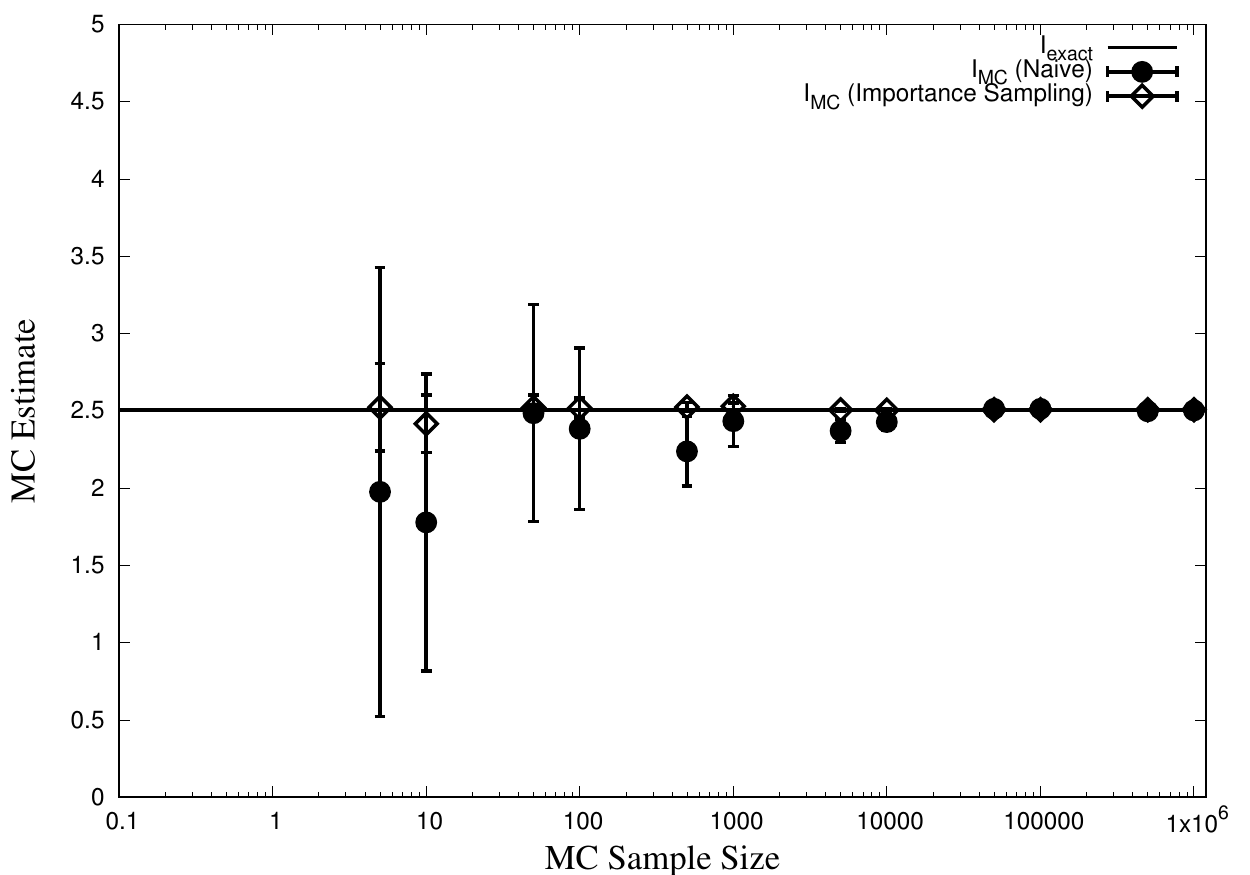}
  \caption{Monte Carlo estimate of the integral $I = \int_{-\infty}^{\infty} dx ~\exp \left( - \hf x^2 + \qtr x - \frac{1}{32} \right)$ using naive sampling (circles) and importance sampling (squares). The true value of the integral is $\sqrt{2 \pi} \simeq 2.5066$. In both cases the integral converges to the analytical value as the sample size is increased. However, statistical uncertainties using importance sampling are dramatically smaller for a given sample size.}
  \label{fig:mc-neutral-imp-samp-data}
  \eec
\end{figure}

Now we write the integral as
\bea
I &=& \int_{-10}^{10} dx ~ \left[ \exp \left( - \hf x^2 + \qtr x - \frac{1}{32} \right) \sqrt{2 \pi} \exp \left( \hf x^2 \right)\right]  \nn \\
&& ~~~~~~~~~~ \times \left\{ \frac{1}{\sqrt{2 \pi}} \exp\left( - \hf x^2 \right) \right\},
\eea
where the part in the square brackets is the quantity whose expectation is being calculated and the part in curly brackets is the density being integrated against. Fig. \ref{fig:imp-sample-fun} shows the functions $f(x)$ and $p(x)$. In Fig. \ref{fig:mc-neutral-imp-samp-data} we show the simulation data, comparing naive sampling and importance sampling estimates of the integral, and in Table \ref{table:mc-method-naive-imp} we show the corresponding numerical data. We provide a C++ program that produces the data with naive sampling in Appendix \ref{sec:num-int-monte-carlo-naive}, and a program that produces the importance sampling data in Appendix \ref{sec:num-int-monte-carlo-importance}.

\begin{table}[h]
\bec
\begin{tabular}{ c c c }
\hline
$n$ & $I_{\rm naive}(n) \pm \delta I(n)$ & $I_{\rm imp.~samp.}(n) \pm \delta I(n)$ \\ \hline \hline
5  &  $1.9756 \pm 1.4516$ & $2.5228 \pm 0.2832$  \\
10  &  $1.7779 \pm 0.9623$ & $2.4164 \pm 0.1850$  \\
50  &  $2.4844 \pm 0.7022$ & $2.5209 \pm 0.0803$  \\
100  &  $2.3842 \pm 0.5227$ & $2.5181 \pm 0.0656$  \\
500  &  $2.2380 \pm 0.2246$ & $2.5247 \pm 0.0294$  \\
1000  &  $2.4333 \pm 0.1646$ & $2.5299 \pm 0.0206$  \\
5000  &  $2.3700 \pm 0.0741$ & $2.5066 \pm 0.0091$  \\
10000  &  $2.4277 \pm 0.0533$ & $2.5039 \pm 0.0063$  \\
50000  &  $2.5133 \pm 0.0242$ & $2.5071 \pm 0.0029$  \\
100000  &  $2.5131 \pm 0.0171$ & $2.5071 \pm  0.0020$  \\
500000  &  $2.4976 \pm 0.0076$ & $2.5070 \pm 0.0009$  \\
1000000  &  $2.5029 \pm 0.0054$ & $2.5073 \pm 0.0006$  \\ \hline
\end{tabular}
\caption[Table]{Computing the integral given in Eq. \eqref{eq:integral} using naive and importance sampling Monte Carlo. The exact value is around $2.5066$.}
\label{table:mc-method-naive-imp}
\eec
\end{table}

\subsection{When Does Importance Sampling Fail?}
\label{sec:When-does-importance-sampling-fail?}

The tails of the distributions are important and we cannot ignore them. We may be happy with choosing an importance sampling function $p(x)$ that has roughly the same shape as that of $f(x)$. But serious difficulties can arise if $p(x)$ gets smaller much faster than $f(x)$ out in the tails. Realizing a value $x_r$ from the far tails of $p(x)$ is highly improbable. However, if it happens, then the Monte Carlo estimator will take a shock: the value, $f(x_r)/p(x_r)$ for such an improbable $x_r$ may be orders of magnitude larger than the typical values $f(x)/p(x)$ that we encounter. Therefore, conventional importance sampling techniques can turn out to be useless when applied to the problem of calculating tail probabilities for a given density function. (See Ref. \cite{Wessel:1990} for more details.)

Generally, rejection method and importance sampling method fail in higher dimensions. An alternative that works better in higher dimensions is Markov chain Monte Carlo (MCMC). Rejection sampling and importance sampling come under independent Monte Carlo method, while dependent Monte Carlo method consists of MCMC algorithms such as Gibbs sampling, Metropolis sampling and Hybrid (or Hamiltonian) Monte Carlo (HMC).

\begin{figure}[h]
\bec
  \includegraphics[width=9cm]{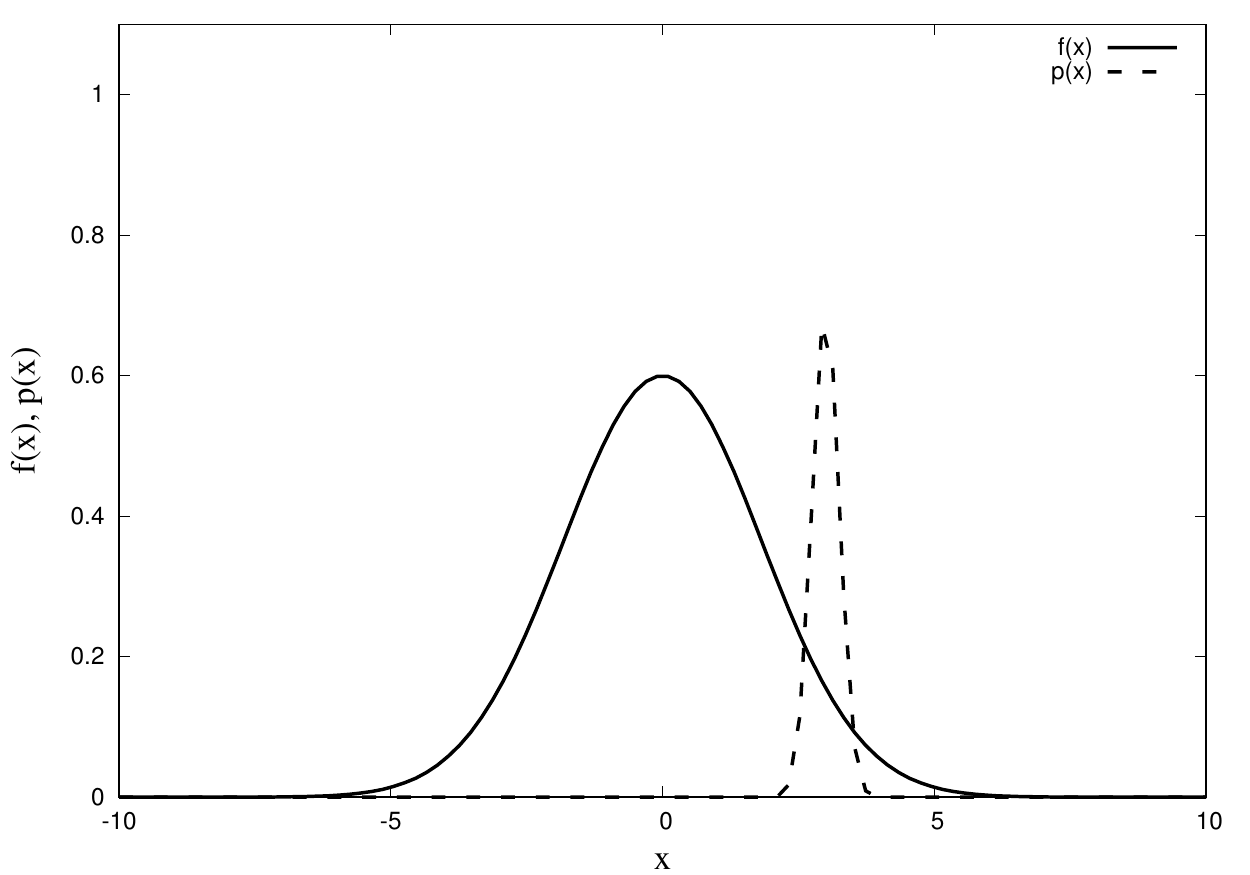}
  \caption{A situation where importance sampling fails. The red dashed curve represents the importance function $p(x)$. It fails to make the ratio $f(x)/p(x)$ nearly flat.}
  \label{fig:imp-sample-fail}
  \eec
\end{figure}

The technique of importance sampling is effective when the weight $p(x)$ approximates $f(x)$ over most of its domain. When $p(x)$ misses high probability regions of $f(x)$ and systematically gives sample points with small weights, importance sampling fails. We illustrate this situation in Fig. \ref{fig:imp-sample-fail}. We end up with high variance since the effective sample size is reduced. MCMC methods such as Metropolis sampling and HMC try to overcome this difficulty by biasing a local random search towards higher probability regions without sacrificing the asymptotic ``fair sampling" properties of the algorithm \cite{Neal:1993, Neal:1996, Schuurmans:2013}

\section{Markov Chains}
\label{sec:Markov-chains}

In the previous Section we looked at Monte Carlo integration methods that employ naive sampling and importance sampling. There, we used a uniform random sampling method with or without a weight function to find the integral of a `well-behaved' function.

Markov chain Monte Carlo (MCMC) is also a random sampling method. Unlike Monte Carlo integration, the goal of MCMC is not to sample a multi-dimensional region uniformly. Instead, the goal is to visit a point $\bx$ with a probability proportional to some given distribution function say, $\pi(\bx)$. The distribution $\pi(\bx)$ is not quite a probability. It is not necessarily normalized to have a unity integral over the sampled region. However, it is proportional to a probability. MCMC ``automatically" puts its sample points preferentially where $\pi(\bx)$ is large, in direct proportion. This is a huge advantage of using MCMC over independent Monte Carlo integration methods.

In a highly multi-dimensional space, or where the distribution $\pi(\bx)$ is expensive to compute, MCMC can be advantageous by many orders of magnitude compared to rejection sampling or naive Monte Carlo sampling.

In order to get an intuitive understanding of MCMC let us look at the {\it Rosenbrock function} (also known as Rosenbrock's banana function). This function is used as a performance test problem for optimization algorithms. The global minimum of the function is inside a long, narrow, parabolic flat valley. It is trivial to find the valley but it is difficult to converge to the global minimum.

\begin{figure}[t]
\bec
  \includegraphics[width=9cm]{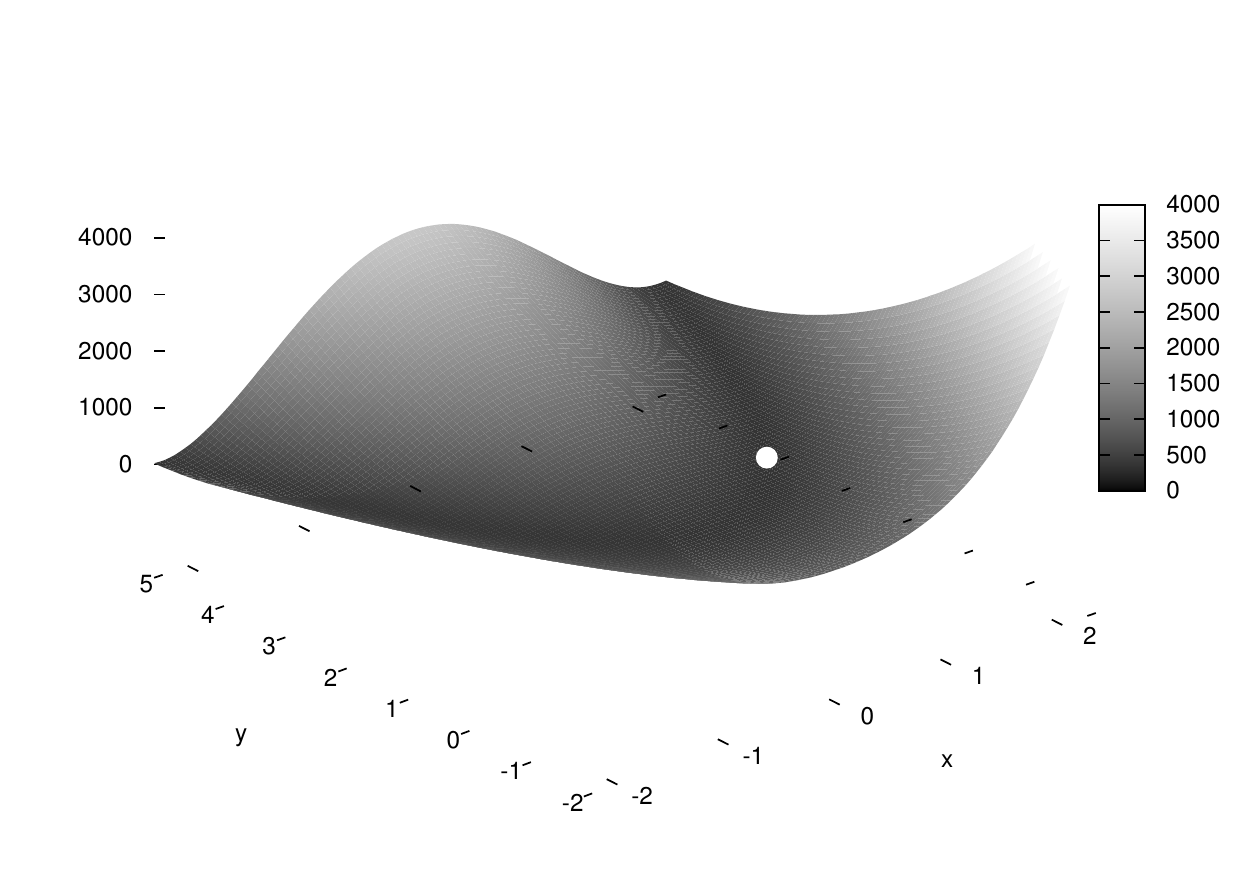}
  \caption{Plot of the Rosenbrock function of two variables, $f(x, y) = (a - x)^2 + b (y - x^2)^2$. The parameters are $a = 1$, $b = 100$. The global minimum, which is at $(x_0, y_0) = (a, a^2) = (1, 1)$, is indicated using a white filled circle.}
  \label{fig:Rosenbrock}
  \eec
\end{figure}

The two-dimensional Rosenbrock function is defined by
\beq
f(x, y) = (a - x)^2 + b (y - x^2)^2,
\eeq
with $a$ and $b$ constants. This function has a global minimum at $(x, y) = (a, a^2)$, where $f(x, y) = 0$. Fig. \ref{fig:Rosenbrock} shows a plot of the Rosenbrock function of two variables. 

\begin{figure}[t]
\bec
  \includegraphics[width=9cm]{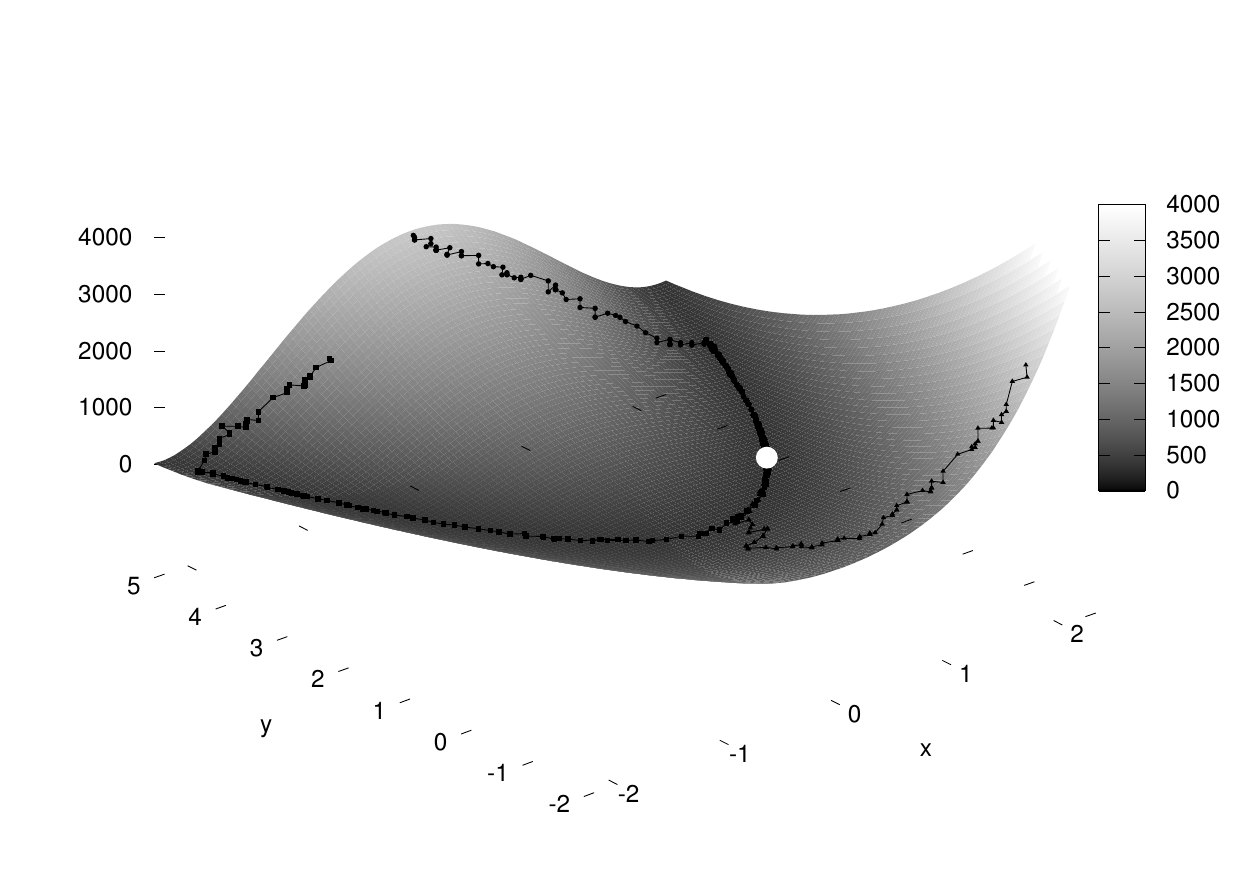}
  \caption{Three Markov chains with different starting points running on the two-dimensional Rosenbrock function with the help of Metropolis sampling algorithm. The white filled circle indicates the position of the global minimum. The sampling algorithm eventually finds the global minimum, irrespective of the position of the starting configuration.}
  \label{fig:MCMC_Rosen}
  \eec
\end{figure}

In Fig. \ref{fig:MCMC_Rosen} we show the simulation data corresponding to three Markov chains running on the two-dimensional Rosenbrock function with the help of Metropolis sampling algorithm. The three chains, though they have different starting points, finally converge to the same equilibrium distribution, which is around the global minimum (indicated by the white filled circle).

In MCMC, we sample distribution $\pi(\bx)$ by a Markov chain\footnote{Markov chains were introduced by the Russian mathematician Andrey Markov (1856 - 1922) in 1906.}. A Markov chain is a sequence of points, $\bx_0, \bx_1, \bx_2, \cdots $, that are {\it locally correlated}, with the {\it ergodic property}. The sequence of points will eventually visit every point $\bx$, in proportion to $\pi(\bx)$. Here Markov means that each point $\bx_i$ is chosen from a distribution that depends only on the value of the immediately preceding point $\bx_{i-1}$. The chain has memory extending only to one previous point and it is completely defined by a {\it transition probability function} of two variables, $p(\bx_i | \bx_{i-1})$. That is, the probability with which $\bx_i$ is picked given a previous point $\bx_{i-1}$.

If $p(\bx_i | \bx_{i-1})$ is chosen to satisfy the detailed balance equation
\beq
\label{eq:detailed-bal}
\pi(\bx_1) p(\bx_2 | \bx_1) = \pi(\bx_2) p(\bx_1 | \bx_2),
\eeq
then the Markov chain will in fact sample $\pi(\bx)$ ergodically.

Equation \eqref{eq:detailed-bal} expresses the idea of physical equilibrium
\beq
\bx_1 ~~ \longleftrightarrow~~ \bx_2.
\eeq

If $\bx_1$ and $\bx_2$ occur in proportion to $\pi(\bx_1)$ and $\pi(\bx_2)$, respectively, then the overall {\it transition rates} in each direction are the same. Transition rate here is a product of a {\it population density} and a {\it transition probability}.

Integrating both sides with respect to $\bx_1$
\beq
\label{eq:transition-prob}
\int d \bx_1 ~ \pi(\bx_1) ~p(\bx_2 | \bx_1) = \pi(\bx_2) \int d \bx_1 ~ p(\bx_1 | \bx_2) = \pi(\bx_2). 
\eeq

The left-hand side of the above equation is the probability of $\bx_2$, computed by integrating over all possible values of $\bx_1$ with the corresponding transition probability. The right-hand side is the desired $\pi(\bx_2)$. Thus Eq. \eqref{eq:transition-prob} says that if $\bx_1$ is drawn from $\pi$, then so is its successor $\bx_2$, in the Markov chain.

\subsection{Properties of Markov Chains}
\label{sec:Properties-of-Markov-chains}

Let us consider a Markov chain consisting of a sequence of random elements, $\bx_0, \bx_1, \bx_2, \cdots$, of some set\footnote{See Refs. \cite{Norris:1997, LevinPeresWilmer:2006} for in-depth introduction and applications of Markov chains.}. As noted before, the chain has the property that the conditional distribution of $\bx_i$, given $\bx_0, \bx_1, \bx_2, \cdots, \bx_{i-1}$ depends only on $\bx_{i-1}$. The set in which $\bx_0, \bx_1, \bx_2, \cdots$ take their values is called the state space $S$ of the Markov chain. A state space could be finite or infinite.

Markov chains exhibit the so-called {\it Markov property} or {\it memoryless property}. Memoryless property in words can be put as: {\it ``The future depends on the past only through the present."}

We are interested in finding a stationary distribution, $\pi(\bx)$, starting from an initial distribution say, $\mu(\bx_0)$. 

\beq
 \mu(\bx_0) \xrightarrow[\text{~~to stationary distribution~~}]{ } \pi(\bx)
\eeq

We achieve this through a series of Markov chain steps, jumping from $\bx_i$ to $\bx_{i+1}$, forming a chain. What is helping us is a transition probability function $p$. It is a matrix in the discrete case and a {\it kernel} in the continuous case.

For a Markov chain to have a stationary distribution, it must be {\it irreducible} and {\it aperiodic}. 

{\bf Irreducibility:} We can reach any other state in finite time (or steps), regardless of the present state we are in. That is, the probability to go from every state to every state, in one or more steps, is greater than zero. 

{\bf Aperiodicity:} If a state has period 1 then we say that it is aperiodic. That is, the state does not repeat after a given time period. If a state $s_i$ has period 2, then the chain can be in $s_i$ every second time.

To illustrate irreducibility and aperiodicity let us consider a Markov chain on a small state space
\beq
S = \{s_1, s_2, s_3, s_4, s_5 \}.
\eeq

The state $s_j$ is {\it accessible} from state $s_i$ if there exists a non-zero probability to reach state $s_j$ from state $s_i$. It is denoted by $s_i \to s_j$. If $s_i \to s_j$ and $s_j \to s_k$ then we must have $s_i \to s_k$. 

We say that the states $s_i$ and $s_j$ {\it communicate} if $s_i \to s_j$ and $s_j \to s_i$, and this property is denoted by $s_i \leftrightarrow s_j$. If we have $s_i \leftrightarrow s_j$ and $s_j \leftrightarrow s_k$ then we must have $s_i \leftrightarrow s_k$. The set of all states that communicate with each other is called a {\it class}. If $C_A$ and $C_B$ are two communicating classes, then either $C_A = C_B$ or $C_A$ and $C_B$ are disjoint. We can partition the set of all states into separate classes.

\begin{figure}[t]
\bec
  \includegraphics[width=9cm]{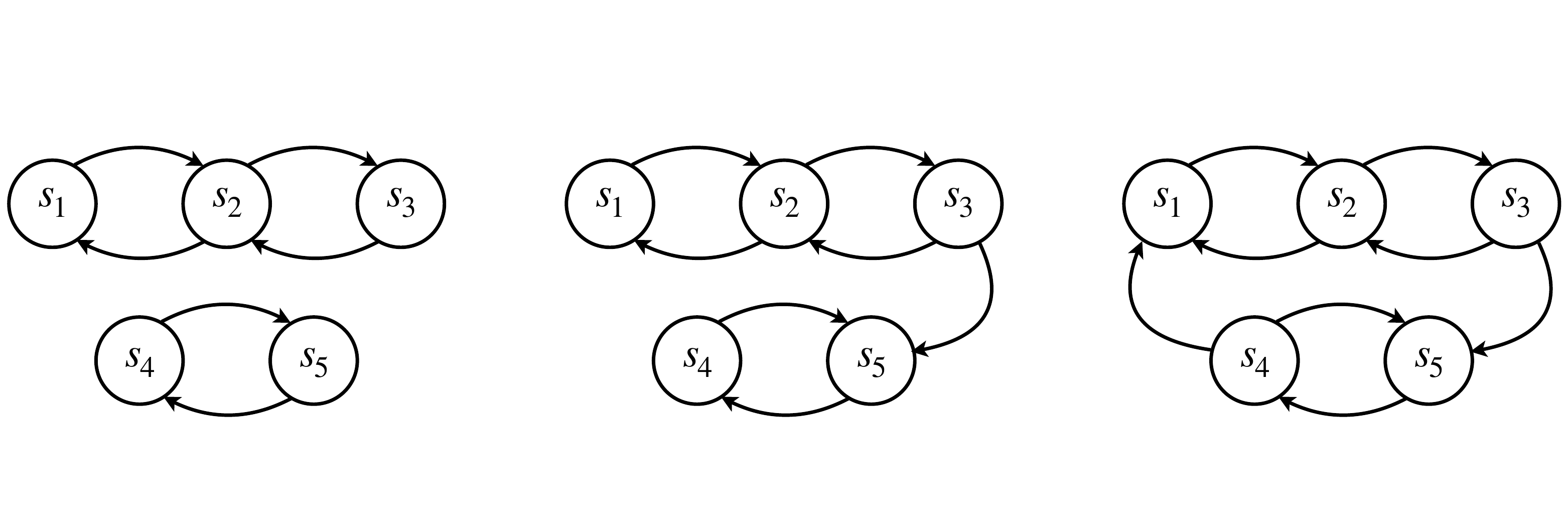}
  \caption{Transition graphs of three Markov chains. The left and middle ones are reducible. They form more than one communicating classes. The chain on the right is irreducible. That is, all states are in one single communicating class.}
  \label{fig:MC_irreducible}
  \eec
\end{figure}

\begin{figure}[h]
\bec
  \includegraphics[width=9cm]{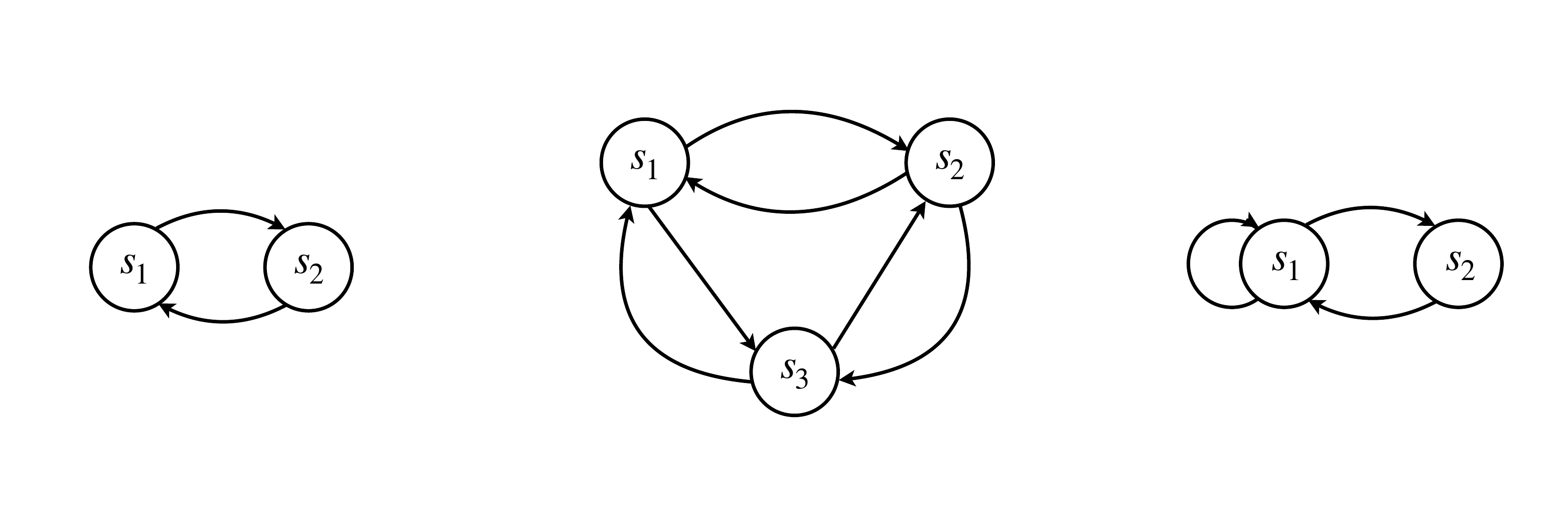}
  \caption{Among  the transition graphs of three Markov chains the one on the left has period 2 while the other two are aperiodic. That is, the state does not repeat after a given time period.}
  \label{fig:MC_aperiodic}
  \eec
\end{figure}

Let us look at the transition graphs of Markov chains within $S$. In Figs. \ref{fig:MC_irreducible} and  \ref{fig:MC_aperiodic} we provide such graphs. An arrow means positive transition probability. No arrow means zero transition probability. Fig. \ref{fig:MC_irreducible} shows example transition graphs of reducible and irreducible Markov chains, and Fig. \ref{fig:MC_aperiodic} shows those of periodic and aperiodic Markov chains.

\subsection{Convergence of Markov Chains}
\label{sec:Convergence-of-Markov-chains}

If both irreducibility and aperiodicity are respected, by a finite state Markov chain, then there exists a {\it stationary distribution}.

Let us denote the stationary distribution by $\pi(\bx)$. Start with an element $\bx_0$, drawn from an initial distribution $\mu(\bx)$. Then distribution of $\bx_n$ will converge to $\pi(\bx)$ in finite time (or number of steps). Thus we say that the Markov chain has reached an {\it equilibrium state}. 

{\bf Theorem:} Uniqueness theorem for Markov chains: Any irreducible and aperiodic Markov chain has exactly one stationary distribution. If we are able to find a candidate stationary distribution, then we have found the unique one.

{\bf Theorem:} Fundamental limit theorem for Markov chains: Let $\bx_0, \bx_1, \cdots, $ be an irreducible and aperiodic Markov chain, with state space
\beq
S = \{ s_1, s_2, \cdots, s_k \}, 
\eeq
and a transition matrix (or Markov matrix) $P$. It also has an arbitrary initial distribution $\mu^{(0)}$. Then, for any distribution $\pi$, which is stationary distribution for $P$, we have
\beq
\mu^{(n)} \longrightarrow \pi
\eeq
with $n \geq 0$.

What if when state space $S$ is uncountable? Then we must think of initial distribution as an {\it unconditional probability distribution}, and the transition probability as a {\it conditional probability distribution}.

A stationary distribution can be represented as a row vector $\pi$ whose entries are probabilities summing to 1. It satisfies
\beq
\label{eq:matrix-eq}
\pi = \pi P,
\eeq
for a given Markov matrix $P$. The above equation is the same as saying that $\pi$ is {\it invariant} by the matrix $P$.

Let us mention a few important properties of the Markov matrix: (i.) the product of two Markov matrices gives rise to another Markov matrix, (ii.) taking $\lambda_i$ as the eigenvalue of a Markov matrix we have $| \lambda_i | \leq 1$ and (iii.) at least one eigenvalue of the Markov matrix must be equal to 1.

Transposing the matrices in Eq. \eqref{eq:matrix-eq} we get
\beq
\pi^T = P^T \pi^T.
\eeq

The transposed transition matrix $P^T$ has eigenvectors with eigenvalue $1$ that are stationary distributions expressed as column vectors. This implies that if the eigenvectors of $P^T$ are known, then so are the stationary distributions of the Markov chain with transition matrix $P$. Thus, the stationary distribution is a left eigenvector of the transition matrix.

When there are multiple eigenvectors associated with the eigenvalue $1$, each such eigenvector gives rise to an associated stationary distribution. This can occur only when the Markov chain is reducible, that is, when it has multiple communicating classes.

A Markov chain with stationary distribution $\pi$ is called {\it reversible} if its transition matrix (or kernel) $P$ is such that it exhibits {\it detailed balance}
\beq
\pi(\bx) p(\bx, \by) = \pi(\by) p(\by, \bx).
\eeq

That is, the chain would look the same if we ran it forwards or backwards in time.

A {\it reversible chain} $\bx_0, \bx_1, \cdots $ has the property that if we start the chain in the stationary distribution and look at a typical realization
\beq
\cdots, \bx_{i-2}, \bx_{i-1}, \bx_i, \bx_{i+1}, \bx_{i+2} \cdots
\eeq
and then reverse the order
\beq
\cdots, \bx_{i+2}, \bx_{i+1}, \bx_i , \bx_{i-1}, \bx_{i-2} \cdots
\eeq
they will have the same probabilistic behavior.

Reversibility is a property of a distribution on $S$, which is related to the transition matrix $P$. Reversibility is a stronger statement than stationarity. Stationarity does not imply reversibility.

Let us look at a discrete example. Consider the following transition matrix $P^T$
\beq
P^T = 
\begin{pmatrix}
2/3 & 1/2 & 1/2 \\
1/6 & 0 & 1/2 \\
1/6 & 1/2 & 0 \\
\end{pmatrix},~~~~ p_{ij} = P(x_i \leftrightarrow x_j).
\eeq

This transition matrix represents the Markov chain as a matrix containing the transition probabilities. The matrix is normalized such that the elements of any given column add up to 1.

The equilibrium distribution $\pi^T$ is
\beq
\pi^T = 
\begin{pmatrix}
3/5 \\
1/5 \\
1/5 \\
\end{pmatrix}.
\eeq

Again note that the elements of $\pi^T$ add up to 1. We can show that  $\pi$ is an {\it invariant distribution} of $P$. That is, $P^T \pi^T = \pi^T$ 
\beq
\sum_x P(x' \leftarrow x) \pi(x) = \pi(x).
\eeq

We can also show that $\pi$ is the equilibrium distribution of $P$. Suppose we start with an initial distribution $\mu$ say,
\beq
\mu^T = \begin{pmatrix}
1 \\
0 \\
0 \\
\end{pmatrix}.
\eeq

Then, successive application of $P^T$ on $\mu^T$ would take us to the equilibrium distribution $\pi^T$. That is,
\beq
\underbrace{\left( P^T \right) \left( P^T \right) \cdots \left( P^T \right)}_{n \text{~times},~n \geq 1} \mu^T = \pi^T.
\eeq

\section{Markov Chain Monte Carlo}
\label{sec:Markov-chain-Monte-Carlo}

As we have seen, in the previous Section, in a Markov chain we have the following two types of distributions, leading to a joint distribution
\begin{enumerate}
\item{The marginal distribution of $\bx_0$, called the {\it initial distribution}.}
\item{The conditional distribution of $\bx_{i+1}$ given $\bx_i$, called the {\it transition probability distribution}.}
\end{enumerate}

If the state space is finite, then the initial distribution can be associated with a vector. Then the transition probabilities can be associated with a matrix $P$ having elements $p_{ij}$.

In naive and importance sampling Monte Carlo the random samples of the integrand used are statistically independent. In MCMC methods the samples are auto-correlated. Correlations of samples introduces the need to use the {\it Markov chain central limit theorem} when estimating the error of mean values.

There exist several algorithms that can create a Markov chain that leads to the unique stationary distribution, which is proportional to the given probability function. The two important ones are:
\begin{enumerate}
\item{Metropolis-Hastings algorithm,}
\item{Hybrid Monte Carlo or Hamiltonian Monte Carlo (HMC) algorithm.} 
\end{enumerate}

\subsection{Metropolis-Hastings Algorithm}
\label{sec:Metropolis-Hastings-algorithm}

Metropolis-Hastings algorithm \cite{Hastings:1970aa} is a general framework of MCMC to obtain a chain of random numbers from a probability distribution. Typically, direct sampling would be difficult from such probability distributions. This algorithm was introduced by W. K. Hastings in 1970 and it includes as a special case, the very first and a simpler MCMC, Metropolis algorithm \cite{Metropolis:1953am}.  

Metropolis-Hastings algorithm can get random samples from any probability distribution $\Pi(\bx)$, given that (i.) we have a function $\pi(\bx)$ in hand, which is proportional to the density of $\Pi$, that is,
\beq
\pi(\bx) = \frac{1}{Z} \Pi(\bx),
\eeq
and (ii.) the values of $\pi(\bx)$ can be calculated. We note that, in practice, computing the required normalization factor $Z$ is often extremely difficult. 

The Metropolis-Hastings algorithm works the following way. By generating a sequence of random samples, in such a way that, as more and more values are generated from the sample, the distribution of these values become more and more close to the the target distribution $\Pi(\bx)$. Since these samples are produced in an iterative manner and the distribution of the next sample depends only on the current sample value, the sequence of samples turn into a Markov chain. To be more specific, at a given step of the iteration, the algorithm selects a candidate for the next sample value based on the current sample value. After that, the candidate sample value is either rejected or accepted with some probability. The probability of acceptance is obtained by making a comparison between the value of the function $\pi(\bx)$ of the current and the candidate sample value with respect to the target distribution $\Pi(\bx)$.

In order to illustrate this process, let us look at the Metropolis algorithm, which is a special case of the Metropolis-Hastings algorithm, where we use a symmetric proposal function.

\subsection{Metropolis Algorithm}
\label{sec:Metropolis-algorithm}

Let  $\pi(\bx)$ be a function that is proportional to the desired (target) probability distribution $\Pi(\bx)$.

\begin{enumerate}
\item{{\bf Initialization.} In this step we choose an arbitrary point $\bx_0$ as a first sample. Let us denote the conditional probability density given $\by$ as $p(\cdot | \by)$. This arbitrary probability density $p(\bx | \by)$ selects a choice for the next sample value $\bx$, given the previous value $\by$. Note that the probability $p$ is required to be symmetric for the Metropolis algorithm. That is, it must satisfy $p(\bx | \by) = p(\by | \bx)$. Commonly, we choose $p(\bx | \by)$ to be a Gaussian distribution with mean at $\by$, such that the points around $\by$ are more likely to be selected next - creating a random walk out of the sequence of samples selected. The function $p$ is commonly called a {\it proposal density} or {\it jumping distribution}.}
\item{For each iteration $i$, starting with $i = 0$:}
\begin{enumerate}
\item{{\bf Generate.} A candidate $\bx'$ for the next sample is generated by selecting from the distribution $p(\bx' | \bx_i)$.}
\item{{\bf Calculate.} Calculate the {\it Metropolis ratio} ({\it acceptance ratio} or {\it odds ratio})}
\beq
r = \frac{\pi(\bx')}{\pi(\bx_i)}.
\eeq
This ratio determines if the candidate should be rejected or accepted. Since $\pi$ is proportional to the density of $\Pi$, the ratio is 
\beq
r = \frac{\pi(\bx')}{\pi(\bx_i)} = \frac{\Pi(\bx')}{\Pi(\bx_i)}
\eeq
\item{{\bf Accept or Reject.}}
\begin{enumerate}
\item{Generate a uniform random number $u$ on $[0, 1]$.}
\item{If $u \leq r$ accept the candidate by setting $\bx_{i+1} = \bx'$,}
\item{If $u > r$ reject the candidate and set $\bx_{i+1} = \bx_i$, instead.}
\end{enumerate}
\end{enumerate}
\end{enumerate}

When the proposal distribution is not symmetric we compute the {\it Hastings ratio}:
\beq
r(\bx_i, \bx') = \frac{\pi(\bx') p(\bx', \bx_i)}{\pi(\bx_i) p(\bx_i, \bx')}.
\eeq

Thus, sometimes accepting the sample choice and sometimes remaining in the same place this algorithm moves forward within the sample space. How probable the new proposed sample value is with respect to the current sample value, according to the distribution $\Pi(\bx)$, is encoded in the acceptance ratio $r$. Suppose we try to move to a point in a higher density region (and thus more probable than the existing point) of $\Pi(\bx)$, we will always accept the move. If the algorithm tries to move to a point with lower density (less probable), it will sometimes accept the move, and the more the relative drop in the probability, the less likely we are to accept the new sample point. In any event, the algorithm tends to remain in and thus generate large numbers of samples from, higher-density parts of $\Pi(\bx)$, while only sporadically encountering the low-density parts. This is the intuitive explanation of why this algorithm works, and generates samples that follow the target distribution $\Pi(\bx)$. 

{\bf Theorem:} Metropolis-Hastings theorem: Metropolis-Hastings update is reversible with respect to the invariant distribution $\pi(\bx)$ of the Markov chain. 

This theorem tells us that the transition probability that describes the update is reversible with respect to the distribution having unnormalized density $\pi(\bx)$. Note that this form of the acceptance probability, $r(\bx_1, \bx_2)$, is not unique. There can be many other possibilities of acceptance probability functions, which can provide a chain with the desired properties. One can show that this form is {\it optimal}, in that suitable candidates are rejected least often and thus the statistical efficiency is maximized.

For the conditional probability $p$, there is an infinite range of available choices. One choice is having a Gaussian. Another simple choice is picking a point from a uniform distribution. That is, consider the random update trial as
\beq
\bx_{i+1} = \bx_i + {\bf e},
\eeq
where ${\bf e} \sim U$, and $U$ is a uniform distribution say, $[-c, +c]$ with $c = 0.5$. We note that for the probability distribution we use, $+c$ and $-c$ should appear with the same probability, otherwise the detailed balance condition would not be respected.

The convergence of the chain will depend on the relationship between $p$ and $\pi$. For practical purposes we should choose $p$ such that it can easily be sampled and evaluated.

\subsection{Worked Example - Metropolis for Gaussian Integral}
\label{sec:Worked-example-Metropolis-for-Gaussian-integral}

Let us consider the Gaussian function $\Pi(x) = \exp(-x^2)$ giving
\beq
\pi(x) = \frac{1}{Z} \Pi(x) = \frac{e^{-x^2}}{\int dx ~ e^{-x^2}},
\eeq
where the denominator just normalizes $\exp(-x^2)$ to a probability.

Let us compute the averages
\beq
\label{eq:simple-model-metro}
\langle x \rangle = \frac{\int dx ~ x e^{-x^2}}{\int dx ~ e^{-x^2}} ~~{\rm and}~~\langle x^2 \rangle = \frac{\int dx ~ x^2 e^{-x^2}}{\int dx ~ e^{-x^2}},
\eeq
using Metropolis sampling\footnote{We note that 
\beq
\int_{-\infty}^\infty dx ~ e^{-a x^2} = \sqrt{\frac{\pi}{a}}, ~~ \int_{-\infty}^\infty dx ~ x ~e^{-a x^2} = 0, ~~{\rm and}~~\int_{-\infty}^\infty dx ~ x^2 ~e^{-a x^2} = \frac{1}{2a} \sqrt{\frac{\pi}{a}}. \nn
\eeq
}.

Metropolis algorithm to compute $\langle x \rangle$ and $\langle x^2 \rangle$ can be constructed in the following way. 

Start from a random initial state say, $x = x_0$. Then choose a move $x_1 = x_0 + \epsilon R$, where $\epsilon$ is the step size and $R$ is a random number between $[-0.5, 0.5]$. In the next step, construct the Metropolis ratio $r = \exp(- \Delta x)$ with $\Delta x = x^2_1 - x^2_0$. In order to perform Metropolis test, throw in a random number $u$ in the interval $[0, 1]$. If $u < r$ accept the trial move $x_1$ as the new $x_0$, otherwise keep the $x_0$ as the new $x_0$. Repeat the process and eventually the attempts will converge to the equilibrium distribution.

\begin{figure}[h]
\bec
  \includegraphics[width=9cm]{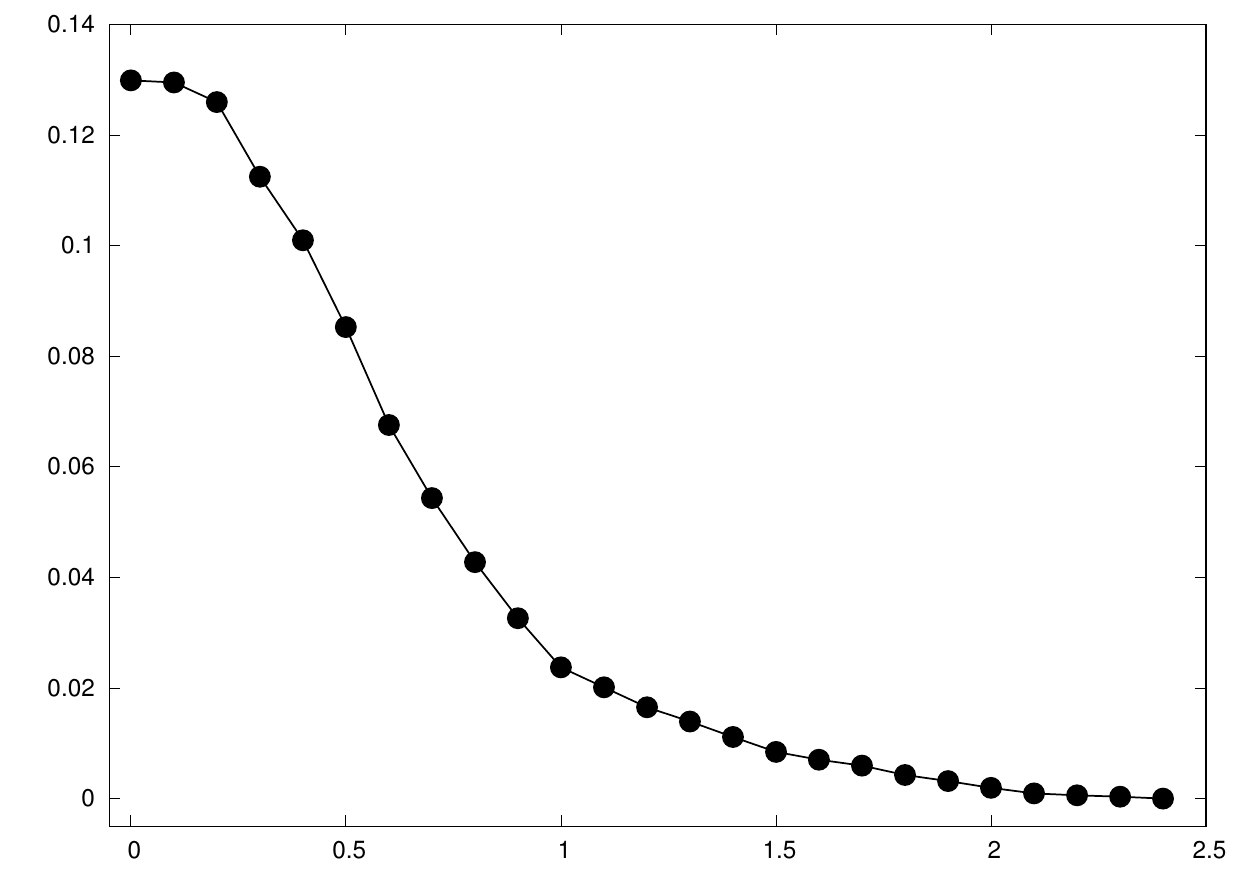}
  \caption{The distribution of $x^2$ from applying Metropolis sampling to compute $\langle x^2 \rangle = Z^{-1}\int dx ~ x^2 e^{-x^2}$, with $Z = \int dx ~ e^{-x^2}$. We see that the distribution is half of a Gaussian, as expected.}
  \label{fig:dist_x_sq}
  \eec
\end{figure}

A C++ program to compute $\langle x \rangle$ and $\langle x^2 \rangle$ is provided in Appendix \ref{sec:simple-metro-x-xsq}. Running this program with $N = 10^6$ samples and step size $\epsilon = 0.75$ gives the results $\langle x \rangle = 0.0045 \pm 0.0023$ and $\langle x^2 \rangle = 0.5147 \pm 0.0023$, and they are close to the exact results, 0 and $\hf$, respectively. In Fig. \ref{fig:dist_x_sq} we provide the distribution of $x^2$, which is a Gaussian, as expected.

The parameters of the algorithm are the sample size $N$ and the step size $\epsilon$. In order to improve the efficiency of the algorithm we need to choose a suitable step size such that the simulations have a reasonable acceptance rate for the proposed state values. What we mean by ``reasonable" depends on the algorithm. For random walk Metropolis, a reasonable step size would be the one that gives about $25\%$ acceptance rate. We can increase the sample size $N$ to bring down the Monte Carlo error to a desired accuracy say, $1\%$ to $5\%$. We will discuss error reduction techniques, including the one related to the so-called {\it auto-correlation} of the observables, in Section \ref{sec:Reliability-of-simulations}.    

\begin{figure}[h]
\bec
  \includegraphics[width=9cm]{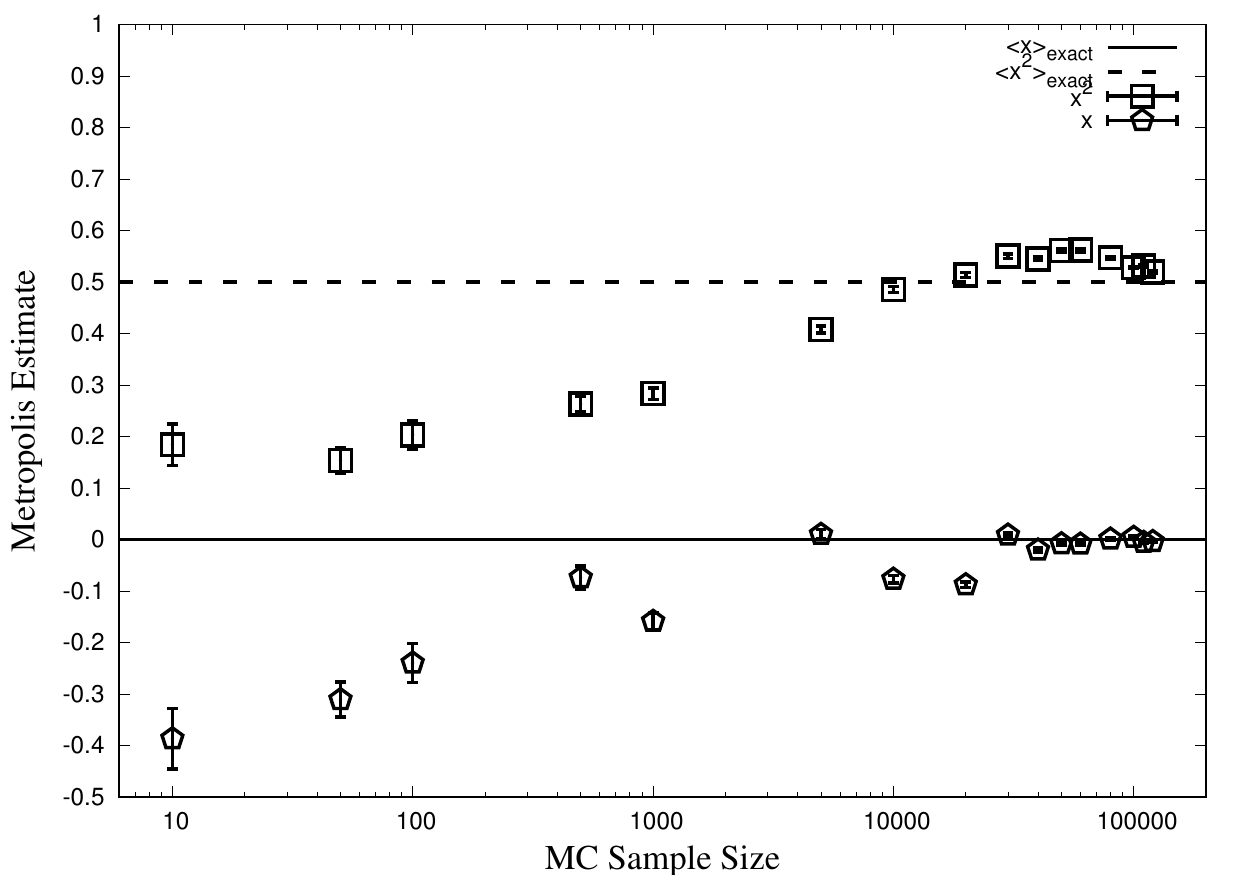}
  \caption{The quantities $x$ and $x^2$, computed using Metropolis sampling, against the Monte Carlo sample size for the Gaussian model. As the sample size is increased  $x$ and $x^2$ approach their exact values, 0 and 0.5, respectively.}
  \label{fig:x_x_sq}
  \eec
\end{figure}

In Fig. \ref{fig:x_x_sq} we show how $x$ and $x^2$ approach their exact values as the sample size is increased.

\subsection{Thermalization in Markov Chain Monte Carlo}
\label{sec:Thermalization-in-MCMC}

Thermalization (or burn-in) is a term that describes the practice of throwing away some part of the iterations at the beginning of an MCMC run. In other words, it is the process of bringing the Markov chain into equilibrium from a random starting probability vector. The Markov process of generating one state of configuration after other is referred to as {\it updating}. The starting point of the simulations can be arbitrary and it can be far away from the equilibrium values. The chain explores the configuration space through a series of updating process and eventually binds to the equilibrium configurations. We then discard some iterations from the beginning of the MCMC run to the time around which it merges on to the equilibrium configurations. 

Suppose $M$ iterations are used for thermalization. Then the ergodic average of an observable $O$ is calculated in the following way
\beq
\langle O \rangle = \frac{1}{N-M} \sum_{k = M+1}^N O(\bx_k).
\eeq

\begin{figure}[h]
\bec
  \includegraphics[width=9cm]{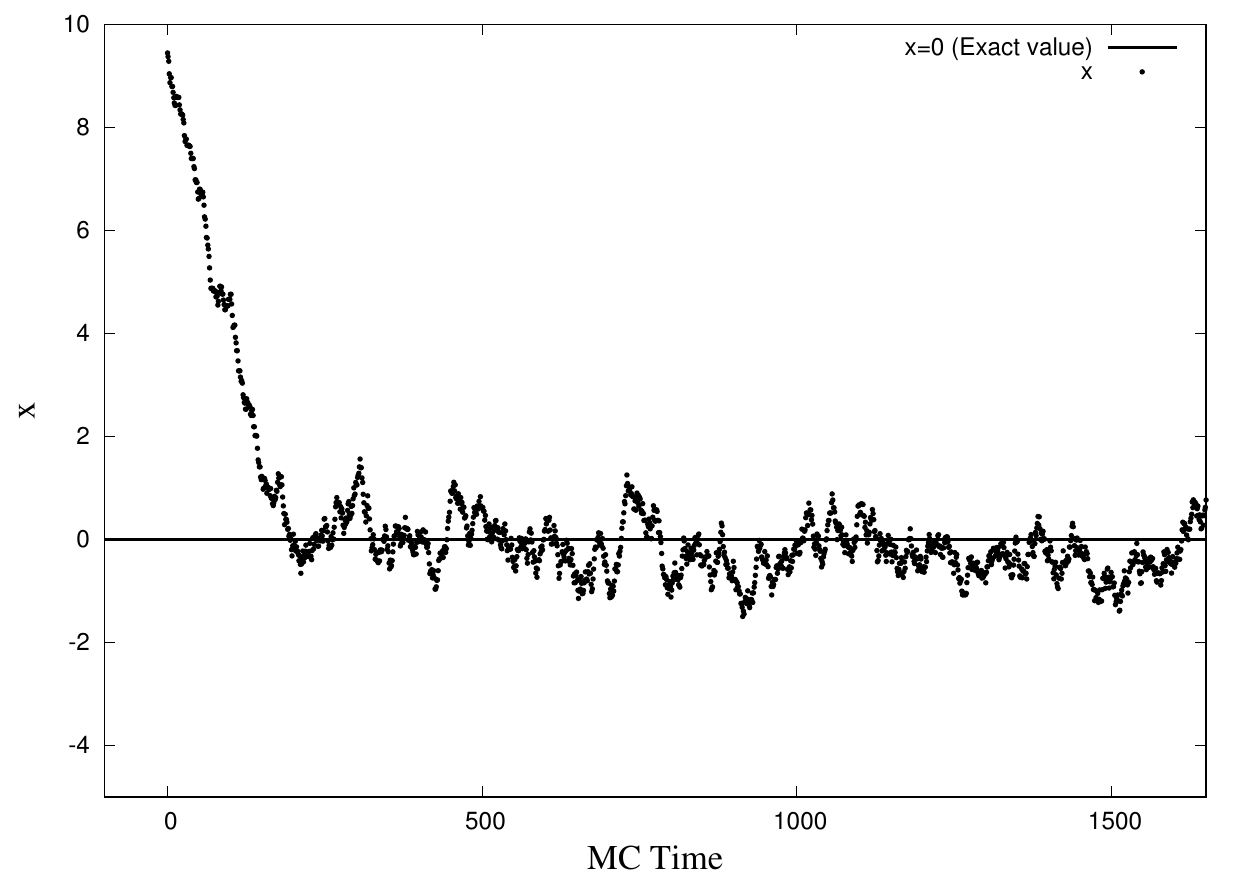}
  \caption{Thermalization or burn-in time history of the observable $x$ of the simple Gaussian model. See Eq. \eqref{eq:simple-model-metro}. The Metropolis step size used is $\epsilon = 0.5$ and starting value is $x_0 = 9.5$.}
  \label{fig:burn_in}
  \eec
\end{figure}

As an illustration of thermalization, we show in Fig. \ref{fig:burn_in} the thermalization time history for the observable $x$ of the simple Gaussian model.  

\section{MCMC and Feynman Path Integrals}
\label{sec:Monte-Carlo-and-Feynman-path-integrals}

This Section reveals the connection between Feynman path integrals in Euclidean quantum field theory and Markov chain Monte Carlo. We begin with transition amplitudes in quantum field theory and then introduce Feynman path integral in Euclidean spacetime as a way to extract the observables in a quantum field theory. We also look at physics examples such as supersymmetry breaking in a zero-dimensional quantum field theory with a square-well potential, two-point correlation function in a one-dimensional simple harmonic oscillator, and a matrix model with $U(N)$ symmetry, that undergoes a phase transition as the coupling parameter is varied. 

\subsection{Transition Amplitudes}
\label{sec:Transition-amplitudes}

Let us consider a system where a small particle of mass $m$ is constrained to move only along the $x$-axis. The trajectory of the particle is described by its location $x$ as a function of time $t$, which we denote as $x(t)$. 

When we consider the quantum mechanics of such a system we are interested in a quantity known is the {\it transition amplitude}. It is the probability amplitude for the particle to go from point $x_i$ at time $t_i$ to point $x_f$ at time $t_f$. We denote it as
\beq
M(f, i) \equiv \langle x_f(t_f) | x_i(t_i) \rangle
\eeq 

Let us work in the Heisenberg picture of quantum mechanics where the state vectors $| \psi \rangle$ are time independent, and operators and their eigenvectors evolve with time
\beq
\phi(t) = e^{i H t} \phi(0) e^{-i Ht} ~~~{\rm and}~~~| \phi(t) \rangle = e^{i H t} |\phi_n(0) \rangle.
\eeq

Inserting a complete and discrete set of Heisenberg-picture eigenstates $|\phi_n(0) \rangle$ of the Hamiltonian into the transition amplitude
\bea
\label{eq:transition-amp}
M(f, i) &=& \sum_n \langle x_f(t_f) | \phi_n(0) \rangle \langle \phi_n(0) | x_i(t_i) \rangle \nn \\
&=& \sum_n \phi_n(x_f) \phi^*_n(x_i) e^{-iE_n(t_f - t_i)/\hbar},
\eea
where $\langle x(0) | \phi_n(0) \rangle \equiv \phi_n(x)$ is the wavefunction in coordinate space of the $n$-th stationary state. 
 
Thus we see that the transition amplitude contains information about all energy levels and all wavefunctions.

Let us see how we can compute the expectation values of observables in the ground state (vacuum) of the theory. The transition amplitude given in Eq. \eqref{eq:transition-amp} can provide this information in the limit of very large time $T$, by taking $t_i = -T$ and $t_f = T$. We have 
\beq
\langle x_f(T) | x_i(-T) \rangle = \sum_{n=0}^\infty \langle x_f(0) | \phi_n(0) \rangle \langle \phi_n(0) | x_i(0) \rangle e^{- 2i E_n T/\hbar}.
\eeq 

Let us assume that the vacuum of the theory is non-degenerate. Also using $E_{n+1} > E_n$, for $n = 0, 1, 2, \cdots$, we can explore the properties of the ground state of the model. We get  
\beq
\label{eq:vacc-T}
\langle x_f(T) | x_i(-T) \rangle \simeq \langle x_f(0) | 0 \rangle \langle 0 | x_i(0) \rangle e^{- 2i E_0 T/\hbar}.
\eeq 

We can also apply the limit of large $T$ to find more complicated amplitudes
\bea
&&\langle x_f(T) | x(t_2) x(t_1) | x_i(-T) \rangle \nn \\
&& \hspace{2cm} = \sum_{n, m = 0}^\infty \langle x_f(0) | \phi_n(0) \rangle \langle \phi_n(0) | x(t_2) x(t_1) | \phi_m(0) \rangle \nn \\
&& \hspace{3cm} \times \langle \phi_m(0)    | x_i(0) \rangle e^{- i (E_n + E_m) T/\hbar}.
\eea 

Upon simplification this gives
\beq
\label{eq:two-pt-T}
\langle x_f(T) | x(t_2) x(t_1) | x_i(-T) \rangle \simeq \langle x_f(0) | 0 \rangle \langle 0 | x(t_2) x(t_1) | 0 \rangle \langle 0 |x_i(0) \rangle e^{- 2 i E_0 T/\hbar}.
\eeq 

Taking the ratio of Eq. \eqref{eq:two-pt-T} and \eqref{eq:vacc-T}, we can obtain the vacuum expectation value of $x(t_2) x(t_1)$
\beq
\langle 0 | x(t_2) x(t_1) | 0 \rangle = \lim_{{\rm large}~T} \frac{\langle x_f(T) | x(t_2) x(t_1) | x_i(-T) \rangle }{\langle x_f(T) | x_i(-T) \rangle}.
\eeq

The above result can be generalized to higher products of the position operator $x(t)$. It is interesting to note that all observables of this theory can be extracted from the correlation functions (vacuum expectation values) of the position operator. 

The energies of the stationary states, for example, are contained in the two-point correlator
\beq
\label{eq:oscillatory-E}
\langle 0 |x(t) x(0) | 0 \rangle = \sum_n | \langle 0 | x(0) | \phi_n(0) \rangle |^2 e^{-i E_n t / \hbar}  
\eeq

In a similar manner we can also extract more complicated correlation functions.

From Eq. \eqref{eq:oscillatory-E} we see that the energies $E_n$ are encoded in oscillatory functions and it is very difficult to extract energies from these oscillatory exponentials. Such a task would have been much easier if we had decaying exponentials. 

It is indeed possible to get decaying exponentials. We can make use of a procedure known as Wick rotation: rotate the time axis from the real axis to the imaginary axis. The rotation amounts to $t \to -i \tau$, and the Wick rotated correlation function has the form 
\beq
\langle 0 |x(\tau) x(0) | 0 \rangle = \sum_n | \langle 0 | x(0) | \phi_n(0) \rangle |^2 e^{-E_n \tau / \hbar}.  
\eeq

This imaginary time formalism provides an important advantage for Monte Carlo simulations of quantum field theories.

\subsection{Feynman Path Integrals}
\label{sec:Feynman-path-integrals}

The quantum mechanical transition amplitude, $M(f, i)$, mentioned in the previous section can be computed in several ways. 

In the language of Feynman path integral we can write it as
\beq
M(f, i) \sim \sum_{{\cal P}} \exp \left( i S[x(t)] / \hbar \right),
\eeq
with $S[x(t)]$ denoting the action and ${\cal P}$ representing all paths $x(t)$ from $x_i(t_i)$ to $x_f(t_f)$.

The above expression tells us that the transition amplitude is a sum over histories or a path integral. All paths contribute to the transition amplitude, but with different phases determined by the action. Thus, in this formalism, we only need to compute a multi-dimensional integral in order to evaluate the transition amplitude.

We can compute any correlation function using path integrals. In situations where we have to deal with strongly interacting quantum field theories, such as QCD, we need to numerically evaluate the required path integrals using computers.

Coming back to the case of the single particle of mass $m$, constrained to move only along the $x$-axis, the action is given by
\beq
S = \int L(x, \dot{x}) \; dt = \int (K - U) \; dt,
\eeq
where $L, K, U$ are the Lagrangian, kinetic energy and potential energy, respectively, of the particle.
 
Let us first divide time into small steps of width $\epsilon$, with $N \epsilon = t_f - t_i$ for large integer $N$. 

\begin{figure}[t]
\bec
  \includegraphics[width=9cm]{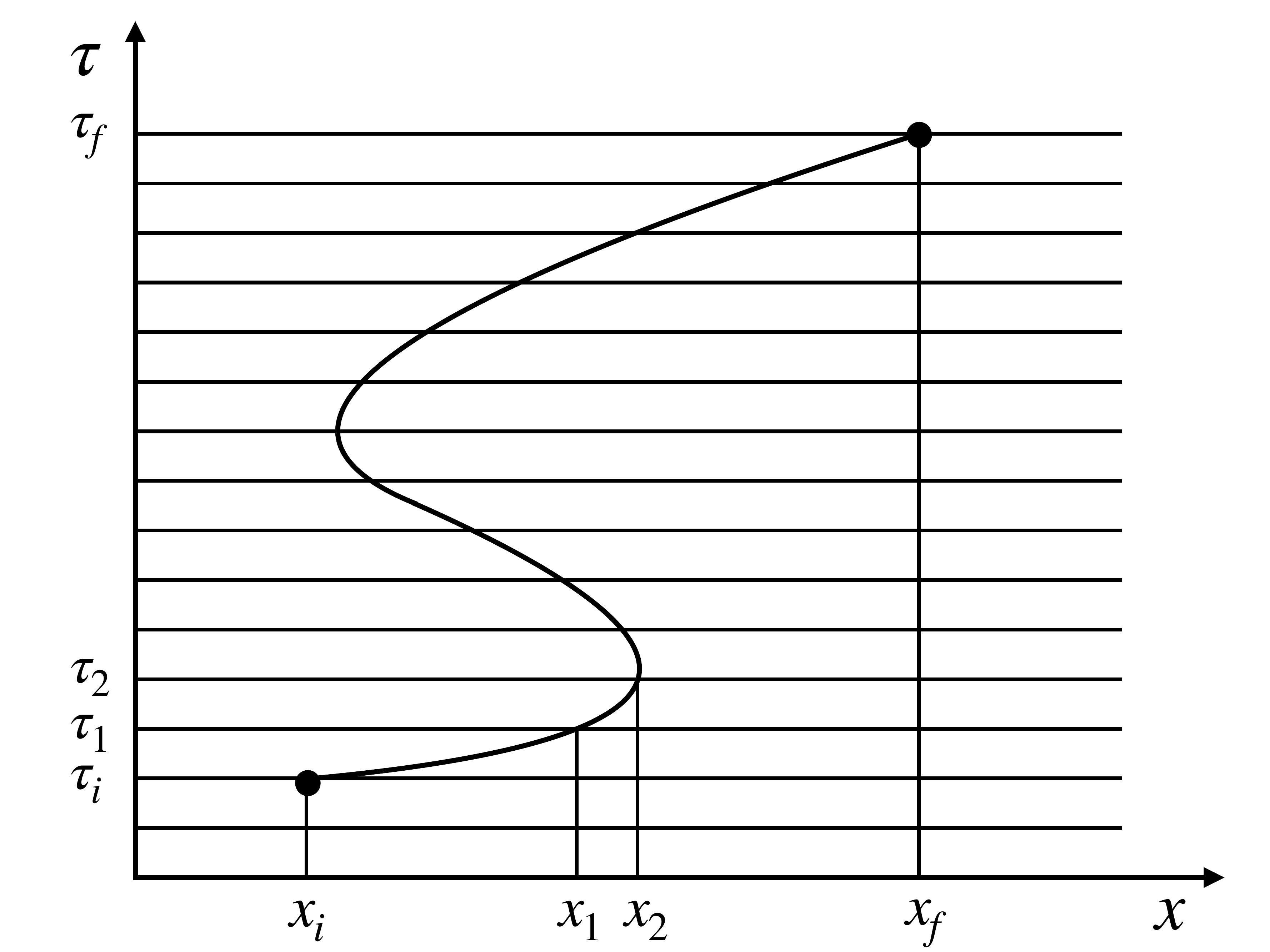}
  \caption{A typical path in the path integral formalism for a non-relativistic free particle of mass $m$ moving in one dimension.}
  \label{fig:plot-path-lattice}
  \eec
\end{figure}

The path integral is now defined as
\beq
M(f, i) = \lim_{N \to \infty} \frac{1}{B} \int \frac{dx_1}{B} \frac{dx_2}{B} \cdots \frac{dx_{N-1}}{B} \exp \left( i S[x(t)] / \hbar \right),
\eeq
where $B$ is a normalization factor depending on $\epsilon = (t_f - t_i)/N$, and chosen so that the path integral is well defined. In a non-relativistic theory it is forbidden for paths to double back in time. In Fig. \ref{fig:plot-path-lattice} we show a typical path.

For a free particle of mass $m$ in one dimension the Lagrangian is 
\beq
L = \hf m \dot{x}^2.
\eeq  

The amplitude for the particle to travel from $x_i$ at time $t_i$ to location $x_f$ at later time $t_f$ has the form
\beq
\langle x_f(t_f) | x_i(t_i) \rangle = \int_i^f Dx(t) \; e^{i S[f, i]/\hbar},
\eeq
and the integration denotes the sum over all allowed paths with $x(t_i) = x_i$ and $x(t_f) = x_f$. 

After performing the path integral we get the final result for the transition amplitude for a free particle in one dimension
\beq
\langle x_f(t_f) | x_i(t_i) \rangle = \sqrt{ \frac{m}{2 \pi i \hbar (t_f - t_i)}}~ \exp \left( \frac{i m (x_f - x_i)^2}{2 \hbar (t_f - t_i)} \right).
\eeq

For a free particle of mass $m$ moving in one dimension with periodic boundary conditions at $x=0$ and $x = L_x$ the transition amplitude takes the form
\beq
\langle x_f(t_f) | x_i(t_i) \rangle = \sqrt{ \frac{m}{2 \pi i \hbar (t_f - t_i)} } \sum_{n = -\infty}^\infty \exp \left( \frac{i m (nL_x + x_f - x_i)^2}{2 \hbar (t_f - t_i)} \right).
\eeq

For a simple harmonic oscillator we have the (Minkowski) Lagrangian
\beq
L = K - U = \hf m \dot{x}^2 - \hf m \omega^2 x^2,
\eeq
with $m$ and $\omega$ denoting the mass and frequency, respectively.

The transition amplitude has the form
\beq
\langle x_f(t_f) | x_i(t_i) \rangle = \sqrt{ \frac{m \omega}{2 \pi i \hbar \sin\omega T} } ~e^{ \frac{i S_{cl}}{\hbar} },
\eeq
where the classical action is  
\beq
S_{cl} = \frac{m \omega}{2 \sin (\omega T)} \left[ (x_i^2 + x_f^2) \cos \omega T - 2 x_i x_f \right], 
\eeq
with $T = t_f - t_i$.

\subsection{Worked Example - Supersymmetry Breaking}
\label{sec:Worked-example-Supersymmetry-breaking}

Let us consider a zero-dimensional quantum field theory consisting of a scalar field $\phi$ and two Grassmann odd variables (fermions) $\psi$ and $\psib$.  The action (which is the same as the Lagrangian) of this model has the form
\beq
\label{eq:susy-model}
S = \hf B^2 + i B W'  - \psib W'' \psi.
\eeq 

The potential $W(\phi)$ is called the superpotential, and $W'$ and $W''$ are its derivatives with respect to $\phi$. 

Let us use the following ``square-well" form for $W'$
\beq
W' = g (\phi^2 + \mu^2),
\eeq
with $g$ and $\mu$ denoting the two parameters of the theory. The Grassmann even field $B$ is an auxiliary field (which can be integrated over) and it satisfies the equation of motion $B = -i W'$. 

This theory has a symmetry known as supersymmetry. The action is invariant under two supersymmetry charges, $Q$ and $\Qb$. That is, $Q S = \Qb S = 0$.

After integrating over the fermions we get the following effective form of the action
\beq
S = \hf \left( W' \right)^2 - \ln W''.
\eeq

It would be interesting to ask if supersymmetry is broken or preserved in this model. Supersymmetry is preserved if the ground state energy $E_0$ of the theory is zero and it is broken otherwise.

When $\mu^2 > 0$ the classical minimum of the theory is given by the field configuration $\phi = 0$ with the energy
\beq
E_0 = \hf g^2 \mu^4 > 0,
\eeq
and thus supersymmetry is broken in this theory. The ground state energy of this theory can be computed as the expectation value of the bosonic action $S_B = \hf (W')^2$ at the classical minimum, which is $\phi = 0$.

\begin{figure}[h]
\bec
  \includegraphics[width=9cm]{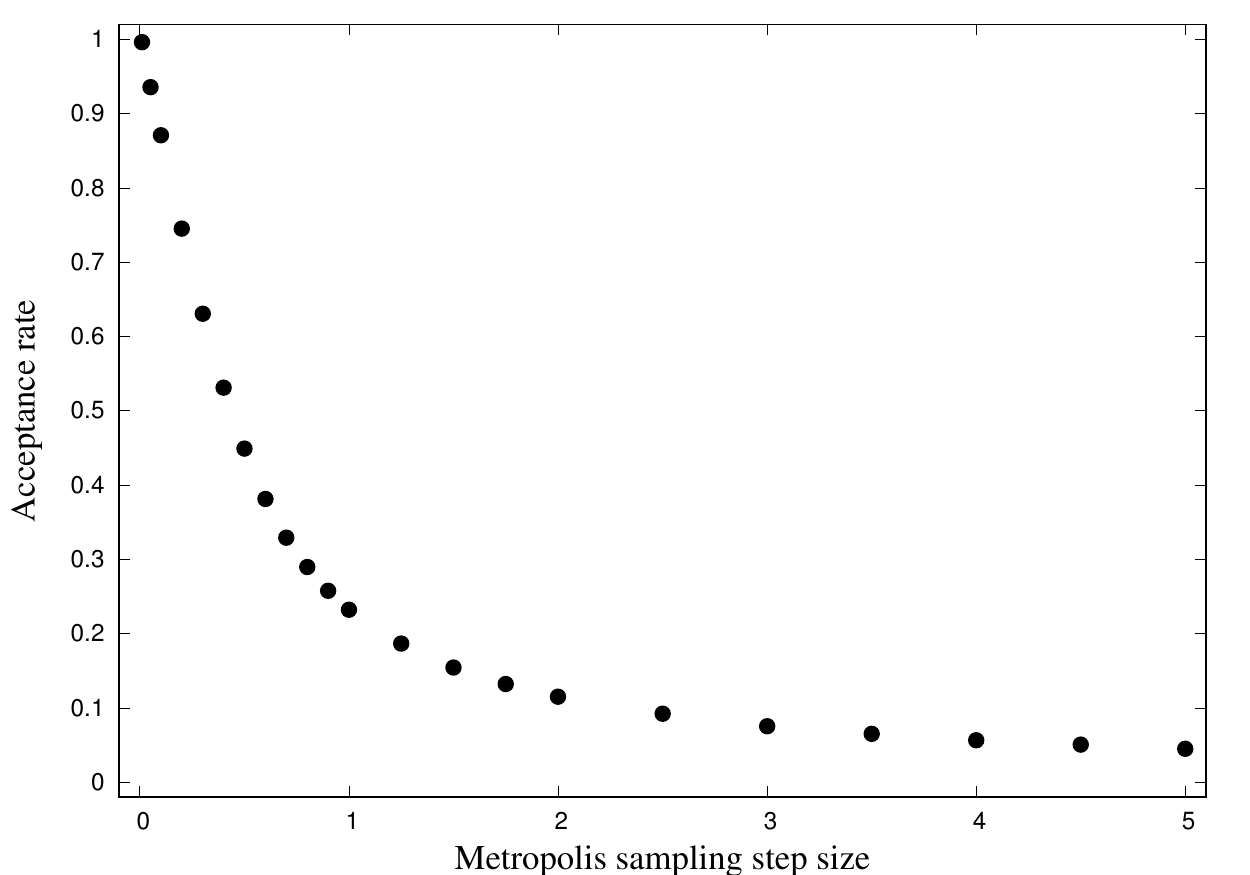}
  \caption{Acceptance rate against Metropolis sampling step size $\epsilon$ for the supersymmetric model given in Eq. \eqref{eq:susy-model}.}
  \label{fig:susy-eps-acc}
  \eec
\end{figure}

We can perform a Metropolis sampling of the action to compute the ground state energy for given values of $g$ and $\mu$. A C++ code to simulate this theory is given in Appendix \ref{sec:susy-metropolis}. In Fig. \ref{fig:susy-sb-history} we show the ground state energy of the theory against Monte Carlo time. In Fig. \ref{fig:susy-eps-acc} we show the acceptance rate against the Metropolis step size for the same model.

\begin{figure}[h]
\bec
  \includegraphics[width=9cm]{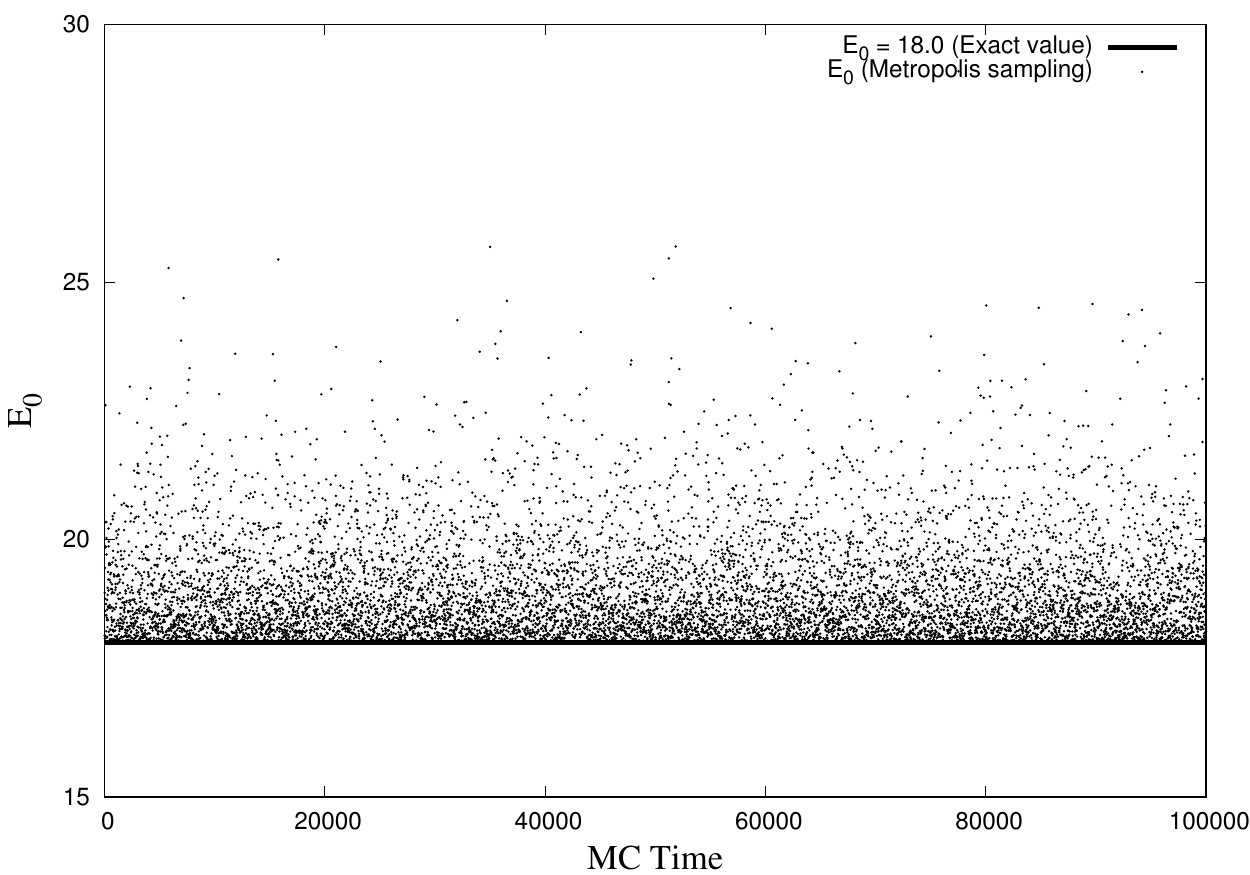}
  \caption{Monte Carlo time history of the ground state energy $E_0$ of the supersymmetric model given in Eq. \eqref{eq:susy-model}. The parameters are $g = 6.0, \mu = 1.0$. The classical value of the ground state energy is $E_0 = \hf g^2 \mu^2 = 18.0$. The simulation gives the value $E_0 = 18.9515(95)$, for a Metropolis step size of $\epsilon = 0.1$ and sample size $N_{\rm gen} = 10^5$. We discarded $N_{\rm therm} = 10^5$ thermalization steps. A gap of $10$ was used between each measurement.}
  \label{fig:susy-sb-history}
  \eec
\end{figure}

\subsection{Worked Example - Simple Harmonic Oscillator}
\label{sec:Worked-example-Simple-harmonic-oscillator}

The Euclidean action of a simple harmonic oscillator takes the following form
\beq
\label{eq:sho-eucl-action}
S \left[ x(\tau) \right] = \int_{\tau_a}^{\tau_b} d\tau \left( \hf m \dot{x}^2 + \hf m \omega^2 x^2 \right),
\eeq
where $\tau$ is the Euclidean time, $x(\tau)$ is position at time $\tau$.

In imaginary time formalism paths contribute to sum over histories with real exponential weights (instead of phases). The two-point function takes the form  
\bea
\langle x_f(\tau_f) | x(\tau_2) x(\tau_1) | x_i(\tau_i) \rangle &=& \int_i^f {\cal D} x \; x(\tau_2) x(\tau_1) \nn \\
&& ~~~~~~~~ \times \exp \left( - \frac{1}{\hbar} \int_{\tau_i}^{\tau_f} d\tau \; L(x, \dot{x}) \right).
\eea

Note that the Euclidean action is real and positive definite and thus the probability weights are real and positive. Reality of the Euclidean action will be crucial for Monte Carlo method for path integrals. We also note that the classical path gets the highest weighting since the action is minimized (or extremized in general) along that path.

The vacuum expectation values of the correlation functions can be obtained from the large $T$ limit, as discussed in the previous section.

For the case of simple harmonic oscillator the correlators we are interested in are
\bea
\langle x(\tau_1) \rangle &=& 0, \\
\langle x(\tau_1) x(\tau_2) \rangle &=& \frac{1}{2 m \omega} e^{-\omega (\tau_2 - \tau_1)}, ~~ \tau_2 \geq \tau_1.
\eea

Let us simulate this model using Metropolis algorithm and compare the simulation results with the above analytical results for the correlators. 

\begin{figure}[t]
\bec
  \includegraphics[width=7cm]{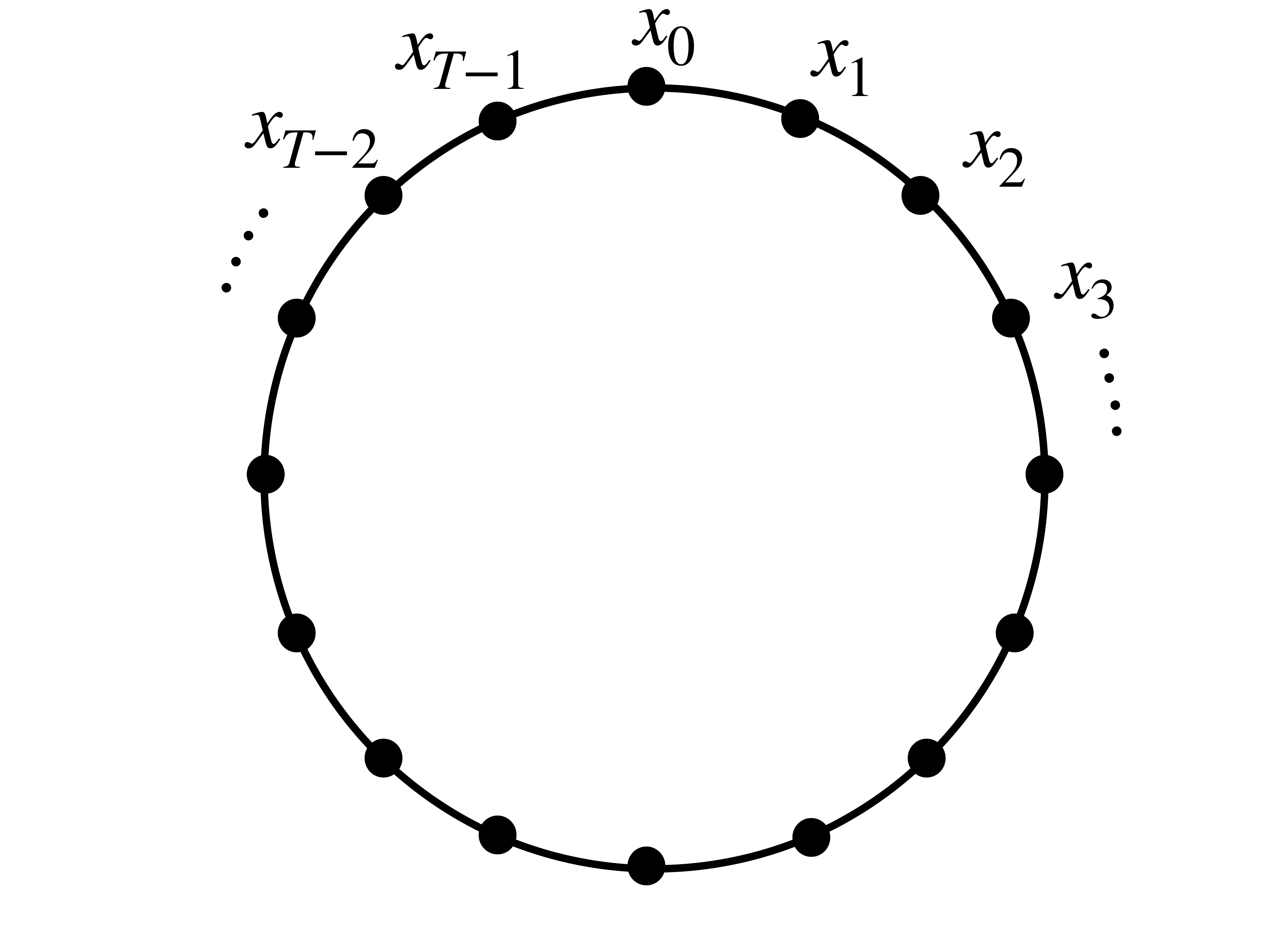}
  \caption{A one-dimensional periodic lattice with $N_\tau$ equally spaced sites. The position variables $x_n$ live at sites $n$. The lattice spacing is $a$. Periodic boundary condition is imposed as $x_{n + N_\tau} = x_n$. The circumference of the circle is $a N_\tau$.}
  \label{fig:plot-lattice-sites}
  \eec
\end{figure}

We discretize time $\tau$ on a one-dimensional lattice, with $N_\tau$ sites, labelled by $n = 0, 1, \cdots, N_\tau-1$. We have $\tau_f - \tau_i = a N_\tau$, with $a$ denoting the lattice spacing, which is the distance between two successive lattice sites. The position $x$ at site $n$ is denoted by $x_n$. We will use periodic boundary conditions on the lattice. That is, $x_{n+N_\tau} = x_n$. (See Fig. \ref{fig:plot-lattice-sites}.)

We can write down the action given in Eq. \eqref{eq:sho-eucl-action} on the lattice in the following way
\beq
S_L = \frac{m a}{2} \sum_{n = 0}^{N_\tau-1} \left[ \left(\frac{x_{n+1} - x_n}{a} \right)^2 + \omega^2 \left( \frac{x_{n+1} + x_n}{2} \right)^2 \right].
\eeq

Upon introducing dimensionless parameters
\bea
\widehat{m} &=& m a, \\
\widehat{\omega} &=& \omega a, \\
\widehat{x}_n &=& \frac{x_n}{a},
\eea
the lattice action takes the form
\beq
S_L = \frac{\widehat{m}}{2} \sum_{n = 0}^{N_\tau-1} \left[ \left(\widehat{x}_{n+1} - \widehat{x}_n \right)^2 + \frac{\widehat{\omega}^2}{4} \left( \widehat{x}_{n+1} + \widehat{x}_n \right)^2 \right].
\eeq

Let us use Metropolis algorithm to sample the configurations $x_n$. We randomly choose a location $x_n$ and update it by proposing a random shift 
\beq
- \Delta \leq \delta \leq \Delta,
\eeq
in position with uniform probability. For example, we could take $| \Delta | = 0.5$. Thus the trial position is
\beq
\widehat{x}_n^{new} = x_n + \delta.
\eeq

We update the locations one at a time, by performing a random scan on the lattice\footnote{In a Markov chain, the updates can be performed using a fixed scan (such as raster scan) or a random scan, depending on the model and the algorithm. It is best to use random scan if we are using Metropolis sampling. As an example, for the case of two-dimensional Ising model, a Gibbs sampler (which is a variant of Metropolis-Hastings algorithm with the property that the proposed moves are always accepted), with any of these scans would produce an irreducible Markov chain. However, using a fixed scan with Metropolis updates fails to produce an irreducible Markov chain for this model.}.

The change in the action (after and before the random shift) is calculated as
\bea
\delta S_L &=& \frac{\widehat{m}}{2} \sum_{n = 0}^{N_\tau-1} \Big\{ \left( \widehat{x}_{n+1} - \widehat{x}^{new}_n \right)^2 - \left( \widehat{x}_{n+1} - \widehat{x}_n \right)^2 \nn \\
&& + \frac{ \widehat{\omega}^2}{4} \left( \widehat{x}_{n+1} + \widehat{x}^{new}_n \right)^2 - \frac{ \widehat{\omega}^2}{4} \left( \widehat{x}_{n+1} + \widehat{x}_n \right)^2 \Big\}.
\eea

The Metropolis update is carried out by accepting the proposed change in position with a probability 
\beq
{\rm min} \left(1, e^{-\delta S_L} \right).
\eeq

The above update steps are repeated for each $x_n$ for $n = 0, 1, \cdots, N_\tau-1$ by randomly choosing the $n$ values. To start the Markov chain we can either choose a random path, where all $x_n$ are initialized to random values (called a {\it hot start}) or we can choose a path where all $x_n = 0$ (called a {\it cold start}). We also need to make sure that the Markov chain has been thermalized before taking any measurements.

\begin{figure}[t]
\bec
  \includegraphics[width=9cm]{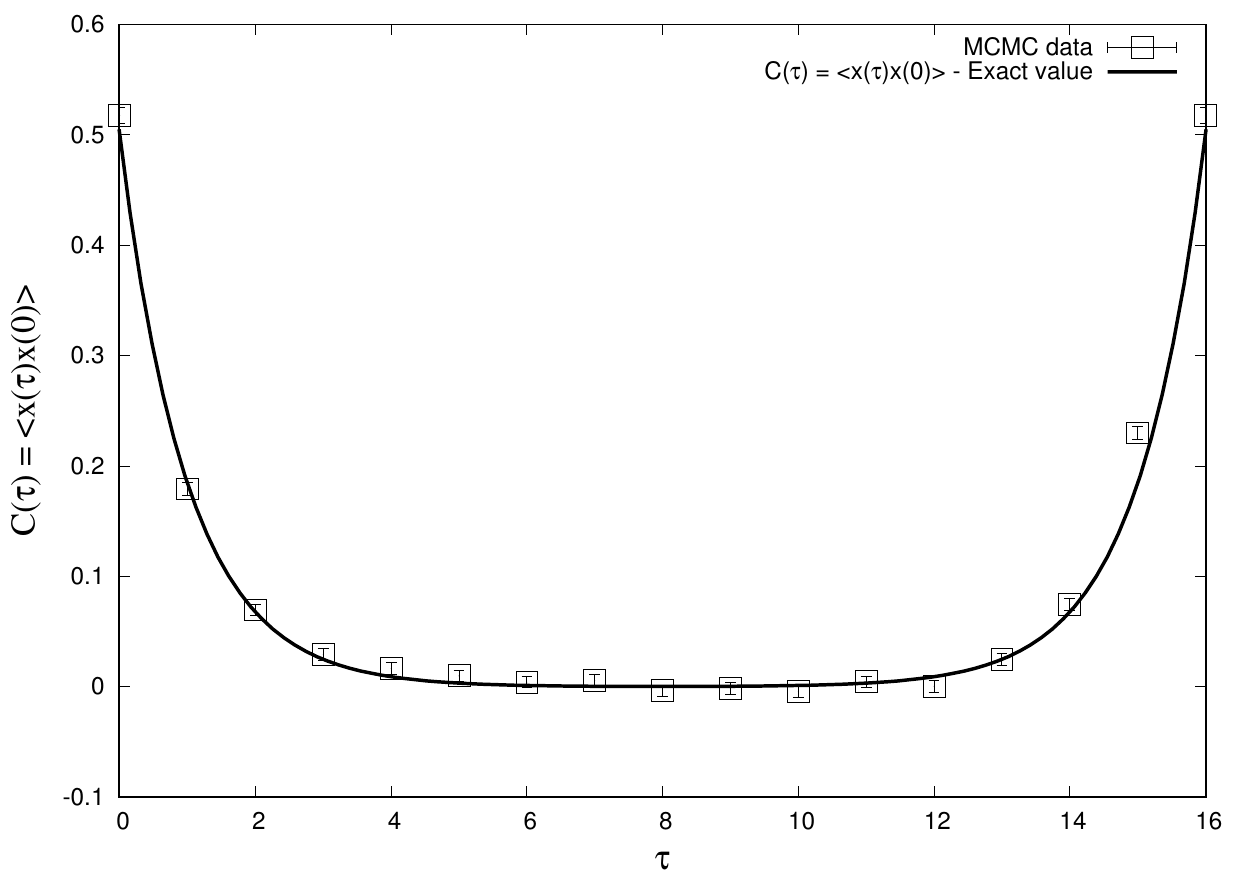}
  \vspace{0.5cm}
  \includegraphics[width=9cm]{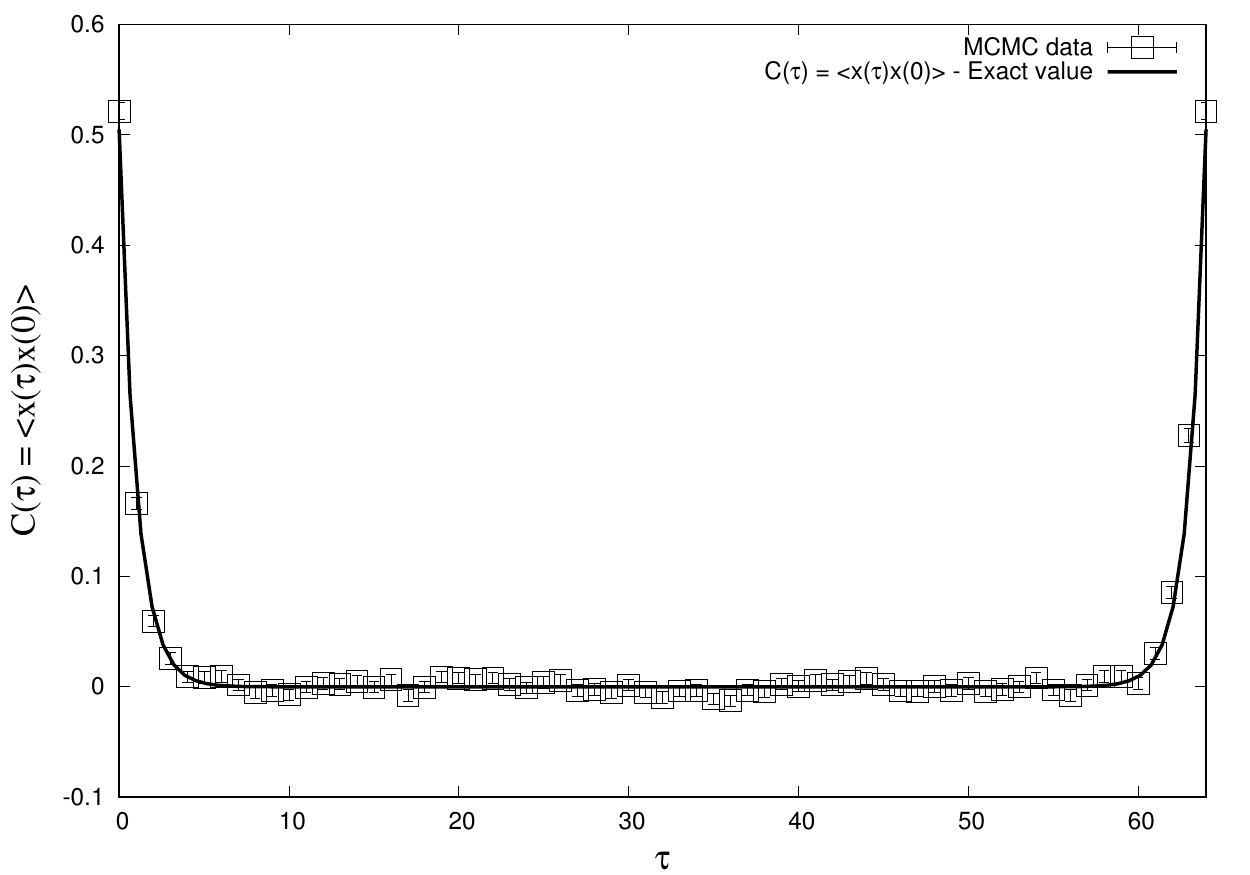}
  \caption{The correlation function $C(\tau) = \langle x(\tau) x(0) \rangle$ against Euclidean time $\tau$ on a lattice with 16 sites (Top) and 64 sites (Bottom). The data are produced using Metropolis algorithm. $C(\tau)$ is periodic due to periodic boundary condition on the lattice. We can fit the correlator data to the analytic form, $C(\tau) = (2 m \omega)^{-1} [ e^{-\omega \tau}  +  e^{\omega \tau} e^{-\omega N_\tau}]$, where $\omega$ is the frequency and $N_\tau$ is the number of lattice sites (time slices).}
  \label{fig:plot-correlator}
  \eec
\end{figure}

In Fig. \ref{fig:plot-correlator} we show the Monte Carlo data for correlator $C(\tau)$ against time $\tau$. We can fit the correlator data to the following analytic expression 
\beq
C(\tau) = \left \langle x(\tau) x(0) \right \rangle  = \frac{1}{2 m \omega} \left[ e^{-\omega \tau}  +  e^{\omega \tau} e^{-\omega N_\tau} \right].
\eeq

In Appendix \ref{sec:sho-metro} we provide a C++ program for simulating the simple harmonic oscillator.

\subsection{Worked Example - Unitary Matrix Model}
\label{sec:Worked-example-Unitary-matrix-model}

Let us consider another example - a unitary matrix model that undergoes the so-called Gross-Witten-Wadia phase transition. Analytic solution is available for this model and it is discussed in Refs. \cite{Gross:1980he, Wadia:1980cp}.

The degrees of freedom are encoded in a unitary matrix of rank $N$ and the action of the model is given by
\beq
\label{eq:GWW-action}
S[U] = - \frac{N g}{2} \left( {\rm Tr} \; U + {\rm Tr} \; U^\dagger \right),
\eeq
where $g$ is a coupling parameter.

The partition function is given by
\beq
Z_g = \int dU e^{-S[U]}. 
\eeq

Instead of the $U$ variable we can directly work with the angle variables $\theta_i$, $i = 1, 2, \cdots, N$. We have
\beq
U = {\rm diag} \; (\theta_1, \theta_2, \cdots, \theta_N).
\eeq

The change of variables introduces a Jacobian, which is the {\it Vandermonde determinant}. Thus, the action of our interest takes the form
\beq
S[\theta] = - \frac{Ng}{2} \sum_{k=1}^N \left( e^{i \theta_k} + e^{-i \theta_k} \right) - \sum_{j, k = 1,\; j \neq k}^N \log \sin \left| \frac{\theta_j - \theta_k}{2} \right|.
\eeq

The Polyakov loop observable, $P$, in this model can be computed analytically. We have
\beq
P = \frac{1}{N} {\rm Tr} \; U = \frac{1}{N} \sum_{k = 1}^N e^{i \theta_k}.
\eeq

The analytical value is
\beq
P = \left\{
  \begin{array}{lr}
    \frac{g}{2} & {\rm ~for} ~g < 1, \\
    1 - \frac{1}{2g} & {\rm ~for} ~g > 1.
  \end{array}
\right.
\eeq

\begin{figure}[t]
\bec
  \includegraphics[width=9cm]{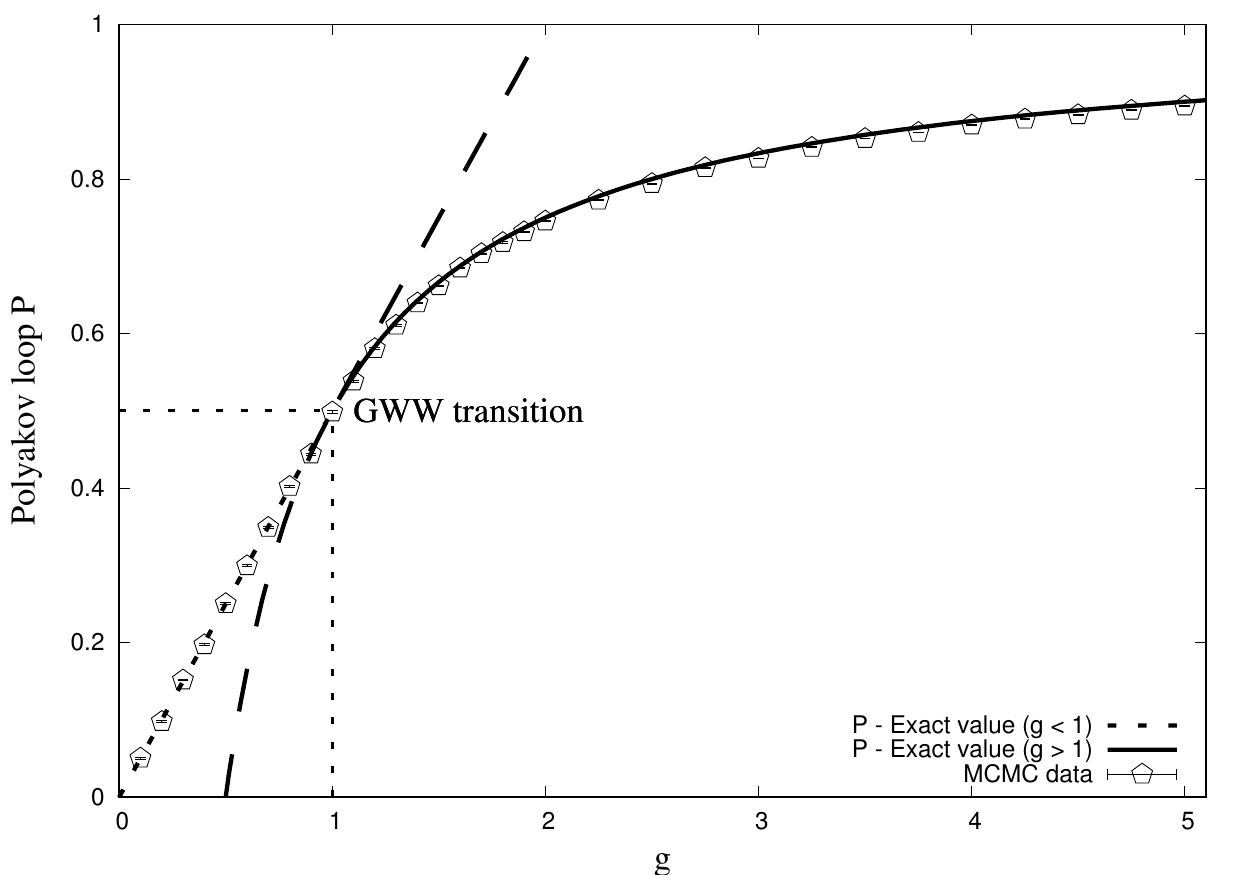}
  \caption{The Polyakov loop $P$ against the coupling $g$ for a unitary matrix model given in Eq. \eqref{eq:GWW-action}. The rank of the matrix $U$ is taken as $N=50$. Solid curves represent the analytical results and the filled circles with error bars denote the Monte Carlo data. Gross-Witten-Wadia transition occurs at $g=1$.}
  \label{fig:poly-g-GWW}
  \eec
\end{figure}

In Fig. \ref{fig:poly-g-GWW} we show the plot of Polyakov loop against the coupling $g$, for $N=50$, comparing the Monte Carlo data with the analytical result. In Appendix \ref{sec:GWW-metro} we provide a C++ program for simulating the unitary matrix model.

\section{Reliability of Simulations}
\label{sec:Reliability-of-simulations}

Since Monte Carlo methods are based on random sampling, it is crucial to understand the reliability of the data we get from Monte Carlo simulations. 

Any starting configuration of the simulation is a good configuration to help us find the desired unique stationary distribution. If we generate two chains of simulations say, one with a cold start and the other with a hot start, they both should approach the {\it one and only one} stationary distribution. There can be a situation in which the system we are probing, with a certain set of physics parameters, can be around the boundary of a phase transition. The simulation chains can exhibit very long autocorrelation times (to be discussed in Sec. \ref{sec:Auto-correlation-time}) and thus appear to be stuck in two different invariant distributions. Long autocorrelations can also appear if the system we are simulating is too large. The Monte Carlo time history of observables might show that the simulation has been converged but in reality the chain may be exploring just around the vicinity of the starting point in phase space. In order to avoid this pitfall we could start with systems with smaller size and then gradually extend simulations to larger sized systems.  

In order to avoid the bias from the starting configuration we need to remove the part of the chain where the system has not been thermalized. As a rule of thumb the thermalization time history should be about 10\% of the total Monte Carlo time history. Also care must be taken to choose the appropriate value of gap between each measurement steps during the simulations. The gap should be determined based on the value of the integrated auto-correlation time $\tau_{\rm int}$, which is given in Eq. \eqref{eq:int-auto-corr-time} below. 

\subsection{Auto-correlation Time}
\label{sec:Auto-correlation-time}

The configurations generated through a Markov process depend on the previous elements in the chain. This dependence is known as auto-correlation, and this quantity can be measured in the simulations. The configurations are highly correlated if the value of auto-correlation is near 1. The configurations are independent of each other if the value is near 0. We should look at ways to decrease auto-correlations since it has the benefit of decreasing the Monte Carlo error for a given length of the Markov chain. The dependence between the elements in a chain decreases as the distance between them is increased. 

In practice, due to auto-correlations, we should not consider every element in the chain for measurements. We need to skip some number of elements between measurements. The auto-correlation length, and thus the number we use to skip, depends on the details of the theory, the algorithm and the parameters of choice. 

Let $O$ be some observable we compute in the model with 
\beq 
O_k = O(\phi^{(k)})
\eeq
denoting the observable made out of the $k$-th configuration $\phi^{(k)}$. The average value $\langle O \rangle$ and the statistical error $\delta O$ are given by 
\beq
\langle O \rangle = \frac{1}{N} \sum_{k = 1}^N O_k,~~{\rm and}~~\delta O = \frac{\sigma}{\sqrt{N}}.
\eeq

The variance is $\sigma^2 = \langle O^2 \rangle - \langle O \rangle^2$. 

We note that the error estimate given above is valid provided the thermalized configurations $\phi^{(1)}$, $\phi^{(2)}$, $\phi^{(3)}$, $\cdots$, $\phi^{(N)}$ are statically uncorrelated. However, in real simulations, as mentioned above, this is certainly not the case. In general, two consecutive configurations will be dependent, and the average number of configurations which separates two ``really uncorrelated" configurations is called the {\it auto-correlation time}. The correct estimation of the error of the observable will depend on the auto-correlation time.

Let us take a non-zero positive integer $a$ as the {\it lag time}. Then we can define the {\it lag-$a$ auto-covariance function} $\Gamma_a$ and the auto-correlation function (which is the normalized auto-covariance function) $\rho_a$ for the observable $O$ as
\bea
\Gamma_a &=& \frac{1}{(N - a)} \sum_{k = 1}^{(N - a)} \left( O_k - \langle O \rangle \right)  \left( O_{k+a} - \langle O \rangle \right), \\
\rho_a &=& \frac{\Gamma_a}{\Gamma_0}. 
\eea

These functions vanish if there is no auto-correlation. Also we have $\Gamma_0 = \sigma^2$.

In the generic case, where the auto-correlation function is not zero, the statistical error in the average $\langle O \rangle$ is given by
\beq
\delta O = \frac{\sigma}{\sqrt{N}} \sqrt{2 \tau_{\rm int}},
\eeq
where $\tau_{\rm int}$ is called the integrated auto-correlation time. It is given in terms of the normalized auto-correlation function $\rho_a$ as
\beq
\tau_{\rm int} = \hf + \sum_{a = 1}^\infty \rho_a.
\eeq

The auto-correlation function $\Gamma_a$ for large $a$ cannot be precisely determined, and hence one must truncate the sum over $a$ in $\tau_{\rm int}$ at some cutoff $M$, in order not to increase the error $\delta \tau_{\rm int}$ in $\tau_{\rm int}$, as a result of simply summing up the noise.

The integrated auto-correlation time $\tau_{\rm int}$ should then be defined by
\beq
\label{eq:int-auto-corr-time}
\tau_{\rm int} = \hf + \sum_{a = 1}^M \rho_a.
\eeq

The value of $M$ is chosen as the first integer between $1$ and $N$ such that 
\beq
M \geq 4 \tau_{\rm int} + 1.
\eeq

The error $\delta \tau_{\rm int}$ in $\tau_{\rm int}$ is given by
\beq
\delta \tau_{\rm int} = \sqrt{\frac{4M + 2}{N}} \tau_{\rm int}.
\eeq

\begin{figure}[h]
\bec
  \includegraphics[width=9cm]{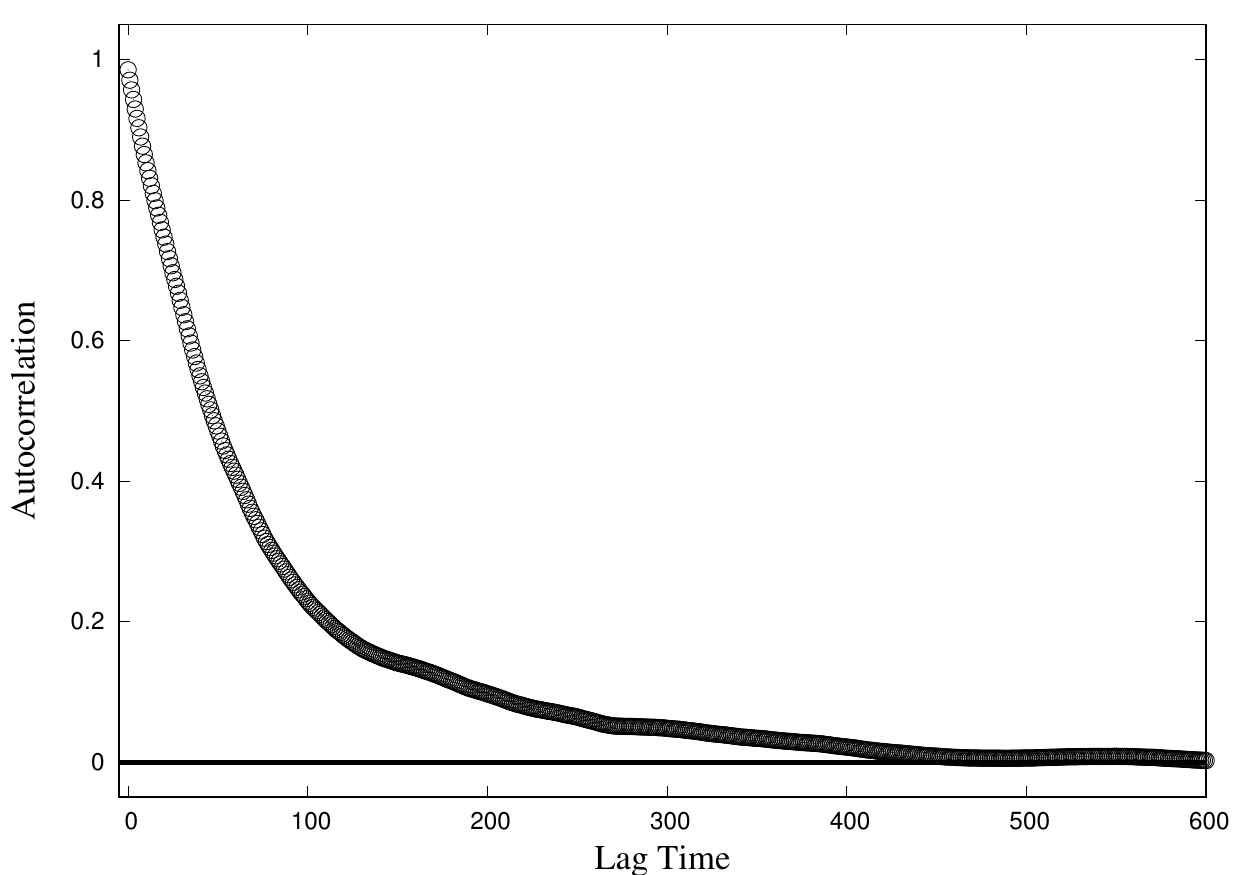}
  \caption{Auto-correlation against lag time in the supersymmetric model, Eq. \eqref{eq:susy-model}. The figure indicates that we should skip the configurations with an interval of about $M = 298$ to reduce the Monte Carlo error estimate. The auto-correlation analysis shows that $M = 298$, $\tau_{\rm int} = 74.08$ and $ \delta \tau_{\rm int} = 25.60$.}
  \label{fig:auto-corr-susy}
  \eec
\end{figure}

In Fig. \ref{fig:auto-corr-susy} we show the auto-correlation against lag time for the supersymmetric model given in Eq. \eqref{eq:susy-model}. The figure shows that we should skip the configurations with an interval of about $M = 298$ to reduce the Monte Carlo error estimate.

In Appendix \ref{sec:auto-corr} we provide a C++ program that computes the auto-correlation against lag time. It also provides the value of $\tau_{\rm int}$.

For details on the advantages of explicitly analyzing auto-correlation functions rather than binning strategies see Ref. \cite{Wolff:2003sm} 

\subsection{Error Analysis}
\label{sec:error-analysis}

It is important to have a control on the error on the simulation data. Whenever possible we must check the simulation results with the corresponding analytical or experimental results. The Monte Carlo time history of the observables should show convergence to a value. The convergence to each measured quantity can be checked using {\it binning}. We can use {\it binning} or {\it jackknife} (or {\it bootstrap}) to obtain reliable errors. We could also try to reproduce the simulation results using a different random number generator.    

\section{Hybrid (Hamiltonian) Monte Carlo}
\label{sec:Hybrid-(Hamiltonian)-Monte-Carlo}

The classic 1953 paper of Metropolis et al. \cite{Metropolis:1953am} introduced to us the world of Markov Chain Monte Carlo (MCMC). In their work, MCMC was used to simulate the distribution of states for a system of idealized molecules. In 1959, another approach to molecular simulation was put forward by Alder and Wainwright \cite{Alder:1959}. In their work, they used a deterministic algorithm for the motion of the molecules. This algorithm followed Newton's laws of motion, and it can be formalized in an elegant way using Hamiltonian dynamics. The two approaches, statistical (MCMC) and deterministic (molecular dynamics), coexisted peacefully for a long time. In 1987, an extraordinary paper by Duane, Kennedy, Pendleton, and Roweth \cite{Duane:1987de} combined the MCMC and molecular dynamics approaches. This combined approach is now known as {\it Hamiltonian Monte Carlo} or {\it Hybrid Monte Carlo} (HMC)\footnote{In their work, Duane, et al. applied HMC to lattice field theory simulations of QCD, not to molecular simulation.}.

Let us see how we can use Hamiltonian dynamics to construct an MCMC algorithm. Suppose we wish to sample from a probability distribution. We begin by constructing a Hamiltonian in terms of this probability distribution. On top of the variables we are interested in (they are the {\it position} variables in HMC language), we should also introduce some auxiliary field variables ({\it momentum} variables in HMC). These auxiliary variables typically would have have independent distributions that are Gaussian. The HMC method draws these momentum variables from a Gaussian distribution and then computes a trajectory according to a discretized version of Hamiltonian dynamics. At the end of the trajectory the new proposed state is accepted or rejected by a Metropolis step. The advantage of this method is that it can propose a new state that is far away from the current state, with a high probability of acceptance. This would be a huge gain compared to the random walk Metropolis.

\subsection{Hamilton's Equations}
\label{sec:Hamilton's-equations}

Let us consider a $d$-dimensional position vector $\bq$ and a $d$-dimensional momentum vector $\bp$ such that the state space is a $2d$-dimensional space. Hamiltonian dynamics operates on this state space. We can describe the system by a function of $\bq$ and $\bp$, known as the Hamiltonian, $H(\bq, \bp)$. The time evolution of $\bq$ and $\bp$ is determined by the partial derivatives of the Hamiltonian. We have the Hamilton's equations
\bea
\frac{dq_i}{dt} &=& \frac{\partial H}{\partial p_i}, \\
\frac{dp_i}{dt} &=& - \frac{\partial H}{\partial q_i},
\eea
for $i = 1, 2, \cdots, d$; and $t$ is the time variable.

For HMC, we are usually interested in Hamiltonian functions that can be written as a sum of potential energy $U(\bq)$ and kinetic energy $K(\bp)$: $H(\bq, \bp) = U(\bq) + K(\bp)$.

\subsection{Properties of Hamiltonian Dynamics}
\label{sec:Properties-of-Hamiltonian-dynamics}

While constructing MCMC updates, several properties of Hamiltonian dynamics come into play. Some of them are

\begin{enumerate}
\item{{\bf Reversibility.} Hamiltonian dynamics is reversible. We need to make use of this property to show that MCMC updates that use Hamiltonian dynamics leave the target distribution invariant. We can prove this by showing that Markov chains constructed using states proposed by this dynamics are reversible.}
\item{{\bf Conservation of the Hamiltonian.} The Hamiltonian dynamics keeps the Hamiltonian invariant
\beq
\frac{dH}{dt} = 0.
\eeq

In HMC we use Metropolis algorithm to  accept or reject a proposal found by Hamiltonian dynamics and the acceptance probability is 1 if $H$ remains invariant. In practice this is difficult to achieve since we can only make $H$ approximately invariant due to the discretized nature of the evolution equations.}

\item{{\bf Volume preservation.} Hamiltonian dynamics preserves volume in $(\bq, \bp)$ space. This tells us that in MCMC we need not account for any change in volume in the acceptance probability for Metropolis updates. Suppose we are using some arbitrary, non-Hamiltonian, dynamics, to propose new states. Then we would need to compute the determinant of the Jacobian matrix for the mapping defined by the dynamics, and computing this might be a highly non-trivial task.}
\end{enumerate}

\subsection{Leapfrog Method}
\label{sec:Leapfrog-method}

In order to implement the differential equations of Hamiltonian dynamics on a computer we must discretize the time evolution equations. There exist several schemes to discretize the Hamiltonian dynamics. The simplest among them is the {\it leapfrog method}.

In order to implement the algorithm on a computer, Hamilton's equations should be approximated by discretizing the time variable. We can use a small step size say, $\epsilon$ as the discrete time step. Starting with the state at time zero, we can iteratively compute (of course with some numerical errors) the state at times $\epsilon$, $2 \epsilon$, $3 \epsilon$, etc. 

The steps involved in the leapfrog method, to go from $t$ to $t + \epsilon$ are the following. We begin with a half step for the position variables, using $q_i (t)$ and $p_i(t)$. Then perform a full step for the momentum variables, using the new position variables. Finally, we perform another half step for the position variables using the new values for the momentum variables. These steps are summarized in the equations
\bea
q_i (t + \hf \epsilon) &=& q_i (t) + \hf \epsilon~ p_i(t), \\
p_i (t + \epsilon) &=& p_i(t) - \epsilon q_i(t + \hf \epsilon), \\
q_i (t + \epsilon) &=& q_i (t + \hf \epsilon) + \hf \epsilon ~p_i(t + \epsilon).
\eea

\begin{figure}[h]
\bec
  \includegraphics[width=9cm]{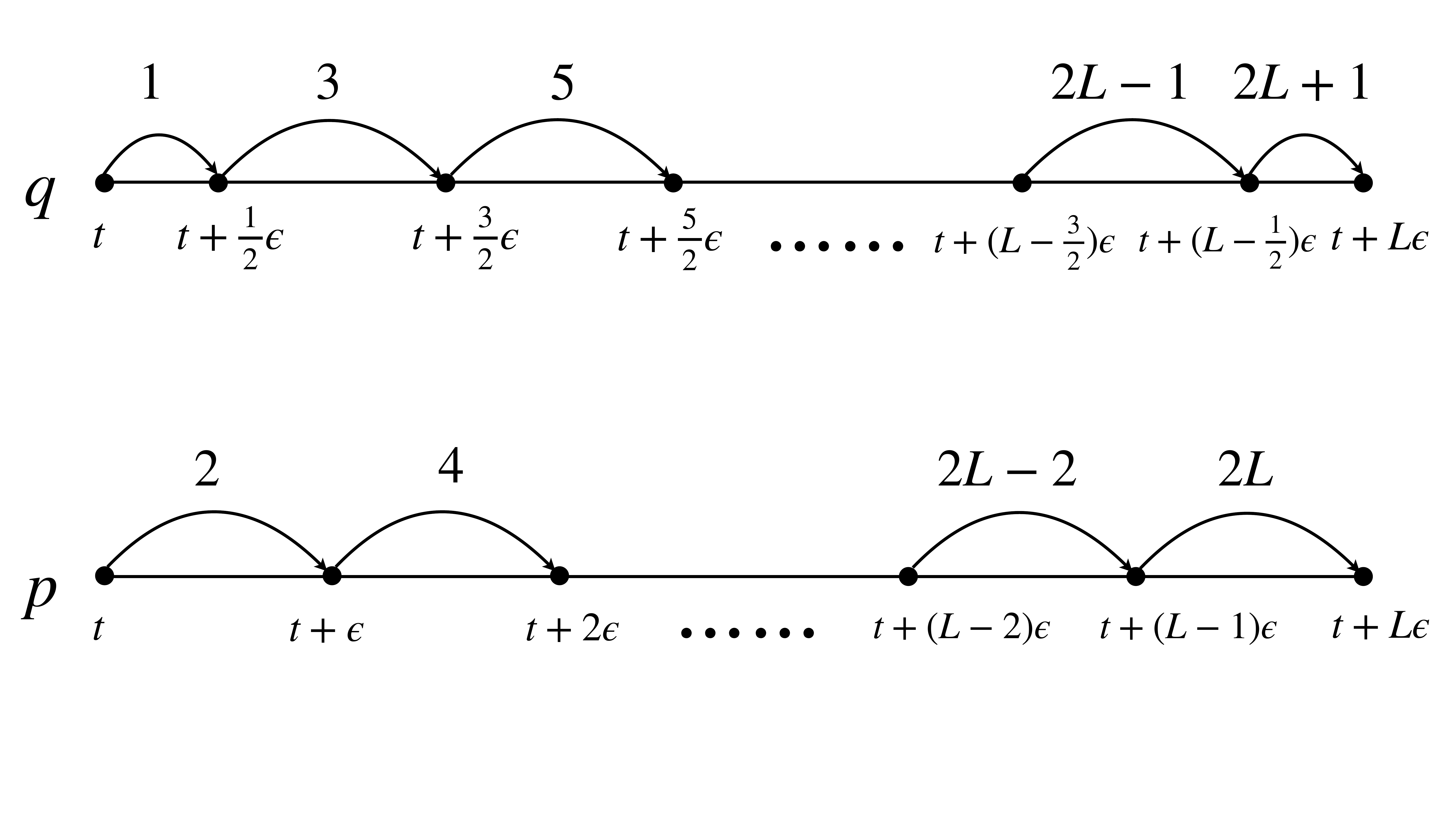}
  \caption{Steps of the leapfrog algorithm. The starting time is $t$ and the ending time is $t + L \epsilon$. Leapfrog step size is denoted by $\epsilon$, $L$ denotes the number of leapfrog steps, and $L \epsilon = l$ is called the trajectory length.}
  \label{fig:leapfrog}
  \eec
\end{figure}

In Fig. \ref{fig:leapfrog} we illustrate the leapfrog algorithm. The leapfrog method preserves volume exactly. It is also a reversible method.

\subsection{MCMC from Hamiltonian Dynamics}
\label{sec:MCMC-from-Hamiltonian-dynamics}

\subsubsection{Joint Probability Distribution}
\label{sec:Joint-probability-distribution}

We can relate the distribution we wish to sample to a potential energy function through the concept of a canonical distribution from statistical mechanics. 

Suppose we have a physical system with an energy function, $E(\bx)$, for the state, $\bx$. Then the canonical distribution over the states has the following probability density function
\beq
P(\bx) = \frac{1}{Z} e^{- \beta E(\bx)}.
\eeq
Here, $\beta = 1/T$ denotes the inverse temperature of the system. The quantity $Z$ plays the role of a normalization constant (the partition function). It is required for the function to integrate (or sum) to one.

The Hamiltonian is an energy function for the joint state of {\it position} $\bq$ and {\it momentum} $\bp$. It encodes a the following joint distribution
\beq
\label{eq:const-prob-density}
P(\bq, \bp) \equiv \frac{1}{Z} e^{- \beta H(\bq, \bp)}.
\eeq

The invariance of $H$ under Hamiltonian dynamics implies that a Hamiltonian trajectory will, if simulated without any numerical errors, move within a hypersurface of constant probability density given by Eq. \eqref{eq:const-prob-density}.

With energy functions $U(\bq)$ and $K(\bp)$ we have the Hamiltonian $H(\bq, \bp) = U(\bq) + K(\bp)$, and the joint density is
\beq
\label{eq:joint-prob}
P(\bq, \bp) = \frac{1}{Z} e^{- \beta U(\bq)} e^{- \beta K(\bp)}.
\eeq
This tells us that $\bq$ and $\bp$ are independent, and each have their own canonical distributions. Our variables of interest are represented in ``position variables" $\bq$. The ``momentum variables" $\bp$ are introduced just to allow Hamiltonian dynamics to operate.

Thus, HMC samples from the joint probability distribution of $\bq$ and $\bp$ defined by Eq. \eqref{eq:joint-prob}, which is also a canonical distribution for these two variables. The distribution of interest is encoded in $\bq$ and it is specified using the potential energy function $U(\bq)$. The distributions of the momentum variables $\bp$ are independent of $\bq$, and they are specified through the kinetic energy function, $K(\bp)$.  

Almost all of the HMC simulation methods use a quadratic kinetic energy, that is, $\bp$ has a zero-mean multivariate Gaussian distribution. Also, the components of $\bp$ are specified to be independent. The kinetic energy function generating this distribution has the form
\beq
K(\bp) = \sum_{i = 1}^d \frac{p_i^2}{2},
\eeq
for unit mass $m_i$.

During each iteration of the HMC algorithm the canonical distributions of $\bq$ and $\bp$ stay invariant. Thus, under each iteration, their combination also leaves the joint distribution invariant.

In the final step, new values for the momentum variables $\bp$ are randomly chosen from their Gaussian distribution, independently of the current values of the position variables $\bq$. Starting with the current state $(\bq, \bp)$, Hamiltonian dynamics is performed for $L$ steps, with the help of the leapfrog algorithm\footnote{We could also use some other reversible and volume preserving method.}, with a step size of $\epsilon$. There are two parameters of the algorithm: $L$ and $\epsilon$. As like any parameters of the algorithm, a good performance can be guaranteed only when they are tuned. At the end of this $L$-step trajectory, the momentum variables are negated, giving a proposed state $(\bq^*, \bp^*)$. See Fig. \ref{fig:leapfrog}. A Metropolis update is performed next, and according to that, the proposed state is rejected or accepted.

In the case when the proposed state is rejected, the algorithm takes the next state as the same one as the current state. In other words, it is counted again while estimating the expectation of some function of state by its average over states of the Markov chain.

Note that we can use HMC to sample only from continuous distributions on ${\mathbf R}^d$. In this case, up to a normalization constant, the density function can be evaluated.

The ergodicity of the algorithm ensures that HMC algorithm will not be trapped in some subset of the state space. It will asymptotically converge to its unique equilibrium distribution. Since any value can be sampled for the momentum variables, during the HMC iterations, we see that this can affect the position variables in arbitrary ways.

\subsubsection{Tuning HMC Algorithm}
\label{sec:Tuning-HMC-algorithm}

We need to tune the leapfrog step size $\epsilon$ and the number of leapfrog steps $L$ that determine the length of the trajectory $l = L \epsilon$ in fictitious time. In general, tuning HMC algorithm is more difficult than tuning Metropolis algorithm.

It is advisable to perform preliminary runs with trial values of $\epsilon$ and $L$ and monitor the simulation time history of the observables for thermalization time and auto-correlation time. A common observable to monitor is the value of the potential energy function, $U(q)$. The auto-correlation for observables indicates how well the Markov chain is exploring the state space. Ideally, we should aim to land at a state, that is nearly independent of the current state, after one HMC iteration.

It is important to select a suitable leapfrog step size, $\epsilon$. If $\epsilon$ is too large, then the acceptance rate for states proposed by simulating trajectories will be very low. If the step size is too small, then the exploration in state space will be too slow, and in addition, we will waste computation time. The choice of step size is almost independent of $L$. The error in the value of the Hamiltonian, which in turn determines the rejection rate, usually does not increase with $L$, provided that $\epsilon$ is small enough that the dynamics is stable. 

When the $\epsilon$ used generates trajectories that are unstable, $H$ grows exponentially with $L$, and as a result the acceptance probability becomes very small. Taking too large a value of $\epsilon$ can affect the performance of HMC very badly. Thus, compared to random-walk Metropolis, HMC is more sensitive to tuning.

It seems necessary to tune the HMC trajectory length $L \epsilon$ by trial and error. If preliminary runs with a suitable $\epsilon$ results in HMC with a nearly independent point after only one iteration, then we could try next with a smaller value of $L$. On the other hand, if instead, there is high auto-correlation in the run with the given $L$ then we should try again with a larger $L$ value. For random walk Metropolis, we should aim for an acceptance rate of about $25\%$ \cite{Gupta:1990, Roberts:1997} for optimal performance. For HMC the optimal performance happens at an acceptance rate of about $65\%$ \cite{neal2012mcmc}.

\subsubsection{HMC Algorithm - Step by Step}
\label{sec:HMC-algorithm-Step-by-step}

Let us reiterate the HMC algorithm. Our goal is to generate a set of configurations, starting from an arbitrary configuration say, $\phi^{(0)}$. The chain created through HMC 
\beq
\phi^{(0)} \to \phi^{(1)} \to \phi^{(2)} \to \cdots \to \phi^{(k-1)} \to \phi^{(k)} \to \phi^{(k+1)} \to \cdots 
\eeq
will eventually reach the (unique) invariant distribution.
 
Once $\phi^{(k)}$ is obtained, $\phi^{(k+1)}$ is obtained in the following way.

\begin{enumerate}
\item{The auxiliary momenta $ p_{{\phi}^{(k)}} $, that are conjugate to $\phi^{(k)}$ are generated randomly from a Gaussian probability
\beq
\frac{1}{\sqrt{2 \pi}} \exp \left( - \hf \left[ p_{{\phi}^{(k)}} \right]^2 \right). \nn 
\eeq
}

\item{The next step is to calculate the initial Hamiltonian
\beq
H_i = S[\phi^{(k)}] + \hf \left( p_{{\phi}^{(k)}} \right)^2,
\eeq
where $S[\phi^{(k)}]$ is the potential energy function. For instance, it can be the Euclidean action of the quantum field theory. 
}

\item{The Hamiltonian dynamics ({\it molecular dynamics evolution}) is performed next. The trajectory length is taken as $l = L \epsilon$. The leapfrog method is applied $L$ times with step size $\epsilon$.

In the first step of leapfrog we make a half step for the ``position"
\beq
\phi^{(k)}(t) \Big|_{t = 0 + 0.5 \epsilon} = \phi^{(k)} (t) \Big|_{t = 0} + 0.5 \epsilon~ p_{\phi^{(k)}}(t) \Big|_{t = 0}.
\eeq

After this, for $n = 1, 2, \cdots, L-1$, we repeat the following 
\bea
p_{\phi^{(k)}}(t) \Big|_{t = n \epsilon} &=& p_{\phi^{(k)}}(t) \Big|_{t = (n-1) \epsilon} - \epsilon \left. \frac{\partial}{\partial \phi^{(k)}} S[ \phi^{(k)} ] \right|_{t = (n - 0.5)\epsilon}, \\
\phi^{(k)}(t) \Big|_{t = (n + 0.5) \epsilon} &=& \phi^{(k)}(t)\Big|_{(n - 0.5) \epsilon} + \epsilon ~p_{\phi^{(k)}}(t)\Big|_{t = n \epsilon}.
\eea

At $n = L$ we make the steps
\bea
p_{\phi^{(k')}}(t) \Big|_{t = n \epsilon} &=& p_{\phi^{(k)}}(t) \Big|_{t = (n-1) \epsilon} - \epsilon \left. \frac{\partial}{\partial \phi^{(k)}} S[ \phi^{(k)} ] \right|_{t = (n - 0.5)\epsilon}, \\
\phi^{(k')}(t) \Big|_{t = n \epsilon} &=& \phi^{(k)}(t)\Big|_{(n - 0.5) \epsilon} + 0.5 \epsilon ~p_{\phi^{(k)}}(t)\Big|_{t = n \epsilon}.
\eea
}

\item{At the end of the trajectory the final Hamiltonian is calculated  
\beq
H_f = S[\phi^{(k')}] + \hf \left( p_{{\phi}^{(k')}} \right)^2.
\eeq
}

\item{Now we are in a place to accept or reject the proposed state $\phi^{(k')}$. This is done through a Metropolis test. Generate a uniform random number $r$ between 0 and 1. If $r < e^{-\Delta H}$ with $\Delta H = H_f - H_i$, then $\phi^{(k+1)} = \phi^{(k')}$. That is, the new configuration is accepted. Otherwise $\phi^{(k+1)} = \phi^{(k)}$. That is, the proposal is rejected.}

\end{enumerate}

\subsection{Worked Example - HMC for Gaussian Model}
\label{sec:Worked-example-HMC-for-Gaussian-model}

Let us use HMC to simulate a model with the Gaussian action 
\beq
S[\phi] = \hf \phi^2.
\eeq 

The Hamiltonian for this model takes the form
\beq
H (\phi, p_\phi) = \hf \phi^2 + \hf p_\phi^2.
\eeq

The gradient of the action, which is needed in the molecular dynamics evolution, is just the field itself
\beq
\frac{\partial S[\phi]}{\partial \phi} = \phi.
\eeq

\begin{figure}[h]
\bec
  \includegraphics[width=9cm]{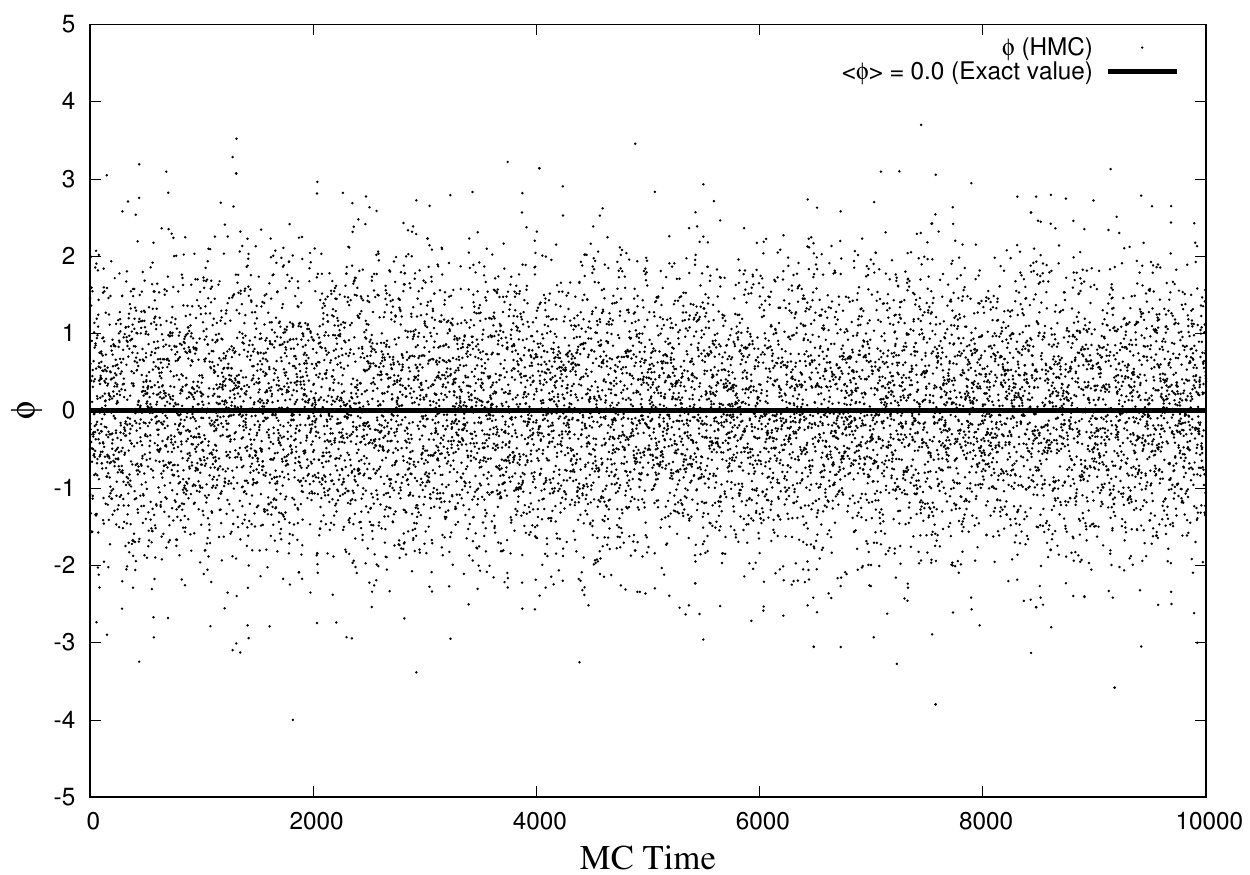} \\
   \vspace{0.5cm}
   \includegraphics[width=9cm]{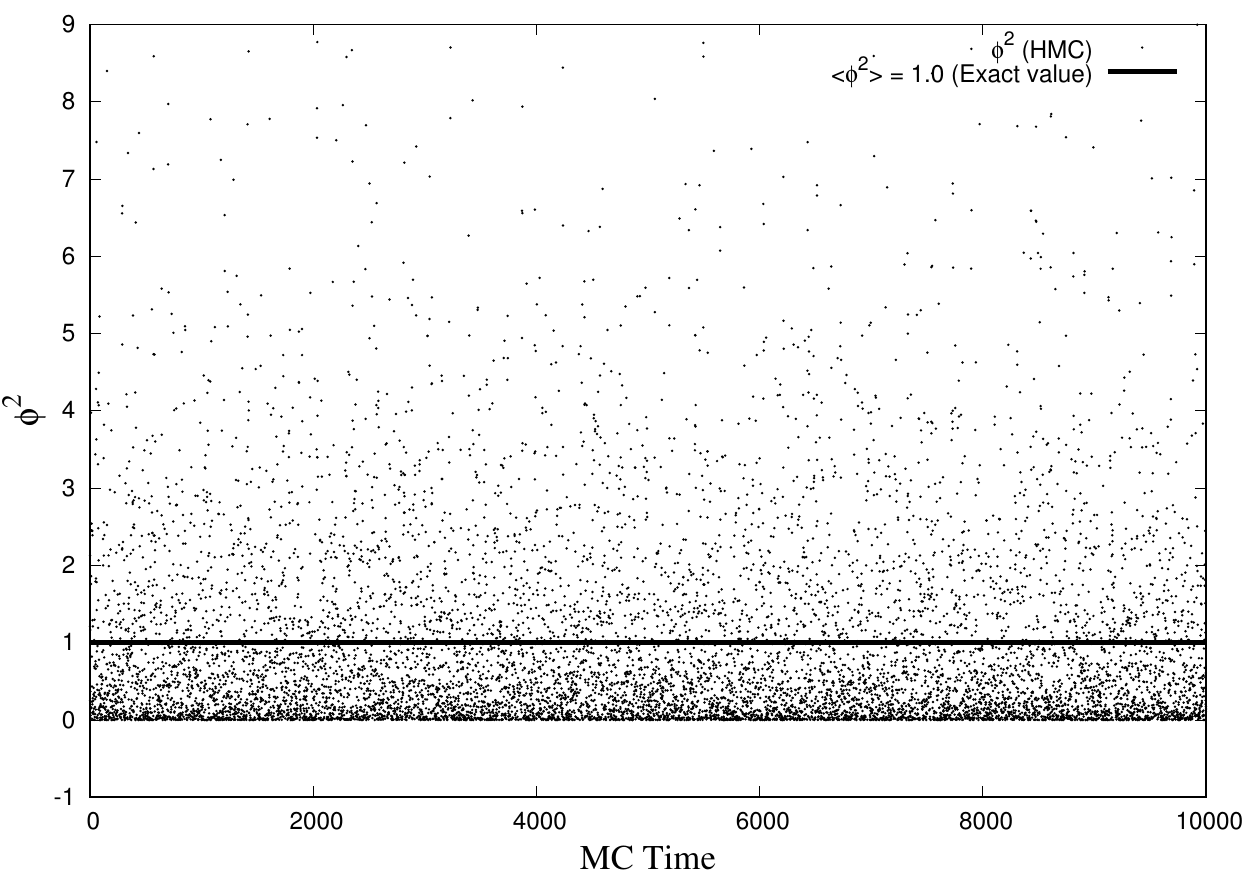}
  \caption{Monte Carlo time history of $\phi$ (Left) and $\phi^2$ (Right) for the Gaussian model with the action $S[\phi] = \hf \phi^2$. We used HMC algorithm to simulate the model, with the leapfrog step size $\epsilon = 0.2$ and the number of leapfrog steps $L = 20$. The exact values are $\langle \phi \rangle = 0$ and $\langle \phi^2 \rangle = 1.0$. The simulation gives $\langle \phi \rangle =  0.0002 \pm 0.0031$ and $\langle \phi^2 \rangle = 0.9921 \pm 0.0044$.}
  \label{fig:hmc-phi-phisq-history}
  \eec
\end{figure}

The molecular dynamics and subsequent Metropolis test are performed following the steps described in the previous section. In Fig. \ref{fig:hmc-phi-phisq-history} we show the Monte Carlo time history of $\phi$ and $\phi^2$. A C++ program to simulate this Gaussian model using HMC is provided in Appendix. \ref{sec:gaussian-hmc}. 

\begin{figure}[h]
\bec
  \includegraphics[width=9cm]{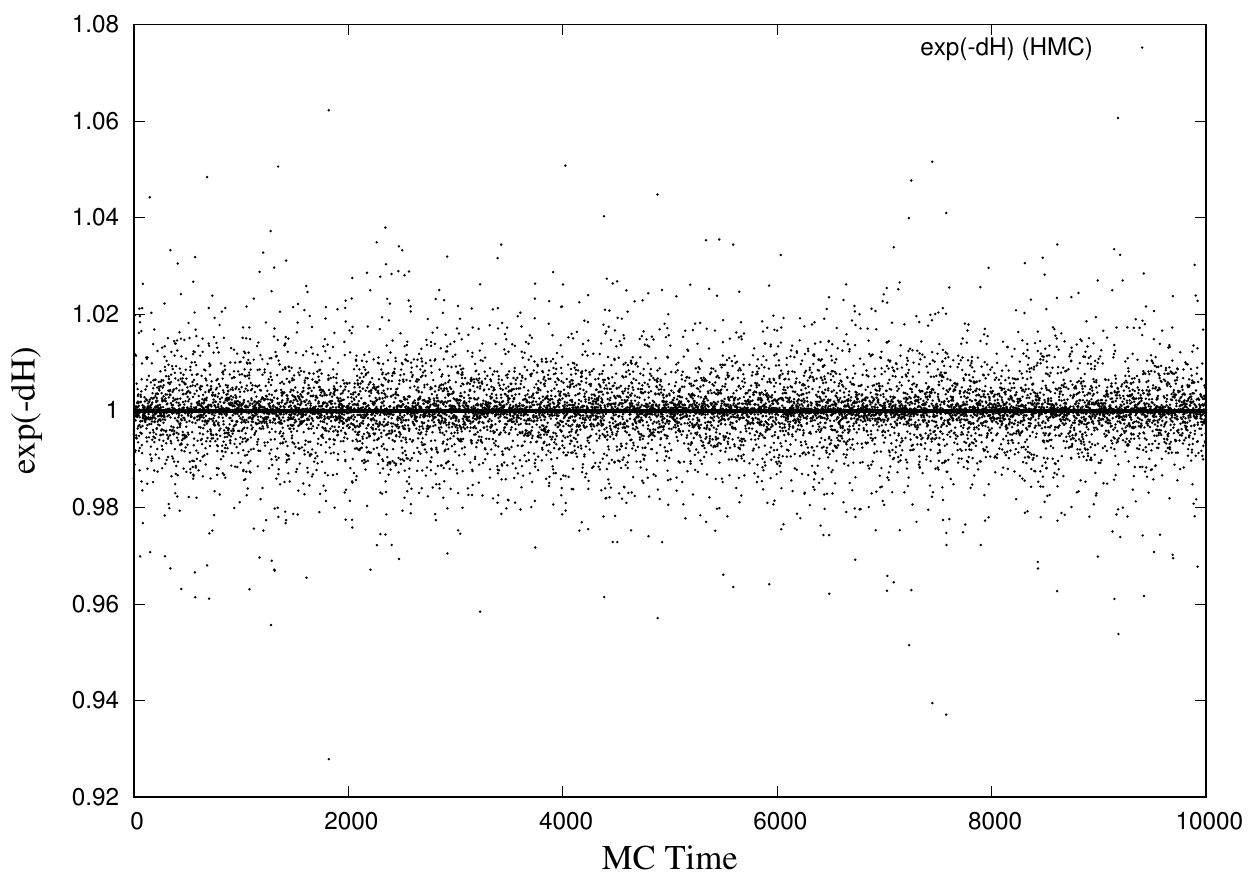}
  \caption{$\exp(-\Delta H)$ against Monte Carlo time history for the Gaussian model with the action $S[\phi] = \hf \phi^2$. We used HMC algorithm to simulate the model, with the leapfrog step size $\epsilon = 0.2$ and the number of leapfrog steps $L = 20$. Ideally, $\exp(-\Delta H)$ should fluctuate around 1 in the simulations.}
  \label{fig:hmc-gauss-dH}
  \eec
\end{figure}

In Fig. \ref{fig:hmc-gauss-dH} we show $\exp(-\Delta H)$ against Monte Carlo time history. As a rule of thumb $\exp(-\Delta H)$ should fluctuate around 1 if everything works fine in the simulations.

\subsection{Worked Example - HMC for Supersymmetric Model}
\label{sec:Worked-example-HMC-for-supersymmetric-model}

Let us simulate the supersymmetric model given in Eq. \eqref{eq:susy-model} using HMC. We need to change the action in the previous code to the action given in Eq. \eqref{eq:susy-model}. The gradient of the action is
\beq
\frac{\partial S[\phi]}{\partial \phi} = g^2 \phi (\phi^2 + \mu^2) - (2 g \phi)^{-1}.
\eeq

\begin{figure}[h]
\bec
  \includegraphics[width=9cm]{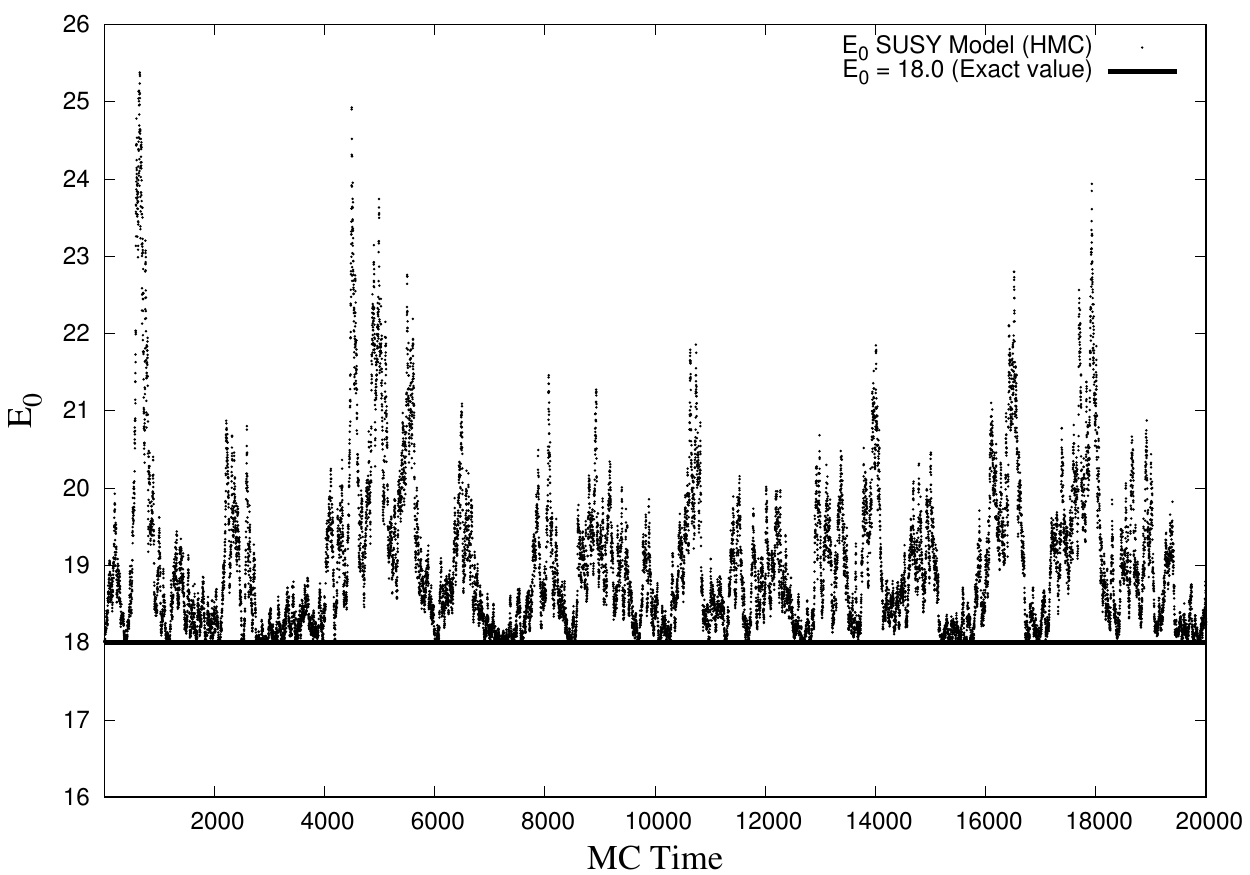}
  \caption{Monte Carlo time history of the ground state energy $E_0$ of the supersymmetric model. In the simulation we used thermalization steps $N_{\rm therm} = 10,000$, generation steps $N_{\rm gen} = 50,000$ (up to $N_{\rm gen} = 20,000$ is shown in the figure above), leapfrog step size $\epsilon = 0.0005$, number of leapfrog steps $L = 20$, coupling parameter $g = 6.0$, mass parameter $\mu = 1.0$ and starting point of field configuration $\phi_0 = 2.0$. The simulation gives $E_0 = 19.0434 \pm 0.0045$, while the classical value of the ground state energy is $E_0 = 18$.}
  \label{fig:hmc-susy-history}
  \eec
\end{figure}

\begin{figure}[h]
\bec
  \includegraphics[width=9cm]{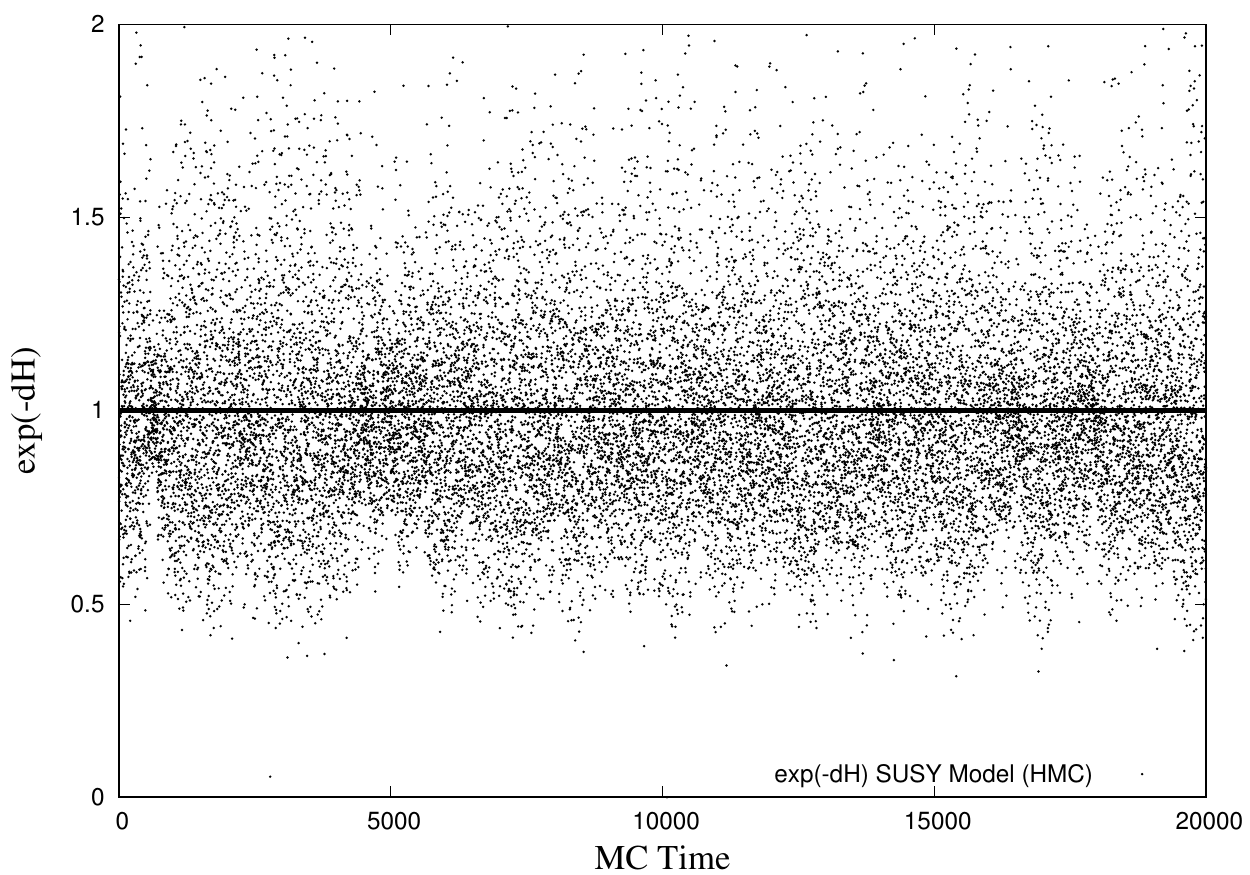}
  \caption{$\exp(-\Delta H)$ against Monte Carlo time history. In the simulation we used thermalization steps $N_{\rm therm} = 10,000$, generation steps $N_{\rm gen} = 50,000$, leapfrog step size $\epsilon = 0.0005$, number of leapfrog steps $L = 20$, coupling parameter $g = 6.0$, mass parameter $\mu = 1.0$ and starting point of field configuration $\phi_0 = 2.0$.}
  \label{fig:hmc-susy-dH}
  \eec
\end{figure}

The Monte Carlo time history of the ground state energy $E_0$ of this model is shown in Fig. \ref{fig:hmc-susy-history}. In Fig. \ref{fig:hmc-susy-dH} we show $\exp(-\Delta H)$ against MC time history. 

\begin{figure}[h]
\bec
  \includegraphics[width=9cm]{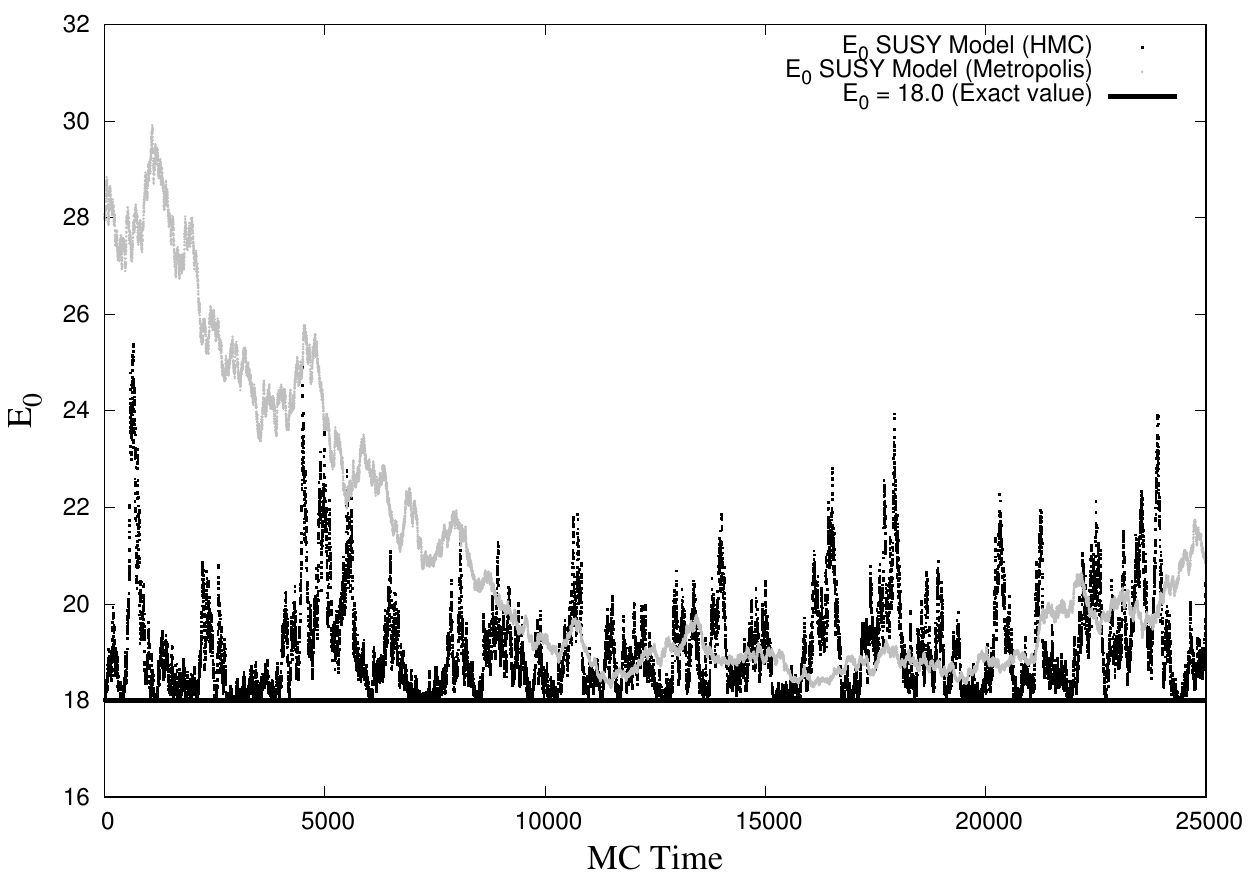}
  \caption{Comparing the Monte Carlo time histories of the ground state energy $E_0$ of the supersymmetric model, with $g = 6.0$ and $\mu = 1.0$, for Metropolis and HMC algorithms. In both the simulations we used a step size $\epsilon = 0.005$ and $\phi_0 = 0.5$ as the starting point of the simulation. For HMC we used the number of leapfrog steps $L = 15$ for each molecular dynamics trajectory. The classical result for the ground state energy is $E_0 = 18$. It is evident that the simple Metropolis algorithm gives data that suffer from large auto-correlation.}
  \label{fig:hmc-metro-susy-history}
  \eec
\end{figure}

In Fig. \ref{fig:hmc-metro-susy-history} we compare the Monte Carlo time histories of the ground state energy $E_0$ of the supersymmetric model, with $g = 6.0$ and $\mu = 1.0$, for Metropolis and HMC algorithms. In both the simulations we used 50000 Monte Carlo steps with a step size $\epsilon = 0.005$ and $\phi_0 = 2.0$ as the starting point of the simulation. For HMC we used the number leapfrog steps $L = 15$ for each molecular dynamics trajectory. The classical value for the ground state energy is $E_0 = 18$. Clearly, the Metropolis algorithm gives the simulation data with a large thermalization time and it also suffers from large auto-correlation. The HMC data shows a very short thermalization history and auto-correlation.

\section{MCMC and Quantum Field Theories on a Lattice}
\label{sec:MCMC-and-quantum-field-theories-on-a-lattice}

We can consider relativistic quantum field theory as the quantum mechanics of fields defined on a spacetime. A field has infinite number of degrees of freedom since it can take a value at every point in spacetime.

We can think of defining a quantum field theory by starting from another quantum field theory with a finite number of degrees of freedom. We can consider field variables taking values at a finite set of discrete points within a finite volume. A quantum field theory made out of these fields will have finite number of degrees of freedom. The points in finite volume can be taken as lattice sites of a hypercubic lattice with periodic boundary conditions (a hyper-torus). 

For a four-dimensional theory, the lattice $\Lambda$ can be defined as
\bea
&& \Lambda = \Big\{ n = (n_1, n_2, n_3, n_4) ~|~ n_1, n_2, n_3 = 0, 1, 2, \cdots, N_s - 1, \nn \\
&& ~~~~~~~~ n_4 = 0, 1, 2, \cdots, N_\tau - 1 \Big\},
\eea
where $N_s$ and $N_\tau$ are the total number of sites along spatial and temporal directions, respectively. The separation between two neighboring sites gives the lattice spacing $a$. The fundamental elements of a lattice are the {\it sites} (points) and the {\it links} connecting neighboring sites. In lattice gauge theories, such as lattice QCD, the {\it plaquettes} consisting of an oriented closed path of four links play a crucial role. 

In order to define the quantum field theory in spacetime continuum we need to perform the {\it continuum limit} (the spacing of the lattice points goes to zero) and {\it infinite volume limit} (the extensions of the hyper-torus grow to infinity).

From a mathematical point of view we can simplify a lot of our calculations if we consider the time variable to be purely imaginary, instead of considering it as real, and work with the resulting Euclidean spacetime. As a result, the Lorentz symmetry of the original theory becomes the compact symmetry of four-dimensional rotations.

Quantum field theory with imaginary time is equivalent to the (classical) statistical physics of the fields. The correspondence is more clear when we consider quantum field theories in the Feynman path integral formalism. There, the Euclidean lattice action becomes the exponent in the Boltzmann factor. 

The definition of QFT on a Euclidean spacetime lattice provides a {\it non-perturbative regularization} of the theory. We need not have to worry about the infinities in quantum field theories since the lattice spacing acts as a UV cutoff in the theory. In perturbation theory we need to take care of the infinities using the renormalization procedure. Note that we can also define perturbation theory on the lattice. Thus lattice also gives an alternative regularization for perturbation theory.

The expectation value of an observable $O$, which is made out of the microscopic fields $\Phi$ of the theory, takes the following form in the Euclidean path integral formalism 
\beq
\label{eq:exp-val-latt}
\langle O \rangle =  \int [d\Phi] e^{- S[\Phi]} O(\Phi),
\eeq    
where the partition function is defined as
\beq
Z = \int [d\Phi] e^{- S[\Phi]},
\eeq 
with $S[\Phi]$ denoting the lattice action, which is assumed to be a real function of the field variables. 

The above expressions show that the Euclidean path integral formalism of lattice field theory is indeed equivalent to the statistical physics of fields.

A typical lattice action contains a summation over the lattice sites $n$. Typically, in a theory like QCD, the number of lattice points would be large and thus there would be a large number of integration variables. Note that Eq. \eqref{eq:exp-val-latt} corresponds to a statistical system with a large number of degrees of freedom. Looking at it from the view point of path integrals, we see that only a small vicinity of the minimum of the ``free energy" density will predominantly contribute to the integral. Such a situation calls for the need to compute the integrals using Monte Carlo method.

As an illustration let us look at QCD on a lattice. (See Refs. \cite{Montvay:270707, Rothe:1492203, DeGrand:2006zz, Gattringer:2010zz} for a few standard text books on lattice field theories, including lattice QCD.)

The Lagrangian density for continuum Euclidean QCD has the form
\beq
{\cal L} = - \frac{1}{2} {\rm Tr}~F_{\mu \nu} F^{\mu \nu} + \sum_{k=1}^{N_f} {\rm Tr} ~\left\{ \psib_k(x) (\slashed{D} + m_k) \psi_k(x) \right\}, 
\eeq
with $\psi_k(x)$ denoting the fermion field corresponding to a quark flavor $k$, with mass $m_k$; $\slashed{D} = \gamma^\mu (\partial_\mu - i g A_\mu(x))$ with $A_\mu$ denoting the gluon field; and $g$ is the coupling parameter. 

In terms of the generators $\lambda^a$ of $SU(3)$, with the normalization, ${\rm Tr} (\lambda^a \lambda^b) = \hf \delta^{ab}$, the gauge field can be decomposed as
\beq
A_\mu(x) = \sum_{a=1}^8 \lambda^a A_\mu^a(x).
\eeq 

The field strength tensor has the form
\beq
F_{\mu\nu}(x) = \sum_{a=1}^8 \lambda^a F_{\mu\nu}^a(x).
\eeq
In terms of the gauge field it takes the form
\beq
F_{\mu \nu}^a = \partial_\mu A_\nu^a - \partial_\nu A_\mu^a + g f_{abc} A_\mu^b A_\nu^c,
\eeq
with $f_{abc}$ denoting the structure constants of $SU(3)$.

On a hypercubic lattice $\Lambda$, the fermionic degrees of freedom are placed on the lattice sites
\beq
\psi(n), ~\psib(n) , ~~n \equiv (n_1, n_2, n_3, n_4) \in \Lambda.
\eeq

The gluons live on the links and they are represented by the group valued link field
\beq
U_\mu(n) = \exp \left[ i g a A^b_\mu(n)\lambda_b \right],
\eeq
with $A_\mu(n)$ denoting the algebra valued lattice gauge fields. The link field $U_\mu(n)$ lives on an oriented link starting from site $n$ and ending at site $n + \hat{\mu}$ along the $\mu$-th direction. The link variables are considered as the fundamental variables, which are integrated over in the path integral.

\begin{figure}[h]
\bec
  \includegraphics[width=9cm]{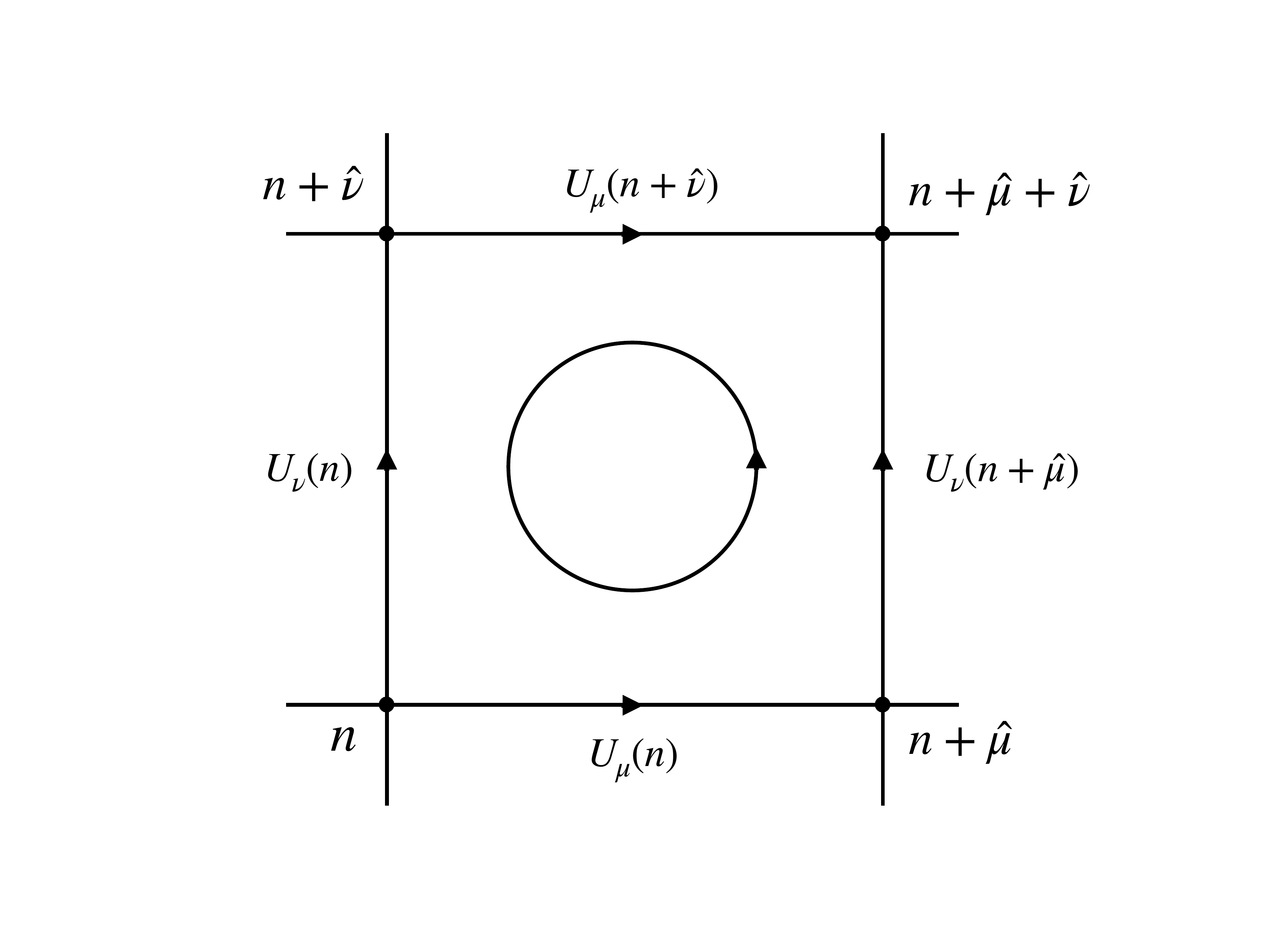}
  \caption{The shortest closed oriented loop on the lattice, the plaquette, is constructed out of four link variables.}
  \label{fig:plaquette}
  \eec
\end{figure}

For the gluon action we can use the shortest nontrivial closed loop on the lattice, called the {\it plaquette}. The plaquette variable $U_{\mu\nu}(n)$ is a product of four link variables defined as (see Fig. \ref{fig:plaquette})
\bea
U_{\mu\nu}(n) &=& U_\mu(n) U_\nu(n + \hat{\mu}) U^\dagger_\mu(n + \hat{\nu}) U^\dagger_\nu(n).
\eea

The gauge action, originally proposed by Wilson has the form
\bea
S_G[U] &=& \frac{2}{g^2} \sum_{n \in \Lambda} \sum_{\mu < \nu} {\rm Re} ~{\rm Tr} \left[ 1 - U_{\mu\nu}(n) \right] \nn \\
& \xrightarrow[a \to 0] { } & \hf \int_0^\beta d\tau \int d^3x \; {\rm Tr} ~ \left( F_{\mu\nu}^2 \right) + {\cal O}(a^2),
\eea
with $\beta$ denoting the circumference of the temporal circle of the 4-torus.
 
The lattice action has the generic form
\beq
S[U, \psi, \bar{\psi}] = S_G[U] + S_F(U, \psi, \bar{\psi}).
\eeq

The fermionic part of the action $S_F$ is quadratic in the Grassmann variables of the fermion fields
\beq
S_F = \sum_{m, n} \bar{\psi}_n M^F_{nm} \psi_m,
\eeq
with $M^F_{nm}$ denoting the fermion operator.

The expectation value of an observable $O$ has the general form
\beq
\langle O \rangle = \frac{1}{Z} \int [DU] [D\psib] [D \psi] e^{-S_G - S_F} O[U, \psi, \psib],
\eeq
where the partition function $Z$ is
\beq
Z = \int [DU] [D\psib] [D \psi] e^{-S_G - S_F}.
\eeq

In Lattice QCD, we compute the expectation values of observables given in Eq. \eqref{eq:exp-val-latt} by Monte Carlo integration
\beq
\langle O \rangle \approx \frac{1}{N} O(\Phi_i).
\eeq 
The sequence of field configurations $\cdots \to \Phi_{i-1} \to \Phi_i \to \Phi_{i+1} \to \cdots$ is drawn from a distribution in which the exponential of the Euclidean action plays the role of the probability weight.

\section{Machine Learning and Quantum Field Theories}
\label{sec:Machine-learning-and-QFT}

The intriguing connection between Euclidean quantum field theory and statistical mechanics has opened up a vast array of new insights and results benefitting both the research communities. This connection can be revealed through the Euclidean path integral formalism of quantum field theory. We can show that:
\bea
&& \textrm{Feynman weight for amplitudes } \exp(-S/\hbar) \nn \\
&& \hspace{4cm} \longleftrightarrow \textrm{Boltzmann factor } \exp(-\beta H) \nn \\
&& \textrm{Vacuum-to-vacuum amplitude } \int [D\phi] \exp(- S[\phi]/\hbar) \nn \\
&& \hspace{4cm} \longleftrightarrow \textrm{Partition function } Z = \sum_{\rm configurations} \exp(- \beta H) \nn \\
&& \textrm{Vacuum expectation value of observable } \langle 0 | O | 0 \rangle \nn \\
&& \hspace{4cm} \longleftrightarrow \textrm{Canonical ensemble average } \langle O \rangle \nn \\
&& \textrm{Changes in the field theory vacuum } \nn \\
&& \hspace{4cm} \longleftrightarrow \textrm{Phase transitions} \nn
\eea
and so son.

The Euclidean formulation of quantum field theory also allows us to use the machinery of Monte Carlo methods to extract non-trivial physics observables that are otherwise difficult to compute using analytical methods. Lattice regularized quantum field theories and their Monte Carlo simulations revealed to us many non-perturbative aspects; such as bound states, symmetry breaking, phase structure; of many interesting quantum field theories, including QCD. 

It turns out that Machine Learning also exhibits an interesting resemblance to quantum field theory and statistical mechanics. A dictionary would involve definitions such as: 
\bea
\textrm{Hamiltonian } &\longleftrightarrow& \textrm{Self-information or Surprisal} \nn \\
\textrm{Locality } &\longleftrightarrow& \textrm{Sparsity } \nn \\
\textrm{Irrelevant operator } &\longleftrightarrow& \textrm{Noise } \nn \\
\textrm{Relevant operator } &\longleftrightarrow& \textrm{Feature} \nn \\
\textrm{Effective theory } &\longleftrightarrow& \textrm{Nearly loss-less data distillation} \nn
\eea
and so on. (See Ref. \cite{Tegmark:2016} for more details.)    

The arena of Machine Learning (ML) could provide a promising way to find the order parameters of systems where they are hard to identify. It would be remarkable if it was possible to identify phases without prior knowledge of their existence or the underlying Hamiltonian. The order parameter can be determined by symmetry considerations of the underlying Hamiltonian in many models. However, there exist states of matter where such a parameter can only be defined in a complicated or nonlocal manner. These systems include topological states such as topological insulators, quantum spin hall states and quantum spin liquids. Therefore, we are in need for developing new novel methods to identify parameters capable of describing phase transitions. 

Recent developments in the implementation of Artificial Intelligence (AI) for physical systems, particularly those that can be formulated on a lattice, show promising evidence in identifying the underlying phase structures \cite{Carrasquilla_2017, 2017NatPh..13..435V, Wang_2016, Broecker_2017, Ch_ng_2017, JMLR:v11:3371, Wetzel_2017, Hu_2017, Kim_2018, Wetzel_2017a, Zhang_2019}. The methods such as the Principal Component Analysis \cite{2017NatPh..13..435V, Wang_2016, Hu_2017, Wetzel_2017a}, Supervised Machine Learning \cite{Broecker_2017, Zhang_2019, JMLR:v18:17-527} and auto-encoders \cite{JMLR:v11:3371, Wetzel_2017, Hu_2017} are shown to be able to identify different phases of classical statistical systems, such as the two-dimensional Ising model. These techniques have also been applied on quantum statistical systems, such as the Hubbard model \cite{Ch_ng_2017}, which describes the transition between conducting and insulating systems. 

Similar investigations were carried out in the context of quantum field theory on a lattice, such as the $SU(2)$ gauge theory \cite{Wetzel_2017a}. Recently, it has been shown that there exists a very interesting connection between renormalization group flow in quantum field theories \cite{Beny:2013, Stoudenmire:2016} and deep neural networks \cite{mehta2014exact}. In statistical mechanics problems we use  renormalization methods to classify various phases of matter. In the context of ML, the renormalization group  flow can be thought of as solving the pattern-recognition problem of classifying the long-range behavior of various statistical systems. Exploring such connections further may lead us to some positive surprises.


\appendix

\section{C++ Codes}
\label{sec:codes}

\subsection{Random Numbers from a Uniform Distribution}
\label{sec:uniform-m-one-p-one}

A C++ program to generate a sequence of random numbers that are uniformly distributed in the interval $[-1, +1)$, is provided below. The code uses the function drand48() to generate random numbers with the default seed, which is 1. See Fig. \ref{fig:data-random} for the plots generated using this code.

\begin{lstlisting}
// Generating random numbers between -1 and +1
// using drand48()

#include <iostream>
#include <math.h>
#include <stdlib.h>
#include <fstream>
#include <iomanip>

using namespace std;

int main()
{
  cout.precision(6);
  cout.setf(ios::fixed | ios::showpoint);

  int i, n;
  double x, rand, I, stder, f_val, f_val2;

  static int first_time = 1;
  static ofstream f_data;

  if (first_time)
  {
    f_data.open("data.txt");
    if (f_data.bad())
    {
      cout << "Failed to open data file\n" << flush;
    }
    first_time = 0;
  }

  cout << "Number of sample points" << endl;
  cin >> n;

  f_val  = 0.0;
  f_val2 = 0.0;

  for (i = 1; i <= n; i++)
  {
    rand = drand48(); // random number between 0 and 1
    // default seed is 1
    x = 2.0 * (rand - 0.5);
    f_data << i << "\t" << x << endl;

    f_val = x;
    f_val2 = f_val2 + x*x;
  }

  I = f_val*2.0/n;

  // evaluate standard deviation error
  f_val = f_val/n;
  f_val2 = f_val2/n;

  stder = 2.0*sqrt((f_val2 - f_val*f_val)/n);

  cout << setw(8) << n << setw(12)
  << I << setw(12) << stder << endl;

  return 0;
}
\end{lstlisting}

\subsection{Random Numbers with a Seed}
\label{sec:uniform-m-one-p-one-seed}

A C++ program to generate random numbers from a uniform distribution in the interval $[-1, +1)$ using the function drand48(), and with the seed function srand48() is provided below. We used the seed value 41 to generate the sequence of random numbers. See Fig. \ref{fig:data-random-w-seed} for the plots generated using this program.

\begin{lstlisting}
// Generating random numbers between -1 and +1
// using drand48() and seed function srand48().

#include <iostream>
#include <math.h>
#include <stdlib.h>
#include <fstream>
#include <iomanip>

using namespace std;

int main()
{
  cout.precision(6);
  cout.setf(ios::fixed | ios::showpoint);

  int i, n;
  double seed, x, rand, I;
  double f_val, f_val2, stder;

  static int first_time = 1;
  static ofstream f_data;

  if (first_time)
  {
    f_data.open("data.txt");
    if (f_data.bad())
    {
      cout << "Failed to open data file\n" << flush;
    }
    first_time = 0;
  }

  cout << "Random seed" << endl;
  cin >> seed;

  cout << "Number of sample points" << endl;
  cin >> n;

  f_val  = 0.0;
  f_val2 = 0.0;
  // Initilized seed for random number generator
  srand48(seed);

  for (i = 1; i <= n; i++)
  {
    rand = drand48(); // random number between 0 and 1
    x = 2.0 * (rand - 0.5);
    f_data << i << "\t" << x << endl;

    f_val = x;
    f_val2 = f_val2 + x*x;
  }

  I = f_val*2.0/n;

  // evaluate standard deviation error
  f_val = f_val/n;
  f_val2 = f_val2/n;

  stder = 2.0*sqrt((f_val2 - f_val*f_val)/n);

  cout << setw(8) << n << setw(12) << I
  << setw(12) << stder << endl;

  return 0;
}
\end{lstlisting}

\subsection{Random Numbers from a Gaussian Distribution}
\label{sec:gauss-random}

A C++ program to generate random numbers from a Gaussian distribution with mean 0 and width 1, using the Box-Muller method is provided below. See Fig. \ref{fig:data-gauss-random-w-seed} for the plots generated using this program.

\begin{lstlisting}
// Generating Gaussian random numbers
// using gaussian_rand() and
// seed function srand().

#include <iostream>
#include <math.h>
#include <stdlib.h>
#include <fstream>
#include <iomanip>

using namespace std;

double gaussian_rand();

int main()
{
  cout.precision(6);
  cout.setf(ios::fixed | ios::showpoint);

  int i, n;
  double seed, x, rand, I;
  double f_val, f_val2, stder;

  static int first_time = 1;
  static ofstream f_data;

  if (first_time)
  {
    f_data.open("data.txt");
    if (f_data.bad())
    {
      cout << "Failed to open data file\n" << flush;
    }
    first_time = 0;
  }

  cout << "Random seed" << endl;
  cin >> seed;

  cout << "Number of sample points" << endl;
  cin >> n;

  f_val  = 0.0;
  f_val2 = 0.0;
  // initilized seed for random number generator
  srand(seed);

  for (i = 1; i <= n; i++)
  {
    rand = gaussian_rand(); // call to function
    x = rand;
    f_data << i << "\t" << x << endl;

    f_val = x;
    f_val2 = f_val2 + x*x;
  }

  I = f_val*2.0/n;

  // evaluate standard deviation error
  f_val = f_val/n;
  f_val2 = f_val2/n;

  stder = 2.0*sqrt((f_val2 - f_val*f_val)/n);

  cout << setw(8) << n << setw(12) << I
  << setw(12) << stder << endl;

  return 0;
}

// double gaussian_rand()
// Gaussian distributed random number
// Probability distribution is exp( -x*x/2 ),
// so that < x^2 > = 1
// Uses rand() random number generator

double gaussian_rand(void)
{
  static int iset = 0;
  static double gset;
  double fac, rsq, v1, v2;
  if (iset == 0)
  {
    do
    {
      v1 = 2.0*rand()/(double)RAND_MAX - 1.0;
      v2 = 2.0*rand()/(double)RAND_MAX - 1.0;
      rsq = v1*v1 + v2*v2;
    }
    while(rsq >= 1.0 || rsq == 0.0);

    fac = sqrt(-2.0*log(rsq)/rsq);
    gset = v1*fac;
    iset = 1;
    return(v2*fac);
  }
  else
  {
    iset = 0;
    return(gset);
  }
}
\end{lstlisting}

\subsection{Numerical Integration - Composite Midpoint Rule}
\label{sec:num-int-compo-midpoint}

A C++ program that implements the numerical integration using composite midpoint rule is provided below.

\begin{lstlisting}
// Numerical integration
// Composite midpoint rule

#include <iostream>
#include <cmath>
#include <iomanip>

using namespace std;

double f(double);

int main()
{
  cout.precision(4); // set precision
  cout.setf(ios::fixed);

  double h;
  int i, m, a, b;

  cout << "Lower limit of integration" << endl;
  cin >> a;

  cout << "Upper limit of integration" << endl;
  cin >> b;

  cout << "Enter h value " << endl;
  cin >> h;

  double x = a, f_val = 0.0;
  m = (b-a)/h;

  cout << "Number of intervals m = "
  << m << endl;

  for(i = 1; i <= m; i++)
  {
    x = (i - 0.5)*h;
    f_val = f_val + h*f(x);
  }

  cout << "Integral for h = " << h
  << " is " << f_val << "\n";

  return 0;
}

// Function for integration
double f(double x)
{
  double y;
  y = (27.0/(2.0*3.1416*3.1416))*
  (x*exp(-x)*((1 - exp(-x)) /
              (1 + exp(-3*x))));
  return y;
}
\end{lstlisting}

\subsection{Numerical Integration - Composite Simpson's One-third Rule}
\label{sec:num-int-compo-simpsons}

A C++ program that implements the numerical integration using composite Simpson's one-third rule is provided below.

\begin{lstlisting}
// Numerical integration
// Composite Simpson's 1/3 rule

#include <iostream>
#include <cmath>
#include <iomanip>

using namespace std;

double f(double);

int main()
{
  cout.precision(4); // set precision
  cout.setf(ios::fixed);

  int i, m, a, b;

  cout << "Lower limit of integration" << endl;
  cin >> a;

  cout << "Upper limit of integration" << endl;
  cin >> b;

  cout << "Enter number of intervals " << endl;
  cin >> m;

  double f_val = 0.0, x, h;
  // for n odd - add -1 to interval to make it even
  if((m/2)*2 != m)
  {
    m = m - 1;
  }

  h = (double)(b-a)/m;

  f_val = f_val + ( f(a) + f(b) )*(h/3.0);

  for(i = 1; i <= (m/2); i++)
  {
    x = a + (2*i-1)*h;
    f_val = f_val + 4.0*f(x+h)*(h/3.0);
  }

  for(i = 1; i <= (m/2)-1; i++)
  {
    x = a + 2*i*h;
    f_val = f_val + 2.0*f(x+h)*(h/3.0);
  }

  cout << "Integral for h = "
  << h << " is " << f_val << "\n";

  return 0;
}

// Function for integration
double f(double x)
{
  double y;
  y = (27.0/(2.0*3.1416*3.1416))*
  (x*exp(-x)*((1 - exp(-x)) /
              (1 + exp(-3*x))));
  return y;
}
\end{lstlisting}

\subsection{Numerical Integration - Monte Carlo Method}
\label{sec:num-int-monte-carlo}

The C++ program provided below implements Monte Carlo sampling method to estimate the value of the integral. Note that the program implements the naive (independent) Monte Carlo sampling method. 

\begin{lstlisting}
// Numerical integration
// Monte Carlo method using
// naive sampling

#include <iostream>
#include <cmath>
#include <iomanip>

using namespace std;

double f(double);

int main()
{
  cout.precision(6);
  cout.setf(ios::fixed | ios::showpoint);

  int i, n;
  double a, b, stder;

  cout << "Lower limit of integration" << endl;
  cin >> a;

  cout << "Upper limit of integration" << endl;
  cin >> b;

  cout << "Number of sample points" << endl;
  cin >> n;

  cout << "    Points    "
  << "Integral   " <<  " Error" << endl;

  double I, x, rand;
  double f_val, f_val2;
  // f_val and f_val2 - for error estimation

  f_val  = 0.0;
  f_val2 = 0.0;

  for (i = 1; i <= n; i++)
  {
    rand = drand48(); // random number between 0.0 and 1.0
    x = a + (b-a)*rand; // scale it to the range we are in

    f_val = f_val + f(x);
    f_val2 = f_val2 + f(x)*f(x);
  }
  I = f_val*(b-a)/n;

  // evaluate integration error
  f_val = f_val/n;
  f_val2 = f_val2/n;
  stder = (b-a)*sqrt((f_val2 - f_val*f_val)/n);

  cout << setw(8) << n << setw(12)
  << I << setw(12) << stder << endl;

  return 0;
}

// Function for integration
double f(double x)
{
  double y;
  y = (27.0/(2.0*3.1416*3.1416))*
  (x*exp(-x)*((1 - exp(-x)) /
              (1 + exp(-3*x))));
  return y;
}
\end{lstlisting}

\subsection{Numerical Integration - Naive Monte Carlo Sampling}
\label{sec:num-int-monte-carlo-naive}

The program shown below produces Monte Carlo estimate of the integral
\beq
I = \int_{-\infty}^{\infty} \exp \left( - \hf x^2 + \qtr x - \frac{1}{32} \right) \; dx \nn
\eeq
using naive (independent) sampling.

\begin{lstlisting}
// Numerical integration
// Monte Carlo method using
// naive sampling

#include <iostream>
#include <cmath>
#include <iomanip>

using namespace std;

double f(double);

int main()
{
  cout.precision(6);
  cout.setf(ios::fixed | ios::showpoint);

  int i, n;
  double a, b, stder;

  cout << "Lower limit of integration" << endl;
  cin >> a;

  cout << "Upper limit of integration" << endl;
  cin >> b;

  cout << "Number of sample points" << endl;
  cin >> n;

  cout << "    Points    "
  << "Integral   " <<  " Error" << endl;

  double I, x, rand;
  double f_val, f_val2;
  // f_val and f_val2 - for error estimation

  f_val  = 0.0;
  f_val2 = 0.0;

  for (i = 1; i <= n; i++)
  {
    rand = drand48(); // random number between 0.0 and 1.0
    x = a + (b-a)*rand; // scale it to the range we are in

    f_val = f_val + f(x);
    f_val2 = f_val2 + f(x)*f(x);
  }
  I = f_val*(b-a)/n;

  // evaluate integration error
  f_val = f_val/n;
  f_val2 = f_val2/n;
  stder = (b-a)*sqrt((f_val2 - f_val*f_val)/n);

  cout << setw(8) << n << setw(12)
  << I << setw(12) << stder << endl;

  cout << sqrt(2.0*3.1416) << endl;
  return 0;
}

// Function for integration
double f(double x)
{
  double y;
  y = exp(-0.5*x*x + 0.25*x - (1.0/32));
  return y;
}
\end{lstlisting}

\subsection{Numerical Integration - Importance Sampling Monte Carlo}
\label{sec:num-int-monte-carlo-importance}

The program shown below produces Monte Carlo estimate of the integral
\beq
I = \int_{-\infty}^{\infty} \exp \left( - \hf x^2 + \qtr x - \frac{1}{32} \right) \; dx \nn
\eeq
using importance sampling.

\begin{lstlisting}
// Numerical integration
// Monte Carlo method using
// naive sampling

#include <iostream>
#include <cmath>
#include <iomanip>

using namespace std;

double f(double);

int main()
{
  cout.precision(6);
  cout.setf(ios::fixed | ios::showpoint);

  int i, n;
  double a, b, stder;

  cout << "Lower limit of integration" << endl;
  cin >> a;

  cout << "Upper limit of integration" << endl;
  cin >> b;

  cout << "Number of sample points" << endl;
  cin >> n;

  cout << "    Points    "
  << "Integral   " <<  " Error" << endl;

  double I, x, rand;
  double f_val, f_val2;
  // f_val and f_val2 - for error estimation

  f_val  = 0.0;
  f_val2 = 0.0;

  for (i = 1; i <= n; i++)
  {
    rand = drand48(); // random number between 0.0 and 1.0
    x = a + (b-a)*rand; // scale it to the range we are in

    f_val = f_val + f(x);
    f_val2 = f_val2 + f(x)*f(x);
  }
  I = f_val*(b-a)/n;

  // evaluate integration error
  f_val = f_val/n;
  f_val2 = f_val2/n;
  stder = (b-a)*sqrt((f_val2 - f_val*f_val)/n);

  cout << setw(8) << n << setw(12)
  << I << setw(12) << stder << endl;

  cout << sqrt(2.0*3.1416) << endl;
  return 0;
}

// Function for integration
double f(double x)
{
  double y;
  y = exp(-0.5*x*x + 0.25*x - (1.0/32));
  return y;
}
\end{lstlisting}

\subsection{Metropolis Algorithm for Gaussian Model}
\label{sec:simple-metro-x-xsq}

The C++ program provided below computes 
\beq
\langle x \rangle = \frac{\int dx ~ x e^{-x^2}}{\int dx ~ e^{-x^2}}. 
\eeq
and
\beq
\langle x^2 \rangle = \frac{\int dx ~ x^2 e^{-x^2}}{\int dx ~ e^{-x^2}}. 
\eeq
and their respective Monte Carlo errors using Metropolis sampling.

\begin{lstlisting}
// Code to compute <x> and <x^2> of a Gaussian
// with respective errorbars
// using Metropolis sampling

#include <iostream>
#include <iomanip>
#include <cstdlib>
#include <fstream>

#include <math.h>

using namespace std;

int main()
{
  int i, N;
  double EPS;
  double u, x, dx, x_new;

  double x_val = 0.0, x_sq_val = 0.0;
  double x_val_e = 0.0, x_sq_val_e = 0.0;
  double avg_x_val = 0.0, avg_x_sq_val = 0.0;
  double std_err_x_val = 0.0, std_err_x_sq_val = 0.0;

  // initilize x value
  x = 0.0;
  // simulation parameters
  N = 100000; // number of samples
  EPS = 0.75; // Metropolis step size

  for (i = 0; i < N; i++) {
    dx = drand48() - 0.5; // random jump in x
    x_new = x + EPS*dx; // proposed value of x

    u = drand48();

    // Metropolis update with weight exp(-x^2)
    if (u < exp(-(x_new*x_new - x*x)))
      x = x_new;

    x_val = x_val + x;
    x_val_e = x_val_e + x*x;

    x_sq_val = x_sq_val + (x*x);
    x_sq_val_e = x_sq_val_e + (x*x)*(x*x);
  }

  avg_x_val = x_val/N;
  avg_x_sq_val = x_sq_val/N;

  x_val_e = x_val_e/N;
  x_sq_val_e = x_sq_val_e/N;

  // Standard error
  std_err_x_val = sqrt((x_val_e
                        - pow(avg_x_val, 2))/N);
  std_err_x_sq_val = sqrt((x_sq_val_e
                           - pow(avg_x_sq_val, 2))/N);

  cout << "<x>: " << avg_x_val << "\t"
  << std_err_x_val << endl;
  cout << "<x^2>: " << avg_x_sq_val << "\t"
  << std_err_x_sq_val << endl;

  return 0;
}
\end{lstlisting}

\subsection{Supersymmetric Model - Metropolis Sampling}
\label{sec:susy-metropolis}

The C++ code provided below computes the ground state energy $E_0$ of the supersymmetric model given in Eq. \eqref{eq:susy-model}, and thus the possibility of dynamical supersymmetry breaking in the model.  

\begin{lstlisting}
// Metropolis algorithm for
// a supersymmetric model

#include <iostream>
#include <iomanip>
#include <cstdlib>
#include <fstream>

#include <math.h>

using namespace std;

double RandomNumber();

int main(void)
{
  static int first_time = 1;
  static ofstream f_obs;

  if(first_time)
  {
    f_obs.open("obs.txt");
    if(f_obs.bad())
    {
      cout << "Failed to open observable file\n" << flush;
    }
    first_time = 0;
  }

  int THERM, SWEEPS, GAP, sweep;
  int count = 0, accept = 0;
  double g, mu, u;
  double EPS;
  double W_p = 0.0, W_pp = 0.0;
  double phi, phi_new, S_old, S_new, dS, S_B;

  // Simulation parameters
  EPS = 0.005;
  THERM = 0;
  SWEEPS = 50000;
  GAP = 1;

  double N = SWEEPS/GAP;
  double b_act = 0.0, b_act_e = 0.0, std_err_b_act = 0.0;
  double acc_rate = 0.0, avg_acc_rate = 0.0;
  double tot_count;

  // Physics parameters - coupling and mass
  g = 6.0;
  mu = 1.0;
  // Initilize field
  phi = 0.5;

  for (sweep = 1; sweep <= THERM; sweep++)
  {
    W_p = g * (phi*phi + mu*mu);
    W_pp = 2.0 * g * phi;
    S_old = 0.5 * W_p * W_p - log(W_pp);

    // generate new field at site x from uniform distribution
    phi_new = phi + EPS*(RandomNumber()-0.5);

    W_p = g * (phi_new*phi_new + mu*mu);
    W_pp = 2.0 * g * phi_new;

    S_new = 0.5 * W_p * W_p - log(W_pp);

    dS = S_new - S_old;

    // metropolis sampling
    // update with probability exp(-dS)
    double u = RandomNumber();
    if( exp(-dS) > u )
      phi = phi_new;
  }

  for (sweep = 1; sweep <= SWEEPS; sweep++)
  {
    W_p = g * (phi*phi + mu*mu);
    W_pp = 2.0 * g * phi;
    S_old = 0.5 * W_p * W_p - log(W_pp);

    // generate new field at site x from uniform distribution
    phi_new = phi + EPS*(RandomNumber()-0.5);

    W_p = g * (phi_new*phi_new + mu*mu);
    W_pp = 2.0 * g * phi_new;

    S_new = 0.5 * W_p * W_p - log(W_pp);

    dS = S_new - S_old;

    // metropolis sampling
    // update with probability exp(-dS)
    double u = RandomNumber();
    if( exp(-dS) > u )
    {
      phi = phi_new;
      accept++;
    }
    count++;

    if(count%100 == 0)
    {
      acc_rate = double(accept)/count;
      cout << "Acceptance rate = " << acc_rate << endl;
      avg_acc_rate = avg_acc_rate + acc_rate;
      tot_count++;
      count = 0;
      accept = 0;
    }

    if(sweep%GAP == 0)
    {
      S_B = 0.5 * pow(g * (phi * phi + mu * mu), 2);

      b_act = b_act + S_B;
      b_act_e = b_act_e + S_B*S_B;

      f_obs << sweep << "\t" << S_B << "\n";
    }
  }
  f_obs << endl;

  avg_acc_rate = avg_acc_rate/tot_count;

  b_act = b_act/N;
  b_act_e = b_act_e/N;

  // Standard error
  std_err_b_act = sqrt((b_act_e - pow(b_act, 2))/N);

  cout << "\nStep size and average acceptance:" << endl;
  cout << EPS << "\t" << avg_acc_rate << endl;

  cout << "\nBosonic action S_B and error: " << endl;
  cout << b_act << "\t" << std_err_b_act
  << "\n" << endl;

  return 0;
}

double RandomNumber(void)
{
  double r = rand()/(double)RAND_MAX;
  return(r);
}
\end{lstlisting}

\subsection{Metropolis for Simple Harmonic Oscillator}
\label{sec:sho-metro}

The C++ program provided below implements the simple harmonic oscillator using Metropolis algorithm. 

\begin{lstlisting}
// MCMC code for one-dimensional
// simple harmonic oscillator
// Using Metropolis update

#include <iostream>
#include <math.h>
#include <stdlib.h>
#include <fstream>

using namespace std;

int main()
{
  static int first_time = 1;
  static ofstream f_data, f_site, f_a_rate;

  if (first_time)
  {
    f_data.open("corr.txt");
    if (f_data.bad())
    {
      cout << "Failed to open correlator file\n"
      << flush;
    }

    f_a_rate.open("acceptance.txt");
    if (f_a_rate.bad())
    {
      cout << "Failed to open acceptance rate file\n"
      << flush;
    }

    f_site.open("site.txt");
    if (f_site.bad())
    {
      cout << "Failed to open sites data file\n"
      << flush;
    }

    first_time = 0;
  }

  // simulation parameters
  int THERM = 1000000; // number of thermalization steps
  int SWEEPS = 1000000; // number of generation steps
  int GAP = 100; // interval between measurements
  double DELTA = 0.5; // random shift range
  //-DELTA <= delta <= DELTA
  double shift, u; // random shift, random number value
  double tot = 0.0;
  int accept = 0, no_calls = 0; // for acceptance rate

  // physics parameters
  int T = 64; // number of time slices
  double omega = 1.0; // frequency omega
  double m = 1.0; // mass m

  double dS; // change in action
  double site[T], old_site[T], new_site[T];

  // observables
  double corr[T]; // to store correlator data
  double corr_sq[T], std_err[T];
  double xsq = 0.0, xsq_sq = 0.0;
  double x_val = 0.0, x_val_sq = 0.0;
  double std_err_x_val, std_err_xsq;

  int tau; // to choose a random site

  // write out initially
  cout << "MCMC for Simple Harmonic Oscillator" << endl;
  cout << "Mass  m = " << m << endl;
  cout << "Frequency omega  = " << omega << endl;

  // initilize observables etc
  for (int t=0; t<T; t++)
  {
    site[t] = (drand48()-0.5);
    old_site[t] = 0.0;
    new_site[t] = 0.0;
    corr[t] = 0.0;
    std_err[t] = 0.0;
    corr_sq[t] = 0.0;
  }

  // begin thermalization MC sweeps
  for (int i=1; i<=THERM; i++)
  {
    // loop over time slices
    for (int t=0; t<T; t++)
    {
      // randomly choose a site
      tau = int(T*drand48());

      // store the current position at tau
      old_site[tau] = site[tau];

      // amount of random shift for position at tau
      shift = 2.0*DELTA*(drand48()-0.5);

      // propose a small change in position at tau
      new_site[tau] = site[tau] + shift;

      // compute change in action
      if (tau != (T-1))
      {
        dS = (pow((site[tau+1] - new_site[tau]), 2.0)
              + 0.25*omega*omega*
              pow((site[tau+1] + new_site[tau]),2.0))
        - (pow((site[tau+1] - old_site[tau]), 2.0)
           + 0.25*omega*omega*
           pow((site[tau+1] + old_site[tau]), 2.0));
        dS = (m/2.0)*dS;
      }
      else if (tau == (T-1))
      {
        dS = (pow((site[0] - new_site[tau]), 2.0)
              + 0.25*omega*omega*
              pow((site[tau+1] + new_site[tau]), 2.0))
        - (pow((site[0] - old_site[tau]), 2.0)
           + 0.25*omega*omega*
           pow((site[tau+1] + old_site[tau]), 2.0));
        dS = (m/2.0)*dS;
      }

      // Metropolis update
      u = drand48();
      if(u < exp(-dS))
      {
        site[tau] = new_site[tau];
        accept++;
        cout << "ACCEPTED with dS of " << dS << endl;
      }
      else
      {
        site[tau] = old_site[tau];
        cout << "REJECTED with dS of " << dS << endl;
      }
    }// end loop over time slices
  }// end thermalization MC steps

  // begin generation MC steps
  for (int i=1; i<=SWEEPS; i++)
  {
    // loop over time slices
    for (int t=0; t<T; t++)
    {
      no_calls++;
      if ((no_calls % 100 == 0) && (!first_time))
      {
        cout << "Acceptance rate "
        << (double)accept / (double)no_calls
        << "\n" << flush;

        // write out acceptance rate to a file
        f_a_rate << (double)accept / (double)no_calls << endl;

        no_calls = 0;
        accept = 0;
      }

      // randomly choose a site
      tau = int(T*drand48());

      // store current position at tau
      old_site[tau] = site[tau];

      // amount of shift for position at tau
      shift = 2.0*DELTA*(drand48()-0.5);

      // propose a small change to position at tau
      new_site[tau] = site[tau] + shift;

      // compute change in action
      if (tau != (T-1))
      {
        dS = (pow((site[tau+1] - new_site[tau]), 2.0)
              + 0.25*omega*omega*
              pow((site[tau+1] + new_site[tau]),2.0))
        - (pow((site[tau+1] - old_site[tau]), 2.0)
           + 0.25*omega*omega*
           pow((site[tau+1] + old_site[tau]), 2.0));
        dS = (m/2.0)*dS;
      }
      else if (tau == (T-1))
      {
        dS = (pow((site[0] - new_site[tau]), 2.0)
              + 0.25*omega*omega*
              pow((site[tau+1] + new_site[tau]), 2.0))
        - (pow((site[0] - old_site[tau]), 2.0)
           + 0.25*omega*omega*
           pow((site[tau+1] + old_site[tau]), 2.0));
        dS = (m/2.0)*dS;
      }

      // Metropolis update
      u = drand48();
      if(u < exp(-dS))
      {
        site[tau] = new_site[tau];
        accept++;
        cout << "ACCEPTED with dS of " << dS << endl;
      }
      else
      {
        site[tau] = old_site[tau];
        cout << "REJECTED with dS of " << dS << endl;
      }
    }// end loop over time slices

    if(i%GAP == 0)
    {
      tot++;
      // write out x[0] to a file
      f_site << tot << "\t" << site[0] << endl;

      // compute correlator, etc.
      for (int t=0; t<T; t++)
      {
        corr[t] = corr[t]
        + site[t]*site[0]/(2.0*m*omega);
        corr_sq[t] = corr_sq[t]
        + pow(site[t]*site[0]/(2.0*m*omega), 2.0);

        x_val = x_val + site[t];
        x_val_sq = x_val_sq + site[t]*site[t];

        xsq = xsq + site[t]*site[t]/(2.0*m*omega);
        xsq_sq = xsq_sq
        + pow(site[t], 4.0)/(pow(2.0*m*omega, 2.0));
      }
    }
  }// end generation MC steps

  // evaluate error in observables
  for (int t=0; t<T; t++)
  {
    corr[t] = corr[t]/tot;
    corr_sq[t] = corr_sq[t]/tot;

    std_err[t] = sqrt((corr_sq[t]
                       - corr[t]*corr[t])/tot);
  }

  xsq = xsq/(tot*T);
  xsq_sq = xsq_sq/(tot*T);

  x_val = x_val/(tot*T);
  x_val_sq = x_val_sq/(tot*T);

  std_err_xsq = sqrt((xsq_sq - xsq*xsq)/(tot*T));
  std_err_x_val = sqrt((x_val_sq
                        - x_val*x_val)/(tot*T));

  cout << "\n<x^2> = "
  << xsq << "\t" << std_err_xsq << "\n" << endl;

  cout << "\n<x> = "
  << x_val << "\t" << std_err_x_val << "\n" << endl;

  cout << "\nE_0 = m*omega^2*<x^2> = "
  << m*pow(omega, 2.0)*xsq << "\t"
  << m*pow(omega, 2.0)*std_err_xsq << "\n"
  << endl;

  // write out correlator to a file
  for (int t=0; t<T; t++)
  {
    f_data << t << "\t" << corr[t] << "\t"
    << std_err[t] << endl;
  }
  f_data << T << "\t" << corr[0] << "\t"
  << std_err[0] << endl;

  return 0;
}
\end{lstlisting}

\subsection{Metropolis for Unitary Matrix Model}
\label{sec:GWW-metro}

The C++ program provided below implements a unitary matrix model. It computes the Polyakov loop for a given value $g$ for the coupling parameter and rank $N$ of the unitary matrix.

\begin{lstlisting}
// Metropolis for Unitary Matrix Model
// Computes Polyakov loop for
// a given coupling g

#include <fstream>
#include <iostream>
#include <complex>
#include <stdlib.h>
#include <math.h>

using namespace std;

const int N = 50; // size of U matrix
const double g = 5.0; // coupling
int n, tau, SWEEPS, THERM, GAP;
double EPS, p = 0.0, p_avg = 0.0;
double p_sq = 0.0, p_std_err = 0.0;
complex <double> noise, P, invP;

int main()
{
  static ofstream f_act, f_a_rate, f_eigs;
  static ofstream f_p_line, f_p_inv_line;

  static int first_time = 1;
  if(first_time)
  {
    f_act.open("action.txt");
    if (f_act.bad())
    {
      cout << "Failed to open action file\n"
      << flush;
    }

    f_a_rate.open("acceptance.txt");
    if (f_a_rate.bad())
    {
      cout << "Failed to open acceptance rate file\n"
      << flush;
    }
    f_eigs.open("eigs.txt");
    if(f_eigs.bad())
    {
      cout << "Error opening eigs file\n"
      << flush;
    }
    f_p_line.open("p_line.txt");
    if(f_p_line.bad())
    {
      cout << "Error opening Poly line file\n"
      << flush;
    }
    f_p_inv_line.open("p_inv_line.txt");
    if(f_p_inv_line.bad())
    {
      cout << "Error opening Inv Poly line file\n"
      << flush;
    }
    first_time = 0;
  }

  SWEEPS = 10000;
  THERM = 5000;
  GAP = 100;
  EPS = 0.5;
  n = 0;
  double u, act, dS; // random number value
  int accept = 0, no_calls = 0; // for acceptance rate

  cout << "GWW MODEL" << endl;
  cout << "NUMBER OF COLORS: " << N << endl;
  cout << "SWEEPS: " << SWEEPS << endl;
  cout << "THERMALIZATION: " << THERM << endl;
  cout << "GAP: " << GAP << endl;
  cout << "STEP SIZE EPS: " << EPS << endl;

  double eta1 = 0.0, eta2 = 0.0;
  double theta_R = 0.0, theta_I = 0.0;
  double re = 0.0, im = 0.0;

  complex<double> theta_old[N], theta_new[N], M[N];
  complex<double> I(0.0,1.0), id(1.0,0.0);

  complex<double> S_old[N], S_new[N];
  complex<double> S_vdm_old[N], S_vdm_new[N];
  complex<double> S_total_old[N], S_total_new[N];

  // initilize angles
  for(int i=0; i<N; i++)
  {
    theta_R = 0.01*drand48();
    theta_I = 0.01*drand48();

    theta_old[i] = complex<double> (theta_R, theta_I);
  }

  // begin thermalization MC steps
  for(int dt=0; dt<THERM; dt++)
  {
    cout << "THERM dt " << "\t" << dt << endl;

    // compute old action
    for(int i=0; i<N; i++)
    {
      S_old[i] = complex<double> (0.0, 0.0);
      S_vdm_old[i] = complex<double> (0.0, 0.0);
      S_total_old[i] = complex<double> (0.0, 0.0);
    }

    for(int i=0; i<N; i++)
    {
      S_old[i] = (exp(I*theta_old[i])
                  + exp(-I*theta_old[i]));

      for(int j=0; j<N; j++)
      {
        if (j != i)
        {
          S_vdm_old[i] = S_vdm_old[i]
          - log(sin(0.5*(theta_old[i]
                         - theta_old[j])));
        }
      }
      S_total_old[i] = (N*g/2.0)*S_old[i]
      + S_vdm_old[i];
    }
    // end compute old action

    act = 0.0;
    for(int i=0;i<N;i++)
    {
      act = act + real(S_total_old[i]);
    }
    cout << "act_old = " << act/N << endl;

    // compute new angle
    tau = (int)(N*drand48()); // randomly select angle
    eta1 = 2.0*(drand48() - 0.5);
    eta2 = 0.0;
    noise.real(eta1);
    noise.imag(eta2);
    for(int i=0; i<N; i++)
    {
      theta_new[i] = theta_old[i];
      if(i==tau)
        theta_new[tau] = theta_old[tau] + EPS*noise;
    }

    // compute new action
    for(int i=0; i<N; i++)
    {
      S_new[i] = complex<double> (0.0, 0.0);
      S_vdm_new[i] = complex<double> (0.0, 0.0);
      S_total_new[i] = complex<double> (0.0, 0.0);
    }

    for(int i=0; i<N; i++)
    {
      S_new[i] = (exp(-I*theta_new[i])
                  + exp(I*theta_new[i]));

      for(int j=0; j<N; j++)
      {
        if (j != i)
        {
          S_vdm_new[i] = S_vdm_new[i]
          - log(sin(0.5*(theta_new[i]
                         - theta_new[j])));
        }
      }
      S_total_new[i] = (N*g/2.0)*S_new[i]
      + S_vdm_new[i];
    }
    // end compute new action

    act = 0.0;
    for(int i=0;i<N;i++)
    {
      act = act + real(S_total_new[i]);
    }
    cout << "act_new = " << act/N << endl;
    f_act << act/N << endl;

    // change in action
    dS = 0.0;
    for(int i=0;i<N;i++)
    {
      dS = dS + real(S_total_new[i])
      - real(S_total_old[i]);
    }

    // Metropolis update
    u = drand48();
    if(u < exp(-dS))
    {
      for(int i=0; i<N; i++)
      {
        theta_old[i] = theta_new[i];
      }

      accept++;
      cout << "ACCEPTED with dS of " << dS << endl;
    }
    else
    {
      cout << "REJECTED with dS of " << dS << endl;
    }
  } // end thermalization MC steps

  // begin generation MC steps
  n=0;
  for(int dt=0; dt<SWEEPS; dt++)
  {
    no_calls++;
    if ((no_calls % 100 == 0) && (!first_time))
    {
      cout << "Acceptance rate "
      << (double)accept / (double)no_calls
      << "\n" << flush;

      // write out acceptance rate to a file
      f_a_rate << (double)accept / (double)no_calls << endl;

      no_calls = 0;
      accept = 0;
    }

    cout << "SWEEP dt = " << dt << endl;

    // compute old action
    for(int i=0; i<N; i++)
    {
      S_old[i] = complex<double> (0.0, 0.0);
      S_vdm_old[i] = complex<double> (0.0, 0.0);
      S_total_old[i] = complex<double> (0.0, 0.0);
    }

    for(int i=0; i<N; i++)
    {
      S_old[i] = (exp(I*theta_old[i])
                  + exp(-I*theta_old[i]));

      for(int j=0; j<N; j++)
      {
        if (j != i)
        {
          S_vdm_old[i] = S_vdm_old[i]
          - log(sin(0.5*(theta_old[i]
                         - theta_old[j])));
        }
      }
      S_total_old[i] = (N*g/2.0)*S_old[i]
      + S_vdm_old[i];
    }
    // end compute old action

    act = 0.0;
    for(int i=0;i<N;i++)
    {
      act = act + real(S_total_old[i]);
    }
    cout << "act_old = " << act/N << endl;

    // compute new angle
    tau = (int)(N*drand48()); // randomly select angle
    eta1 = 2.0*(drand48() - 0.5);
    eta2 = 0.0;
    noise.real(eta1);
    noise.imag(eta2);
    for(int i=0; i<N; i++)
    {
      theta_new[i] = theta_old[i];
      if(i==tau)
        theta_new[tau] = theta_old[tau]
        + EPS*noise;
    }

    // compute new action
    for(int i=0; i<N; i++)
    {
      S_new[i] = complex<double> (0.0, 0.0);
      S_vdm_new[i] = complex<double> (0.0, 0.0);
      S_total_new[i] = complex<double> (0.0, 0.0);
    }

    for(int i=0; i<N; i++)
    {
      S_new[i] = (exp(-I*theta_new[i])
                  + exp(I*theta_new[i]));

      for(int j=0; j<N; j++)
      {
        if (j != i)
        {
          S_vdm_new[i] = S_vdm_new[i]
          - log(sin(0.5*(theta_new[i]
                         - theta_new[j])));
        }
      }
      S_total_new[i] = (N*g/2.0)*S_new[i]
      + S_vdm_new[i];
    }
    // end compute new action

    act = 0.0;
    for(int i=0;i<N;i++)
    {
      act = act + real(S_total_new[i]);
    }
    cout << "act_new = " << act/N << endl;
    f_act << act/N << endl;

    // change in action
    dS = 0.0;
    for(int i=0;i<N;i++)
    {
      dS = dS + real(S_total_new[i])
      - real(S_total_old[i]);
    }

    // Metropolis update
    u = drand48();
    if(u < exp(-dS))
    {
      for(int i=0; i<N; i++)
      {
        theta_old[i] = theta_new[i];
      }

      accept++;
      cout << "ACCEPTED with dS of " << dS << endl;
    }
    else
    {
      cout << "REJECTED with dS of " << dS << endl;
    }

    // begin measurements
    if(dt%GAP==0)
    {
      // print out eigenvalues
      P = 0.0;
      invP = 0.0;

      for(int i=0;i<N;i++)
      {
        M[i] = exp(I*theta_new[i]);

        re = real(M[i]);
        im = imag(M[i]);

        f_eigs << re << "\t " << im
        << "\t" << sqrt(re*re+im*im) << endl;

        P = P + exp(I*theta_new[i]);
        invP = invP + exp(-1.0*I*theta_new[i]);
      }
      P = (1.0/N)*P;

      n++;
      p = p + abs(P);
      p_sq = p_sq + abs(P)*abs(P);

      invP = (1.0/N)*invP;

      cout << "P = \t" << abs(P) << endl;
      cout << "invP = \t" << abs(invP) << endl;

      f_p_line << dt << "\t" << abs(P) << endl;
      f_p_inv_line << dt << "\t" << abs(invP) << endl;
    } // end measurements
  } // end generation MC steps

  p_avg = p/(double)n;
  p_sq = p_sq/(double)n;
  p_std_err = sqrt((p_sq - p_avg*p_avg)/(double)n);

  cout << "N = " << N << endl;
  cout << "g = " << g << endl;
  cout << "gN = " << g*N << endl;

  cout << "g, P, dP" << "\t" << g << "\t" << p_avg
  << "\t" << p_std_err << endl;

  return 0;
}
\end{lstlisting}

\subsection{Computing Auto-correlation Time}
\label{sec:auto-corr}

The C++ program provided below computes the auto-correlation time of a given observable. 

\begin{lstlisting}
// Code to compute auto-correlation time,
// tau_int and delta tau_int
// The code reads in initial observable data
// (one column data)
// from a file and writes out auto-correlation
// data to an out file

#include <iostream>
#include <fstream>
#include <cstdlib>

#include <cmath>

using namespace std;

int main(int argc, char * argv[])
{
  ofstream outdata;
  ifstream indata;

  indata.open(argv[1]);
  if( !indata )
  {
    cerr << "Error: file could not be opened"
    << endl;
    exit(1);
  }

  outdata.open(argv[2]);
  if( !outdata )
  {
    cerr << "Error: file could not be opened"
    << endl;
    exit(1);
  }

  int i, l, m;
  double tau_int;
  double num;
  int LEN, lag;

  cout << "Enter length of the file" << endl;
  cin >> LEN;

  double data[LEN], autocorr[LEN];

  for(i=0; i<LEN; i++)
  {
    data[i] = 0.0;
    autocorr[i] = 0.0;
  }

  for(i=0; i<LEN; i++)
  {
    indata >> num;
    data[i] = num;
  }
  indata.close();


  // returns auto-correlation
  for(lag=1; lag<LEN; lag++)
  {
    double avg = 0;
    double Gamma = 0, rho = 0;

    for(i=0; i<LEN; i++)
    {
      avg += data[i];
    }

    avg = avg/LEN;

    for(i=0; i<(LEN-lag); i++)
    {
      Gamma += (1.0/(LEN-lag))*(data[i] - avg)*
      (data[i+lag] - avg);
    }

    for(i=0; i<LEN; i++)
    {
      rho += (1.0/LEN)*(data[i] - avg)*
      (data[i] - avg);
    }

    autocorr[lag] = Gamma/rho;
  }

  for(m=1; m<LEN; m++)
  {
    tau_int = 0.5;
    for(l=1; l<m; l++)
    {
      tau_int = tau_int + autocorr[l];
    }

    if (m > (int)(4.0*tau_int + 1.0))
    {
      cout << "m = " << m << endl;
      cout << "tau_int = " << tau_int << endl;
      cout << "delta tau_int = "
      << sqrt((4.0*m + 2.0)/LEN)*tau_int << endl;
      break;
    }
  }

  int(cut) = 0.5*lag;
  for (i=0; i<cut; i++)
    outdata << autocorr[i] << endl;

  outdata.close();

  return 0;
}
\end{lstlisting}

\subsection{HMC for a Gaussian Model}
\label{sec:gaussian-hmc}

The C++ program provided implements HMC for a Gaussian model.

\begin{lstlisting}
// HMC for a Gaussian model

#include <iostream>
#include <iomanip>
#include <cstdlib>
#include <fstream>
#include <math.h>

using namespace std;

const int SWEEPS = 100000;
const int L = 20;
const double EPS = 0.2;

double gauss(void);
double action(const double);
double hamiltonian(const double, const double);
double force(const double);
int evolve(double&, double&, double&);

int main()
{
  double seed;
  seed = 41;
  cout << "Using random seed: " << seed << endl;

  // Initilize random seed
  srand48(seed);

  double phi;
  double phi_old;
  double H_i, H_f, r, phi_sq;
  double dH, expmdH;
  int sweep, count = 0, accept = 0;

  double obs1 = 0.0, obs1_e = 0.0, std_err_obs1 = 0.0;
  double obs2 = 0.0, obs2_e = 0.0, std_err_obs2 = 0.0;

  double acc_rate = 0.0, avg_acc_rate = 0.0;
  double tot_count;

  // Initial configuration for phi
  phi = 2.0;

  static int first_time = 1;
  static ofstream f_obs;

  if(first_time)
  {
    f_obs.open("obs.txt");
    if(f_obs.bad())
    {
      cout << "Failed to open observable file\n" << flush;
    }
    first_time = 0;
  }

  phi_sq = 0.0;

  for(sweep = 0; sweep != SWEEPS; sweep++)
  {
    phi_old = phi;
    evolve(phi, H_i, H_f);

    r = drand48();
    dH = H_f - H_i;
    expmdH = exp(-dH);
    if(expmdH > r)
    {
      // accept proposal
      accept++;
    }
    else
    {
      // reject proposal
      phi = phi_old;
    }
    count++;

    if(count%100 == 0)
    {
      acc_rate = double(accept)/count;
      cout << "Acceptance rate = " << acc_rate
      << endl;

      avg_acc_rate = avg_acc_rate + acc_rate;
      tot_count++;
      count = 0;
      accept = 0;
    }

    // phi square
    phi_sq = phi*phi;

    obs1 = obs1 + phi;
    obs1_e = obs1_e + phi*phi;

    obs2 = obs2 + (phi*phi);
    obs2_e = obs2_e + (phi*phi)*(phi*phi);

    // Write out phi, phi^2, exp(-dH)
    f_obs << sweep << "\t" << phi << "\t"
    << phi_sq << "\t" << expmdH << endl;
  }

  avg_acc_rate = avg_acc_rate/tot_count;

  obs1 = obs1/SWEEPS;
  obs1_e = obs1_e/SWEEPS;

  obs2 = obs2/SWEEPS;
  obs2_e = obs2_e/SWEEPS;

  // Standard error
  std_err_obs1 = sqrt((obs1_e - pow(obs1, 2))/SWEEPS);
  std_err_obs2 = sqrt((obs2_e - pow(obs2, 2))/SWEEPS);

  cout << "\nStep size and average acceptance:" << endl;
  cout << EPS << "\t" << avg_acc_rate << endl;

  cout << "\nphi and error: " << endl;
  cout << obs1 << "\t" << std_err_obs1 << "\n" << endl;

  cout << "\nphi_sq and error: " << endl;
  cout << obs2 << "\t" << std_err_obs2 << "\n" << endl;

  return 0;
}

// Gauss random
double gauss(void)
{
  static int iset = 0;
  static double gset;
  double fac, rsq, v1, v2;
  if(iset == 0)
  {
    do
    {
      v1 = 2.0*rand()/(double)RAND_MAX-1.0;
      v2 = 2.0*rand()/(double)RAND_MAX-1.0;
      rsq = v1*v1+v2*v2;
    }
    while(rsq>=1.0 || rsq == 0.0);

    fac = sqrt(-2.0*log(rsq)/rsq);
    gset = v1*fac;
    iset = 1;
    return(v2*fac);
  }
  else
  {
    iset = 0;
    return(gset);
  }
}

// Action
double action(const double phi)
{
  double S = 0.5*phi*phi;
  return S;
}

// Hamiltonian
double hamiltonian(const double phi, const double p_phi)
{
  double H;
  H = action(phi);
  H = H + 0.5*p_phi*p_phi;
  return H;
}

// Find force, dS/dphi
double force(const double phi)
{
  double dS_dphi = phi;
  return dS_dphi;
}

// Evolve phi
int evolve(double& phi, double& H_i, double& H_f)
{
  int i;
  double p_phi;
  double dS;
  p_phi = gauss();

  // calculate Hamiltonian
  H_i = hamiltonian(phi, p_phi);

  // first step of Leapfrog
  phi = phi + 0.5*EPS*p_phi;
  // Steps 2, 3, ... , L
  for(i = 1; i != L; i++)
  {
    dS = force(phi);
    p_phi = p_phi - dS*EPS;
    phi = phi + p_phi*EPS;
  }

  // last step of Leapfrog
  dS = force(phi);
  p_phi = p_phi - dS*EPS;
  phi = phi + p_phi*0.5*EPS;

  // calculate Hamiltonian again
  H_f = hamiltonian(phi, p_phi);

  return 0;
}
\end{lstlisting}



\end{document}